\documentclass[11pt]{article}
\usepackage{CJK}
\usepackage{graphicx,psfrag}
\usepackage{fancyhdr}
\usepackage{amsmath,amsthm,amssymb}
\usepackage{mathrsfs}
\usepackage{multirow,booktabs}
\usepackage{indentfirst}
\usepackage{pifont}
\usepackage{color}
\usepackage{amsfonts}
\usepackage{url}
\usepackage{epstopdf}
\def\hDash{\bot\!\!\!\bot}

\newtheorem{theorem}{Theorem}[section]
\newtheorem{prop}{Proposition}

\newtheorem{lemma}{Lemma}
\newtheorem{remark}{Remark}

\numberwithin{equation}{section}

\begin{document}

\title{Estimation and adaptive-to-model testing for regressions with  diverging number of predictors
\footnote{Lixing Zhu is a Chair professor of Department of Mathematics
at Hong Kong Baptist University, Hong Kong, China. He was supported by a grant from the
University Grants Council of Hong Kong, Hong Kong, China.}} 
\author{Falong Tan and Lixing Zhu \\~\\
Department of Mathematics, Hong Kong Baptist University, Hong Kong
}
\date{}
\maketitle

\renewcommand\baselinestretch{1.2}
{\small}

\begin{abstract}
The research described in this paper is motivated by model checking for parametric single-index models with  diverging number of predictors. To construct a test statistic, we first study the asymptotic property of the estimators of involved parameters of interest under the null  and alternative hypothesis when the dimension is divergent to infinity as the sample size goes to infinity. For the   testing problem, we study an adaptive-to-model residual-marked empirical process as the basis for constructing a test statistic. By modifying the approach in the literature to suit   the diverging dimension settings, we construct a martingale transformation. Under  the null, local and global  alternative hypothesis, the weak limits of the  empirical process are derived and then the asymptotic properties of the test statistic  are investigated. Simulation studies are carried out to examine the performance of the test.  \\

\end{abstract}
\newpage
\baselineskip=21pt

\newpage

\setcounter{equation}{0}
\section{Introduction}
Regression modelling is a vital problem in regression analysis. One important step in regression modelling is to check the adequacy of a model that would be used in further analysis to prevent possible wrong conclusions. There are a number of proposals available in the literature, which will be reviewed later. However, there is an important issue that has not been well studied. We notice that in high dimensional data analysis, the dimension $p$ of the predictor vector is often large even though it is still small compared with the sample size $n$. In this case, we often regard $p$ as a diverging number as $n$ goes to infinity. A relevant reference is Huber (1973) who considered a problem where $p$ goes to infinity at the rate of order $O(n^{1/4})$.   

In this paper, we focus on inference for  parametric single-index  models. Although they are in form  generalized linear models, we do not use this name as  generalized linear models have their own definitions in the literature. Let $Y$ be a response variable associated with a $p$-dimensional predictor vector $X \in \mathbb{R}^p$. If $Y$ is integrable, the regression function $g(x)=E(Y|X=x)$ is well-defined. Let $\mathcal{G}=\{ g(\beta^{\top} \cdot, \theta): \beta \in  \mathbb{R}^p, \theta \in  \mathbb{R}^d \}$ be a given parametric family of functions. The study herewith is motivated by checking whether $g(\cdot, \cdot)$ belongs to $\mathcal{G}$ or not. Thus the null hypothesis we want to test is that $(Y, X)$ follows a parametric single-index model as
\begin{equation}\label{1.1}
Y= g(\beta_{0}^{\top}X, \theta_{0}) +\varepsilon  \quad {\rm for \ some} \ \beta_{0} \in  \mathbb{R}^p,\ \theta_{0} \in \mathbb{R}^d,
\end{equation}
where $\varepsilon=Y-E(Y|X)$ is the error term, $d$ is fixed, $p$ diverges as the sample size $n$ tends to infinity, and $\top$ denotes the transposition.

We now review existing methodologies in the literature.  Two major classes of tests are: locally smoothing tests and globally smoothing tests. Locally smoothing tests use nonparametric smoothing estimators to construct test statistics; see H\"{a}rdle and Mammen (1993), Zheng (1996), Fan and Li (1996), Dette (1999), Fan and Huang (2001), Koul and Ni (2004), and Van Keilegom et al. (2008) as examples. Globally smoothing tests construct test statistics based on averages of  functionals of empirical processes and then  avoid nonparametric estimation. They are called globally smoothing tests as averaging is also a globally smoothing step.  Examples include Bierens (1982, 1990),  Stute (1997), Stute, Thies, and Zhu (1998), Stute et al. (1998),  Khmadladze and Koul (2004). 

All existing methods are limited to the fixed dimension settings. The extension to a diverging dimension case is by no means trivial. When the dimension $p$ is large, most existing tests,  especially locally smoothing tests, perform  badly. Stute and Zhu (2002) can be regarded as a dimension reduction-based test. A martingale transformation leads it to be asymptotically distribution-free. This test has been proved to be powerful in many cases, even when $p$ is large. But Stute and Zhu's (2002) test is not  omnibus, i.e., it fails to be consistent against all alternative hypotheses and thus is a directional test. Escanciano (2006) gave some detailed comments on this issue, and proposed, as well as
Lavergne and Patilea (2008, 2012), tests that are based on projected covariates. Guo et al. (2016) did it also and  put forward to a model adaptation notion in hypothesis testing. This innovative notion provides a deep insight into model checking for regressions and the adaptive-to-model approach can fully use the model structures under both the null and alternative hypothesis. Recently, with the help of sufficient dimension reduction techniques, Tan et al. (2017) generalized Stute and Zhu's (2002) method and obtained an omnibus test which is asymptotically distribution-free and inherits the  dimension reduction properties. It performs very well,  but still requires the condition that $p$ is fixed. In this paper,  we develop a consistent diagnostic test for checking the adequacy of a single-index model when the dimension $p$ of the predictor vector diverges to infinity as the sample size $n$ tends to infinity.

To make full use of the model structure under both the null hypothesis and the alternative hypothesis, we consider the following alternative model
\begin{equation}\label{1.2}
  Y=G(B^{\top}X)+\varepsilon.
\end{equation}
where $E(\varepsilon|X)=0$ and $G(\cdot)$ is an unknown smooth function and  $B$ is a $p \times q$ orthonormal matrix with an unknown $q$ with $1\leq q \leq p$. Note that this is a more general model of (\ref{1.2}) than the nonparametric model $Y=G(X)+\varepsilon$ as it is a special case when $B$ is an $p \times p$ orthonormal matrix with $q=p$.


Similarly as  Stute and Zhu (2002), we still use residual-marked empirical process and the martingale transformation to construct a test statistic when projected predictors vector is used. However, when the projected predictors vector under the null hypothesis is used to construct a test statistic as Stute and Zhu (2002) did, it cannot be an omnibus test. Stute et al (1998a) constructed a residual-marked empirical process by using the original predictors vector. When $p$ is divergent, the test severely suffers from the curse of dimensionality in theory. To alleviate these difficulties, we will adopt a model adaptation strategy as Tan et al (2017) did. It can adaptively uses projected predictors under the null and alternative hypothesis. Under the null, only one projected predictor is used like that in Stute and Zhu's construction, while under the alternatives, it can automatically uses all projections on $q$-dimensional unit sphere to guarantee the  omnibus property. Although this idea seems workable, the theoretical investigation, due to the dimensionality divergence, becomes very complicated. There are no no relevant results in the literature about the convergence of residual-marked empirical process with diverging $p$. Even when we can obtain its limiting Gaussian process, the shift term created by  estimating  the parameter of interest has no a simple formula so that we can easily motivate the  martingale transformation construction proposed by Stute, Thies, and Zhu (1998) to make the test asymptotically distribution-free.   This is a typical problem when $p$ is divergent, which does not happen when $p$ is fixed.

Therefore, the paper is then organized as follows.  Section 2 contains the asymptotic properties of the ordinary least squares estimator in the diverging dimension setting. Based on this, we define an adaptive-to-model residual-marked empirical process as the basis of the proposed test statistic. Since sufficient dimension reduction theory plays a crucial role to achieve the adaptive-to-model property, we give a brief review in this section and give the study on the convergence rate of the relevant estimators. In Section 3, we  present the limit  of the adaptive-to-model empirical process under the null hypothesis and give the investigation for its asymptotics. Then we use a modified approach to define a martingale transformation  because the shift term has no close form in the diverging dimension settings.   The asymptotic properties of the martingale transformation-based innovation process under both the null and alternatives are studied. We  also show that when $p$ is fixed, this transformation is equivalent to the Stute and Zhu's (2002) martingale transformation. In Section 4, we give the test statistic for practical use and then several simulation studies are conducted. A real data example is analysed in Section~5 for illustration. Section~6 contains a discussion. Technical proofs are deferred to Appendix.

\section{Adaptive-to-model residual-marked empirical process}
\subsection{Preliminary}
Let $\{ (X_1, Y_1), \cdots, (X_n, Y_n) \}$ be an i.i.d. sample with the same distribution as $(X,Y)$ and let $\varepsilon=Y-E(Y|X)$ be the unpredictable part of $Y$ given $X$.
Recall that $\mathcal{G}=\{ g(\beta^{\top} \cdot, \theta): \beta \in \mathbb{R}^p, \theta  \in \mathbb{R}^d \}$. We want to test whether or not
\begin{eqnarray*}
H_0:\ Y= g(\beta_{0}^{\top}x, \theta_{0}) + \varepsilon  \quad {\rm for \ some} \ \beta_{0} \in  \mathbb{R}^p,\ \theta_{0} \in \mathbb{R}^d.
\end{eqnarray*}
For estimating the unknown $(\beta_{0}, \theta_{0})$, we in this paper restrict ourselves to the ordinary least squares method. Let
\begin{eqnarray*}
(\hat{\beta}_n, \hat{\theta}_n) = \mathop{\rm argmin}\limits_{\beta, \theta} \sum_{i=1}^{n}[Y_i-g(\beta^{\top}X_i, \theta)]^2.
\end{eqnarray*}
To analyze the asymptotic property of $(\hat{\beta}_n, \hat{\theta}_n)$, define
\begin{eqnarray*}
 (\tilde{\beta}_{0}, \tilde{\theta}_{0}) = \mathop{\rm argmin}\limits_{\beta, \theta} E[Y-g(\beta^{\top}X, \theta)]^2.
\end{eqnarray*}
It is easy to see that if $g(\cdot, \cdot) \in \mathcal{G}$, we have $(\tilde{\beta}_{0}, \tilde{\theta}_{0})=(\beta_{0}, \theta_{0})$. If $g \notin \mathcal{G}$, $(\tilde{\beta}_{0}, \tilde{\theta}_{0})$ typically depends on the distribution of $X$. Let $e=Y-g(\tilde{\beta}_{0}^{\top}X, \tilde{\theta}_0)$. Then under the null hypothesis we have $e=\varepsilon$.

To study the asymptotic properties of $(\hat{\beta}_{n}, \hat{\theta}_{n})$ as $p$ is divergent, we first give some notations and the regularity conditions postpone to Appendix. Suppose that $g(\beta^{\top}x, \theta)$ is third differentiable with respective to $(\beta, \theta)$. Let
\begin{eqnarray*}
g'(\beta, \theta, x)=\frac{\partial g(\beta^{\top}x, \theta)}{\partial (\beta, \theta)}, \quad g''(\beta, \theta, x)=\frac{\partial g'(\beta, \theta, x)}{\partial (\beta, \theta)}. 
\end{eqnarray*}
The matrix $g''(\beta, \theta, x)$ is used in the following matrix $\Sigma_n$ which will play a crucial role in deriving the asymptotic properties of $(\hat{\beta}_n, \hat{\theta}_n)$:
\begin{eqnarray*}
\Sigma_n = E[g'(\tilde{\beta}_{0}, \tilde{\theta}_0, X) g'(\tilde{\beta}_{0}, \tilde{\theta}_0, X)^{\top}]-E[ e g''( \tilde{\beta}_{0},
               \tilde{\theta}_0, X)] =: \Sigma_{1n}- \Sigma_{2n}.
\end{eqnarray*}
The next two results give the norm consistency of $(\hat{\beta}_n, \hat{\theta}_n)$ with respective to $(\tilde{\beta}_{0}, \tilde{\theta}_0)$ and the decomposition of
$\left( \begin{array}{c}
\hat{\beta}_n-\tilde{\beta}_{0}\\
\hat{\theta}_n-\tilde{\theta}_0\\
\end{array}
\right)$
into independent and identically distributed summands. This decomposition generalizes the results of White (1981) to the case where the dimension $p$ of the predictor vector diverges. For simplicity, we define hereafter $\hat{\gamma}_n=(\hat{\beta}_n^{\top}, \hat{\theta}_n^{\top})^{\top}$, $\tilde{\gamma}_{0}=(\tilde{\beta}_{0}^{\top}, \tilde{\theta}_0^{\top})^{\top}$ and $\gamma_{0}=(\beta_{0}^{\top}, \theta_{0}^{\top})^{\top}$.

\begin{prop}\label{prop 1}
Suppose that conditions (A1)-(A6) in Appendix hold. If $p^4/n \to 0$, then $\hat{\gamma}_n$ is a norm consistent estimator of $\tilde{\gamma}_0$ in the sense that $\| \hat{\gamma}_n-\tilde{\gamma}_{0} \|=O_p(\sqrt{p/n})$, where $\|\cdot\|$ denotes the Frobenius norm.
\end{prop}

The convergence rate of order $\sqrt{p/n}$ is in line of the results of the M-estimator that was obtained by Huber (1973) and Portnoy (1984) when the number of parameters $p$ diverges. For the asymptotic decomposition, we have the following result.

\begin{prop}\label{prop 2}
If $p^5/n \rightarrow 0$ and conditions (A1)-(A6) in Appendix hold, we then have
\begin{equation}\label{2.1}
\hat{\gamma}_n-\tilde{\gamma}_{0}=\Sigma_n^{-1} \frac{1}{n} \sum_{i=1}^{n}[Y_i-g(\tilde{\beta}_{0}^{\top}X_i, \tilde{\theta}_0)]g'(\tilde{\beta}_{0}, \tilde{\theta}_0, X_i)+o_p(\frac{1}{\sqrt{n}}).
\end{equation}
\end{prop}

\begin{remark}
The rate $p^4/n \to 0$ or $p^5/n \to 0$ as $n \to \infty$ seems slow. According to the arguments for proving Propositions~1 and 2 in Appendix,
we can see that if $g(\beta^{\top}X, \theta)=\beta^{\top}X$ follows a linear model, then $g''(\beta, \theta, x)=0$ and $g'''(\beta, \theta, x)=0$. Thus we can obtain the norm consistency of $\hat{\gamma}_n$ to  $\tilde{\gamma}_{0}$ and the asymptotic decomposition of $\hat{\gamma}_n-\tilde{\gamma}_{0}$ under the  conditions $p^2/n \to 0$ and $p^3/n \to 0$, respectively. This condition is the same as that of Huber (1973) who only considered the linear model therein. Portnoy (1984, 1985) obtained the norm consistency and the asymptotic normality under  weaker conditions again for linear models. However, his conditions are hard to check in practice what kinds of models, other than linear models, can satisfy. Further,  extending their results to handle the parametric single-index models as we consider here is, to the best of our knowledge, still an open question.
\end{remark}

\subsection{Basic test statistic construction}
Recall the null hypothesis:
\begin{eqnarray*}
H_0:\  \mathbb{P}\{E(Y|X)= g(\beta_{0}^{\top}X, \theta_0)\}=1  \quad {\rm for \ some} \ \beta_{0} \in \mathbb{R}^p,\ \theta_0 \in \mathbb{R}^d,
\end{eqnarray*}
against the alternative hypothesis:
\begin{eqnarray*}
H_1: \mathbb{P}\{E(Y|X) = G(B^{\top}X) = g(\beta^{\top}X, \theta)\}<1 \quad \forall \ \beta \in \mathbb{R}^p,\ \theta \in \mathbb{R}^d
\end{eqnarray*}
where $G(\cdot)$ is an unknown smooth function and the $p \times q$ orthonormal matrix $B$ is given in (\ref{1.2}). We assume that $\tilde{\beta}_{0} \in \mathcal{S}_{E(Y|X)}$ under both the null and alternative hypothesis where $\mathcal{S}_{E(Y|X)}$ is the central mean subspace such that $\mathcal{S}_{E(Y|X)}={\rm span}(B)$.
Under the null hypothesis, this is obvious. Under the alternative hypothesis, $\tilde \beta_0$ would not necessarily parallel to $\beta_0$, but reasonably be a linear combination of all columns of the matrix $B$. 
Thus the assumption is not restrictive.

Also recall $\varepsilon=Y-E(Y|X)$ and $e=Y-g(\tilde{\beta}_{0}^{\top}X, \tilde{\theta}_0)$. Under the null hypothesis,  $e=\varepsilon, q=1$ and $B=\kappa \beta_{0}$ with $\kappa= \pm \frac{1}{\|\beta_{0}\|}$. Therefore, we obtain that $E(e|B^{\top}X)=E(e|\beta_{0}^{\top}X)=0$. Under the alternative hypothesis, we have $E(e|B^{\top}X)=G(B^{\top}X)-g(\tilde{\beta}_{0}^{\top}X, \tilde{\theta}_0) \neq 0 $. Then it follows that
under the null hypothesis
\begin{equation}\label{2.2}
  E[eI(B^{\top}X \leq u)]=E[eI(\kappa \beta_{0}^{\top}X \leq u)]=0.
\end{equation}
While under the alternative, by  Lemma 1 of Escanciaco (2006), there exists an $\alpha \in \mathcal{S}_q^{+}$ such that $E(e|\alpha^{\top}B^{\top}X) \neq 0$, where $\mathcal{S}_q^{+}=\{ \alpha=(a_1, \cdots, a_q)^{\top} \in \mathbb{R}^q: \|\alpha\|=1 \ {\rm and} \ a_1 \geq 0 \}$. Then it follows that
\begin{equation}\label{2.3}
  E[eI(\alpha^{\top}B^{\top}X \leq u)] \neq 0
\end{equation}
Note that under the null we have $q=1$ and $\mathcal{S}_q^{+}=\{1\}$. Thus the quantity $E[eI(\alpha^{\top}B^{\top}X \leq u)]$ actually has the same form in both~(\ref{2.2}) and~(\ref{2.3}). 
Define an adaptive-to-model residual marked empirical process $V_n(u)$ in the diverging dimension setting as below 
\begin{equation}\label{2.4}
V_n(\hat{\alpha}, u) = \frac{1}{\sqrt{n}}\sum_{i=1}^n [Y_i- g(\hat{\beta}^{\top}_n X_i,\hat{\theta}_n)]I(\hat{\alpha}^{\top}\hat{B}_n^{\top} X_i \leq u),
\end{equation}
\begin{equation}\label{2.5}
  V_n(u)= \sup_{ \hat{\alpha} \in \mathcal{S}_{\hat{q}}^{+}} |V_{n}(\hat{\alpha}, u)|
\end{equation}
where $\hat{\beta}_n$ and $\hat{\theta}_n$ are defined as before and $\hat{B}_n$ is the sufficient dimension reduction estimator of $B$ with an estimated structural dimension $\hat{q}$ of $q$, which will be specified later. For $V_n(u)$, one can also use the integral over $\mathcal{S}_{\hat{q}}^{+}$ to define a test statistic.

To achieve the model adaptation property of the process, we need sufficient dimension reduction (SDR) techniques to identify the structural dimension $q$ and the matrix $B$, when $p$ diverges to infinity. 
We give a brief review below on this topic.

\subsection{Adaptive-to-model approach}
In this methodology, we need to identify  the dimension $q$ and the matrix $B$. This  can be done by using the methods in sufficient dimension reduction.
We then  give a brief description. Recall under the alternative hypothesis the model is as
\begin{equation}
  Y=G(B^{\top}X)+\varepsilon,
\end{equation}
where $E(\varepsilon|X)=0$ and $G(\cdot)$ is an unknown smooth function and  $B$ is a $p \times q$ orthonormal matrix with $1\leq q \leq p$. We can see that under both the null and alternative hypothesis, the conditional independence holds respectively:
$$
Y \hDash E(Y|X)|\beta_0^{\top}X, \quad \mbox{and } \quad Y \hDash E(Y|X)|B^{\top}X,
$$
where $\hDash$ means statistical independence. Define
$\mathcal{S}_{E(Y|X)}$ as the central mean subspace of $Y$ with respect to $X$ (see, Cook and Li 2002) that is the intersection of all subspaces spanned by the columns of $A$ $\rm{span}(A)$ such that $Y \hDash E(Y|X)|A^{\top}X$.  The dimension of $\mathcal{S}_{E(Y|X)}$ is called the structural dimension, denoted as $d_{E(Y|X)}$. Under mild conditions, such a subspace $\mathcal{S}_{E(Y|X)}$ always exists (see Cook and Li, 2002). If $\mathcal{S}_{E(Y|X)}=\rm{span}(A)$, then  $E(Y|X)=E(Y|A^{\top}X)$.
Under the null hypothesis (\ref{1.1}),  $d_{E(Y|X)}=1$ and $\mathcal{S}_{E(Y|X)}=\rm{span}(\beta_{0}/\|\beta_{0}\|)$. Under the alternative (\ref{1.2}),  $d_{E(Y|X)}=q$ and $\mathcal{S}_{E(Y|X)}=\rm{span}(B)$.
For simplicity, we assume throughout this paper that  $\mathcal{S}_{E(Y|X)}=\mathcal{S}_{Y|X}$. Here $\mathcal{S}_{Y|X}$ is the central subspace of $Y$ with respect to $X$ (see, Cook 1998).

There are several estimation proposals available in the literature. For instance, sliced inverse regression (SIR, Li (1991)), sliced average variance estimation (SAVE,  Cook and Weisberg (1991)), minimum average variance estimation (MAVE, Xia et.al. (2002)), directional regression (DR, Li and Wang, (2007)), discretization-expectation estimation (DEE, Zhu, et al. (2010a)). All these methods assumed that $p$ is fixed. Zhu, Miao, and Peng (2006) first discussed the asymptotic properties of SIR when $p$ diverges to infinity. In this paper, we adapt cumulative slicing estimation (CSE, Zhu, Zhu, and Feng (2010b)) to identify the central subspace, which is similar to discretization-expectation estimation (DEE, Zhu, et al. (2010a)). This is because both of them are very easily implemented and easy to be extended to handle the case where the dimension $p$ grows to  infinity.

The procedure of CSE is as follows. For simplicity, we assume $E(X)=0, Var(X)=I_p$ for a moment. If the linearity condition (see Li, 1991) holds, it is easy to see that $E[Xh(Y)] \in \mathcal{S}_{Y|X}$ for any function $h(\cdot)$. Theoretically, we obtain infinity amount of vectors in $\mathcal{S}_{Y|X}$. Zhu et.al. (2010b) suggested a determining class of indicator functions to replace $h(\cdot)$. Let $h_t(Y)=I(Y \leq t)$. It follows that
$$ Y \hDash X|B^{\top}X  \Longleftrightarrow h_t(Y) \hDash X|B^{\top}X,  \quad \forall \ t \in \mathbb{R}. $$
Define the target matrix
\begin{equation}\label{2.6}
  M=\int E[X h_t(Y)] E[X^{\top} h_t(Y)] dF_{Y}(t),
\end{equation}
where $F_{Y}$ denotes the cumulative distribution function of $Y$. If the rank of $M$ is $q$, then ${\rm span}(M)=\mathcal{S}_{Y|X}$. Based on this, it is easy to obtain the sample version of $M$. Let $Z_i$ be the standardized $X_i$ and $\hat{\alpha}_t=\frac{1}{n} \sum_{i=1}^{n} Z_i I(Y_i \leq t)$. The estimator of $M$ is given by
\begin{equation}\label{2.7}
  \hat{M}_n=\frac{1}{n} \sum_{j=1}^{n} \hat{\alpha}_{Y_j} \hat{\alpha}_{Y_j}^{\top}.
\end{equation}
If the structural dimension $q$ is given, an estimator $\hat{B}_n(q)$ of $B$ consists of the eigenvectors corresponding to the largest $q$ eigenvalues of $\hat{M}_n$. Throughout this paper, we assume that $q$ is fixed.

Yet we  need a consistent estimator $\hat{q}$ of $q$ as $q$ is usually unknown under the alternative hypothesis. Later we will see that even when $q$ is given, we still want a consistent estimator because we wish the test to have model adaptation property to fully use the dimension reduction structure under the null hypothesis.  Inspired by Xia et al. (2015), we suggest a minimum ridge-type eigenvalue ratio estimator (MRER) to determine $q$. Let $\hat{\lambda}_{1}\ge \cdots \ge \hat{\lambda}_{p} $ and ${\lambda}_{1}\ge \cdots \ge {\lambda}_{p} $ be the eigenvalues of the matrix $\hat{M}_{n}$ and $M$ respectively.  Since $rank(M)=q$, it follows that
$$ \lambda_1 \ge \cdots \ge \lambda_q > \lambda_{q+1}=\cdots =\lambda_p =0.$$
Hence we estimate the structural dimension $q$ by
\begin{equation}\label{2.8}
\hat{q}=\arg\min_{1\leq i \leq p}\left\{i: \frac{\hat{\lambda}^2_{i+1}+c}{\hat{\lambda}^2_i+c}\right\}.
\end{equation}
Here $\hat{\lambda}_{p+1}$ is defined as $0$ and the ridge $c$ is a positive constant. The following result shows that the consistency of MRER is adaptive to the underlying models, when $c$ equals to some appropriate constant. Its proof will be given in Appendix.

\begin{prop}\label{prob 3}
Suppose that the regularity conditions of Theorem~3 in Zhu et al. (2010b) hold. Let $\hat{B}_n(q)$ be a matrix whose columns are the eigenvectors that are associated with the largest $q$ eigenvalues of $\hat{M}_n$. If $c=\log{n}/n$, then \\
(1) under $H_0$, we have $\mathbb{P}(\hat{q}=1) \to 1$ and $\| \hat{B}_n(1) - \kappa \beta_{0} \|=O_p(\sqrt{p/n})$; \\
(2) under $H_1$, we have $\mathbb{P}(\hat{q}=q) \to 1$ and $\| \hat{B}_n (q)- B \|=O_p(\sqrt{p/n})$.
\end{prop}

\section{Main results}
\subsection{Basic properties of the process}
First, we discuss the asymptotic properties of the process $V_n(\hat{\alpha}, u)$ under the null hypothesis. Since the distributional limit theory becomes much simpler if we replace the estimators by their true values, we define the following process
\begin{equation*}
V_n^0(u) = \frac{1}{\sqrt{n}}\sum_{i=1}^n [Y_i- g(\beta_{0}^{\top} X_i,\theta_0)]I(\kappa \beta_{0}^{\top} X_i \leq u).
\end{equation*}
Put
\begin{eqnarray*}
  \sigma_n^2(v) &=& Var(Y|\kappa \beta_{0}^{\top}X = v) \\
  \psi_n(u)     &=& E[Var(Y|\kappa \beta_{0}^{\top}X)I(\kappa \beta_{0}^{\top}X \leq u)].
\end{eqnarray*}
Then we have $\sigma_n^2(v)=E(\varepsilon^2|\kappa \beta_{0}^{\top}X = v)$ and $\psi_n(u)=\int_{-\infty}^{u} \sigma_n^2(v) F_{\kappa \beta_{0}}(dv)$ where $F_{\kappa \beta_{0}}$ is the cumulate
distribution function of $\kappa \beta_{0}^{\top}X$. Obviously, $\psi_n(u)$ is a nondecreasing and nonnegative function. Since $V_n^0(u)=\frac{1}{\sqrt{n}}\sum_{i=1}^n \varepsilon_i I(\kappa \beta_{0}^{\top} X_i \leq u)$ is a centered residual cusum process, it is readily seen that
\begin{eqnarray*}
Cov[V_n^0(s), V_n^0(t)]=\psi_n(s \wedge t).
\end{eqnarray*}
By Theorem 2.11.22 in Van Der Vaart and Wellner (1996), we obtain that $V_n^0(u)$ is asymptotically tight.
If $\psi_n(u) \to \psi(u)$ pointwisely in $u$, it follows that
\begin{eqnarray}\label{3.1}
V_n^0(u) \longrightarrow V_{\infty}(u)   \quad \ {\rm in \ distribution},
\end{eqnarray}
in the space $\ell^{\infty}(\overline{R})$, where $V_{\infty}(u)$ is a centred Gaussian process with the covariance function $\psi(s \wedge t)$. Since $\psi(u)$ is also nondecreasing and nonnegative, it follows that $V_{\infty}(u)=B(\psi(u))$ in distribution, where $B(u)$ is a standard Brownian motion.

For composite model checks, the unknown parameters in $V_n^0(u)$ should be replaced by their estimators, so we go back to $V_n(\hat{\alpha}, u)$ as defined in~(\ref{2.4}). By Proposition 3, $\mathbb{P}(\hat{q}=1) \to 1$ under the null hypothesis. Thus we only need to work on the event $\{\hat{q}=1\}$. Consequently, $\mathcal{S}_{\hat{q}}^{+} =\{1\}$ and $V_n(\hat{\alpha}, u)$ can be rewritten as
\begin{equation*}
V_n(\hat{\alpha}, u) = \frac{1}{\sqrt{n}}\sum_{i=1}^n [Y_i- g(\hat{\beta}^{\top}_n X_i,\hat{\theta}_n)]I(\hat{B}_n^{\top} X_i \leq u)
\end{equation*}
Under some regularity conditions stated in  Appendix and on the event $\{\hat{q}=1\}$, we can show that under the null hypothesis
\begin{equation}\label{3.2}
 V_n(\hat{\alpha}, u)= V_n^0(u)- \sqrt{n} (\hat{\gamma}_n-\gamma_{0})^{\top} M_n(u)+ o_p(1)
\end{equation}
uniformly in $u$, where $M_n(u)=E[g'(\beta_{0}, \theta_0, X) I(\kappa \beta_{0}^{\top} X \leq u)]$. A proof of~(\ref{3.2}) will be given in Appendix. Combined~(\ref{3.2}) with Proposition 2 and some elementary calculations, we have
\begin{equation}\label{3.3}
  V_n(\hat{\alpha}, u)= V_n^0(u)- \frac{1}{\sqrt{n}} M_n(u)^{\top} \Sigma_n^{-1} \sum_{i=1}^{n} g'(\beta_{0}, \theta_0, X_i) \varepsilon_i +o_p(1)
\end{equation}
uniformly in $u$. It is easy to see that the second term of the right hand side of (\ref{3.3}) is also asymptotically tight. Altogether we then obtain the following result.

\begin{theorem} \label{Theorem 3.1}
Suppose that the regularity conditions in Appendix hold. when $p^5/n \to 0$, then under the null hypothesis, we have in distribution
\begin{equation*}
  V_n(u) \longrightarrow |V_{\infty}^1(u)|,
\end{equation*}
where $V_{\infty}^1(u)$ is a zero mean Gaussian process with a covariance function $ K(s,t)$ that  is the pointwise limit of $K_n(s,t)$ as
\begin{eqnarray*}
K_n(s, t) &=& E[\varepsilon^2 I(\kappa \beta_{0}^{\top} X \leq s \wedge t)]- M_n(s)^{\top} \Sigma_n^{-1} E[\varepsilon^2 g'(\beta_{0}, \theta_0, X)I(\kappa \beta_{0}^{\top} X \leq t)] \\
&& -M_n(t)^{\top} \Sigma_n^{-1} E[\varepsilon^2 g'(\beta_{0}, \theta_0, X) I(\kappa \beta_{0}^{\top} X \leq s)] \\
&&+ M_n(s)^{\top} \Sigma_n^{-1}  E[\varepsilon^2 g'(\beta_{0}, \theta_0, X) g'(\beta_{0}, \theta_0, X)^{\top}] \Sigma_n^{-1} M_n(t).
\end{eqnarray*}
\end{theorem}

\subsection{Martingale transformation}
If $p$ is fixed, $V_{\infty}^1(u)$ can be rewritten as $V_{\infty}^1(u)=V_{\infty}(u)+M(u)^{\top}V$ in distribution and its covariance function can be specified. The shift term $M(u)^{\top}V$ is brought out from the second term in~(\ref{3.3}). Stute, Thies, and Zhu (1998) first proposed a martingale transformation to eliminate $M(u)^{\top}V$ in $V_{\infty}^1(u)$ and then obtain a tractable limiting distribution of a functional of $V_{\infty}(u)$. This  has become one of the basic methodologies in the area of model checking to derive asymptotically distribution-free tests. It was motivated by the Khmaladze martingale transformation  in constructing convenient goodness of fit tests for hypothetical distribution functions (Khmaladze,  1982).  There are a number of follow-up studies in the literature to extend this methodology to various high-dimensional models  such as Khmadladze and Koul (2004) and Stute, Xu and Zhu (2008). However, when $p$ diverges as $n$ goes to infinity, the form of the shift term that would be a limit of $M(u)^{\top}V$ can not be given specifically,  as stated in the above theorem. The martingale transformation cannot directly target $M(u)^{\top}V$. We then bypass this difficulty by checking its shift term at the sample level. Note that the shift term comes from  the second term in (\ref{3.2}). This is because in the case with the fixed $p$, $M(u)^{\top}V$ is just its weak limit.  Thus, we then target that term directly at the sample level.

Following Stute, Thies, and Zhu (1998) or Stute and Zhu (2002), recall that $M_n(u)=E[g'(\beta_{0}, \theta_0, X) I(\kappa \beta_{0}^{\top} X \leq u)]$ and $\psi_n(u)=\int_{-\infty}^{u} \sigma_n^2(v) F_{\kappa \beta_{0}}(dv)$. Let
$$
a_n(u)=\frac{\partial M_n(u)}{\partial \psi_n(u)}
$$
be the Radon-Nikodym derivative of $M_n(u)$ with respect to $\psi_n(u)$. Next, define a $(p+d) \times (p+d)$ matrix
\begin{equation*}
  A_n(u)= \int_{u}^{\infty} a_n(z) M_n^{\top}(dz)=\int_{u}^{\infty} a_n(z) a_n(z)^{\top} \sigma_n^2(z) F_{\kappa \beta_{0}}(dz).
\end{equation*}
It can also be written as
$$ A_n(u)= E[a_n(\kappa \beta_{0}^{\top} X)g'(\beta_{0}, \theta_0, X)^{\top} I(\kappa \beta_{0}^{\top} X \geq u)]. $$
Mimicking the martingale transformation in Stute and Zhu (2002) at the sample level, we have
\begin{equation}\label{3.4}
  (T_n f_n)(u) = f_n(u) -\int_{-\infty}^{u} a_n(z)^{\top} A_n^{-1}(z) \left(\int_{z}^{\infty} a_n(v) f_n(dv)\right) \psi_n(dz).
\end{equation}
Here we should assume that $A_n(u)$ is nonsingular and the process $f_n(u)$ should be either bounded variation or a Brownian motion.

Some elementary computation concludes that $T_n (\sqrt{n} (\hat{\gamma}_n-\gamma_{0})^{\top} M_n)=0$. Next, we discuss the approximation properties of $T_n V_n^0$. Note that
\begin{equation*}
  (T_n V_n^0)(u)=V_n^0(u) -\int_{-\infty}^{u} a_n(z)^{\top} A_n^{-1}(z) \left(\int_{z}^{\infty} a_n(v) V_n^0(dv)\right) \psi_n(dz)
\end{equation*}
and
\begin{equation*}
 \int_{z}^{\infty} a_n(v) V_n^0(dv) =\frac{1}{\sqrt{n}}\sum_{i=1}^n a_n(\kappa \beta_{0}^{\top} X_i) I(\kappa \beta_{0}^{\top} X_i \geq z) \varepsilon_i.
\end{equation*}
Combining these two formulas, we obtain that
\begin{equation*}
  T_n V_n^0(u) = V_n^0(u)- \frac{1}{\sqrt{n}}\sum_{i=1}^n  \int_{-\infty}^{u} a_n(z)^{\top} A_n^{-1}(z) a_n(\kappa \beta_{0}^{\top} X_i) I(\kappa \beta_{0}^{\top} X_i \geq z) \psi_n(dz) \varepsilon_i.
\end{equation*}
Therefore, $T_n V_n^0$ is also an i.i.d. centered residual cusum process with a covariance function
\begin{equation}\label{3.5}
 Cov[T_n V_n^0(s), T_n V_n^0(t)] = Cov[V_n^0(s), V_n^0(t)]=\psi_n(s \wedge t).
\end{equation}
This means that $T_n V_n^0(u)$ admits the same limiting distribution as that of $V_n^0(u)$, i.e.,
\begin{eqnarray}\label{3.6}
 T_n V_n^0(u) \longrightarrow V_{\infty}(u)   \quad \  {\rm in \ distribution}.
\end{eqnarray}
Consequently, we get rid of the annoying shift term $\sqrt{n} (\hat{\gamma}_n-\gamma_{0})^{\top} M_n$ and obtain the  process $V_{\infty}(u)$ whose supremum over all $u$ has a tractable limiting distribution.
The assertions~(\ref{3.5}) and~(\ref{3.6}) will be justified in Appendix (Lemma 1).

The transformation $T_n$ obviously contains some unknown quantities and therefore needs to be substituted by their empirical analogues. For this, let $g_1'(t, \theta)=\frac{\partial g(t, \theta)}{\partial t}$ and $g_2'(t, \theta)=\frac{\partial g(t, \theta)}{\partial \theta}$. It follows that
$$g'(\beta_{0}, \theta_0, X)=\left(g_1'(\beta_{0}^{\top}X, \theta_0)X^{\top}, g_2'(\beta_{0}^{\top}X, \theta_0)^{\top} \right)^{\top}. $$
Consequently, we have
\begin{equation*}
  M_n(u)= \left(\int_{-\infty}^{u} g_1'(z/\kappa, \theta_0) r_n(z)^{\top} F_{\kappa \beta_{0}}(dz), \int_{-\infty}^{u} g_2'(z/\kappa, \theta_0)^{\top}F_{\kappa \beta_{0}}(dz) \right)^{\top}
\end{equation*}
where $r_n(v)=E(X| \kappa \beta_{0}^{\top}X=v)$. Conclude that
\begin{equation*}
  a_n(u)=\left( \frac{g_1'(u/\kappa, \theta_0) r_n(u)^{\top}}{\sigma_n^2(u)}, \frac{g_2'(u/\kappa, \theta_0)^{\top}}{\sigma_n^2(u)} \right)^{\top}.
\end{equation*}
Since $a_n(u)$ depends on $r_n(u)$ and $\sigma_n^2(u)$ on which we do not make any assumption rather than smoothness, they need to be estimated in a nonparametric way. For instance, we may adopt a standard Nadaraya-Watson estimator for $r_n(v)$:
\begin{equation*}
  \hat{r}_n(v)=\frac{\sum_{i=1}^n X_i K(\frac{v- \hat{\alpha}^{\top} \hat{B}^{\top}_{n} X_i}{h})}{\sum_{i=1}^n
               K(\frac{v-\hat{\alpha}^{\top}\hat{B}^{\top}_{n} X_i}{h})}
\end{equation*}
where $K(\cdot)$ is an univariate kernel function and $h$ is a bandwidth. Similarly for $\sigma_n^2(u)$. Thus we obtain the empirical estimators $\hat{a}_n(u)$ and $\hat{A}_n(u)$ of $a_n(u)$ and $A_n(u)$ respectively:
\begin{eqnarray*}
\hat{a}_n(u)&=& \left( \frac{g_1'(u/\hat{\kappa}_n, \hat{\theta}_n) \hat{r}_n(u)^{\top}}{\hat{\sigma}_n^2(u)}, \frac{g_2'(u/\hat{\kappa}_n,
                \hat{\theta}_n)^{\top}}{\hat{\sigma}_n^2(u)} \right)^{\top},\\
\hat{A}_n(u)&=& \frac{1}{n} \sum_{i=1}^{n} \hat{a}_n(\hat{\alpha}^{\top} \hat{B}_n^{\top}X_i) g'(\hat{\beta}_n, \hat{\theta}_n, X_i)^{\top}
                I(\hat{\alpha}^{\top} \hat{B}_n^{\top}X_i \geq u).
\end{eqnarray*}
Finally, we can give an estimator $\hat{T}_n$ of $T_n$:
\begin{eqnarray*}
\hat{T}_n V_n(\hat{\alpha}, u) &=& V_n(\hat{\alpha}, u)- \int_{-\infty}^{u} \hat{a}_n(z)^{\top} \hat{A}_n^{-1}(z) \left(\int_{z}^{\infty}
                                   \hat{a}_n(v) V_n(\hat{\alpha}, dv)\right) \hat{\sigma}_n^2(z) F_{\hat{\alpha}}(dz) \\
   &=& \frac{1}{n^{1/2}}\sum_{i=1}^n [Y_i- g(\hat{\beta}^{\top}_n X_i,\hat{\theta}_n)]I(\hat{\alpha}^{\top}\hat{B}_n^{\top} X_i \leq u)- \\
   &&  \frac{1}{n^{3/2}} \sum_{i,j=1}^n I(\hat{\alpha}^{\top}\hat{B}_n^{\top} X_i \leq u) \hat{a}_n(\hat{\alpha}^{\top}\hat{B}_n^{\top} X_i)^{\top}
       \hat{A}_n^{-1}(\hat{\alpha}^{\top}\hat{B}_n^{\top} X_i) \hat{a}_n(\hat{\alpha}^{\top}\hat{B}_n^{\top} X_j) \times \\
   &&  I(\hat{\alpha}^{\top}\hat{B}_n^{\top} X_j \geq \hat{\alpha}^{\top}\hat{B}_n^{\top} X_i) [Y_j- g(\hat{\beta}^{\top}_n X_j,\hat{\theta}_n)]
       \hat{\sigma}_n^2(\hat{\alpha}^{\top}\hat{B}_n^{\top} X_i)
\end{eqnarray*}
where $\hat{\kappa}_n$ is the estimator of $\kappa$ and $F_{\hat{\alpha}}$ is the empirical distribution function of $\hat{\alpha}^{\top}\hat{B}^{\top}_{n} X_i, 1 \leq i \leq n$. Making sure the columns of $\hat{B}_n$ have the same direction as $\hat{\beta}_{n}$, we can assume $\kappa=1/\|\beta_0\|$ and $\hat{\kappa}_n=1/\| \hat{\beta}_n\|$.

\begin{theorem}\label{Theorem 3.2}
Suppose that $A_n(u)$ is nonsingular and $\sigma_n^2(u)$ is bounded away from zero for all $u$. If $p^5/n \to 0$, under the null hypothesis   $H_0$ and the regularity conditions in Appendix, we have
\begin{equation*}
  \sup_{ \hat{\alpha} \in \mathcal{S}_{\hat{q}}^{+}} |\hat{T}_n V_{n}(\hat{\alpha}, u)| \longrightarrow |V_{\infty}(u)|
\end{equation*}
in distribution in the space $\ell^{\infty}([-\infty, x_0])$ for any $x_0 \in \mathbb{R}$.
\end{theorem}
Note that we use $\hat{A}_n(u)$ in the process $\hat{T}_n V_{n}(\hat{\alpha}, u)$. In concrete data analysis, these matrices may be unbounded for large $u$ and thus the distributional behavior of the underlying process may become very unstable in the extreme right tails. These may severely damage the approximation accuracy of the test statistic based on all $\hat{T}_n V_{n}$. Therefore, we restrict $\hat{T}_n V_{n}$ to compact intervals $[-\infty, u_0]$ and obtain the convergence of  $\sup_{ \hat{\alpha} \in \mathcal{S}_{\hat{q}}^{+}} |\hat{T}_n V_{n}(\hat{\alpha}, u)|$  in the space $\ell^{\infty}([-\infty, x_0])$.

In a special case where the predictor $X$ follows a spherically contoured distribution or its extension, the elliptically contoured distribution, we can show that the calculations of the martingale transformation will become much simpler. The idea is similar to Stute and Zhu (2002). Without loss of generality, we only consider spherically contoured distributions. Here we shall assume the regression function $g$ does not depend on $\theta$. Let $g'(t)$ be the derivative of $g(t)$ with respective to $t$. It follows that
$$
M_n(u)=E[g'(\beta_{0}^{\top}X) X I(\kappa \beta_{0}^{\top}X \leq u)] =\Gamma^{\top} E[g'(\beta_{0}^{\top}X) \Gamma X I(\kappa \beta_{0}^{\top}X \leq u)],
$$
where $\Gamma$ is an $p \times p$ orthonormal matrix with the first row $\kappa \beta_{0}^{\top}$ (or $ \beta_{0}^{\top}/\|\beta_{0}\|$). Since the conditional expectation of the other components of $\Gamma X$ given the first is zero, it follows that
$$
M_n(u)=\frac{\beta_{0}}{\|\beta_{0}\|^2} E[g'(\beta_{0}^{\top}X) \beta_{0}^{\top}X I(\kappa \beta_{0}^{\top}X \leq u)]=\frac{\beta_{0}}{\|\beta_{0}\|^2} \int_{-\infty}^{u} g'(z/\kappa)z/{\kappa} F_{\kappa \beta_{0}}(dz),
$$
whence,
\begin{eqnarray*}
a_n(u)&=& \frac{\beta_{0}}{\|\beta_{0}\|^2} \frac{g'(u/\kappa)u/{\kappa}}{\sigma_n^2(u)},\\
\quad A_n(u)&=& \frac{\beta_{0} \beta_{0}^{\top} }{\|\beta_{0}\|^4} \int_{u}^{\infty} \frac{[g'(z/\kappa)z/{\kappa}]^2}{\sigma_n^2(z)} F_{\kappa \beta_{0}}(dz).
\end{eqnarray*}
Note that $A_n(z)$ is a matrix with rank $1$ and is singular when $p>1$. Thus the martingale transformation can not apply directly. However, if we go back to~(\ref{3.2}) and set
$$
\tilde{M}_n(u)= E[g'(\beta_{0}^{\top}X) \beta_{0}^{\top}X I(\kappa \beta_{0}^{\top}X \leq u)],
$$
then~(\ref{3.2}) can be rewritten as
\begin{equation}\label{3.7}
 V_n(\hat{\alpha}, u)= V_n^0(u)- \sqrt{n} (\hat{\gamma}_n-\gamma_{0})^{\top} \frac{\beta_{0}}{\|\beta_{0}\|^2} \tilde{M}_n(u)+ o_p(1)
\end{equation}
Conclude that the new $a_n(u)$ and $A_n(u)$ become the real-valued
$$
a_n(u)=\frac{\partial \tilde{M}_n(u)}{\partial \psi_n(u)}= \frac{g'(u/\kappa)u/{\kappa}}{\sigma_n^2(u)} \quad {\rm and} \quad A_n(u)= \int_{u}^{\infty} \frac{[g'(z/\kappa)z/{\kappa}]^2}{\sigma_n^2(z)} F_{\kappa \beta_{0}}(dz).
$$
Clearly, Theorem~\ref{Theorem 3.2} can  be applied to these new functions.

Hall and Li (1993) shown that, if $p \to \infty$ as $n \to \infty$, expectation over a large number of random variables behaves more or less like  expectation over the multivariate normal distribution. Note that $M_n(u)=E[g'(\beta_{0}^{\top}X)X I(\kappa \beta_{0}^{\top}X \leq u)]$ and multivariate normal distribution is elliptically-contoured. Consequently, even when $X$ is not multivariate normal distributed, $M_n(u)$ can be viewed as expectation on multivariate normal distribution and then the martingale transformation $T_n$ can apply to the real-valued $a_n(u)$ and $A_n(u)$ in practice for large $p$.

\subsection{The properties under the alternative hypothesis}
Now we discuss the asymptotic properties of $\sup_{ \hat{\alpha} \in \mathcal{S}_{\hat{q}}^{+}} |\hat{T}_n V_{n}(\hat{\alpha}, u)|$ under a sequence of local alternatives converging to the null hypothesis at a parametric rate $1/\sqrt{n}$. Consider
\begin{equation}\label{3.8}
H_{1n}: Y=g(\beta_{0}^{\top}X, \theta_0)+ \frac{1}{\sqrt{n}} G(X) + \varepsilon,
\end{equation}
where $E(\varepsilon|X)=0$, $G(X)$ is a random variable with zero mean and satifies $\mathbb{P}\{G(X)=0\}<1$. To derive the asymptotic distribution of $\hat{T}_n V_{n}(\hat{\alpha}, u)$ under $H_{1n}$, we need the asymptotic properties of $\hat{q}$ and $\hat{\gamma}_n$, when $p$ diverges to infinity.

\begin{prop}\label{prob 4}
Assume the regularity conditions of Theorem 3 in Zhu et al. (2010b) hold. Let $\hat{B}_n(1)$ be an eigenvector associating with the largest eigenvalues of $\hat{M}_n$, then we have, under $H_{1n}$, $\mathbb{P}(\hat{q}=1) \to 1$ and $\|\hat{B}_n(1) - \kappa \beta_{0}\|=O_p(\sqrt{p/n})$.
\end{prop}

Next, we derive the norm consistency of $\hat{\gamma}_n$ with respective to $\gamma_{0}$ and a asymptotical decomposition of $\hat{\gamma}_n-\gamma_{0}$ under $H_{1n}$. Here $\hat{\gamma}_n=(\hat{\beta}_n^{\top}, \hat{\theta}_n^{\top})^{\top}$ and $\gamma_{0}=(\beta_{0}^{\top}, \theta_{0}^{\top})^{\top}$ as mentioned before.

\begin{prop}\label{prob 5}
Suppose the regularity conditions in Appendix and~(\ref{3.8}) hold. If $p^4/n \to 0$, then $\hat{\gamma}_n$ is a norm consistent estimator for $\gamma_{0}$ with $\| \hat{\gamma}_n-\gamma_{0} \|=O_p(\sqrt{p/n})$. Moreover, if $p^5/n \to 0$, we have
\begin{equation}\label{3.9}
\sqrt{n} (\hat{\gamma}_n-\gamma_{0})=\Sigma_n^{-1} \frac{1}{\sqrt{n}} \sum_{i=1}^{n} \varepsilon_i g'(\beta_{0}, \theta_0, X_i)+\Sigma_n^{-1} E[G(X)g'(\beta_{0}, \theta_0, X)] + o_p(1).
\end{equation}
\end{prop}

The following theorem states the asymptotic results under various alternatives.
\begin{theorem}\label{Theorem 3.3}
Suppose the regularity conditions in Appendix hold. If $p^5/n \to 0$,\\
(1) under the global alternative $H_1$, we have in probability
\begin{eqnarray*}
\frac{1}{\sqrt{n}} \sup_{ \hat{\alpha} \in \mathcal{S}_{\hat{q}}^{+}} |\hat{T}_n V_{n}(\hat{\alpha}, u)| \longrightarrow |L(u)|,
\end{eqnarray*}
where $L(u)$ is some nonzero function;\\
(2) under the local alternative $H_{1n}$, we have in distribution
$$ \sup_{\hat{\alpha} \in \mathcal{S}_{\hat{q}}^{+}} |\hat{T}_n V_{n}(\hat{\alpha}, u)| \longrightarrow |V_{\infty}(u) + G_1(u)-G_2(u)|,$$
where $V_{\infty}(u)$ is a zero-mean Gaussian process given by (\ref{3.1}) and $ G_1(u)$ and $ G_2(u)$ are the uniform limit of $ G_{1n}(u)$, $ G_{2n}(u)$, respectively which are as follows
\begin{eqnarray*}
G_{1n}(u) &=& E[G(X) I(\kappa \beta_{0}^{\top} X \leq u)],\\
G_{2n}(u) &=& E \{ G(X) \int_{-\infty}^{u} a_n(z)^{\top} A_n^{-1}(z) a_n(\kappa \beta_{0}^{\top} X)
    I(\kappa \beta_{0}^{\top} X \geq z) \psi_n(dz)\}.  
\end{eqnarray*}
\end{theorem}

These results show that under the global alternative, the process diverges to infinity at the rate of order $1/\sqrt n$ and under the local alternatives distinct from the null at the rate of order $1/\sqrt n$, the process converges to a stochastic process. Thus, the test that is based on this process can detect such alternatives.

\section{Numerical studies}
\subsection{Test statistics in practical use}
In this subsection, we use the Cram$\rm \acute{e}$r-von Mises (CM) functional to construct   test statistic.
Consider
\begin{equation}\label{4.1}
  CM_n^2 = \int_{-\infty}^{u_0} \sup_{\hat{\alpha} \in \mathcal{S}_{\hat{q}}^{+}}  |\hat{T}_n V_{n}(\hat{\alpha}, u)|^2 F_n(du),
\end{equation}
where $F_n$ is the empirical distribution function of $\beta_{0}^{\top}X_i/\|\beta_{0}\|$, $1 \leq i \leq n$. According to Theroem~\ref{Theorem 3.2} and the Extended Continuous Mapping Theorem (see Theorem 1.11.1 in Van Der Vaart and Wellner (1996)), we obtain, under the null,
$$
CM_n^2  \longrightarrow  \int_{-\infty}^{u_0}  \frac{B^2(\psi(u))}{\sigma^2(u)} \psi(du)   \quad  {\rm in \ distribution},
$$
where $B(t)$ is a standard Brownian motion and $\sigma^2(u)$ is the pointwise limit of $\sigma_n^2(u)$. Since
$B(t \psi(u_0))/\sqrt{\psi(u_0)}=B(t)$ in distribution, it follows that
$$
\int_{-\infty}^{u_0}  B^2(\psi(u)) \psi(du) = \psi^2(u_0) \int_{0}^{1} B(t)^2 dt   \quad  {\rm in \ distribution}.
$$
Consequently, we consider
\begin{equation}\label{4.2}
  ACM_n^2 = \frac{1}{\hat{\psi}_n(u_0)^2} \int_{-\infty}^{u_0} \sup_{\hat{\alpha} \in \mathcal{S}_{\hat{q}}^{+}}  |\hat{T}_n V_{n}(\hat{\alpha}, u)|^2 \hat{\sigma}_n^2(u) F_n(du).
\end{equation}
Here we use $\hat{\psi}_n(u_0)=\frac{1}{n} \sum_{i=1}^{n} (Y_i-g(\hat{\beta}_n^{\top}X_i, \hat{\theta}_n))^2 I(\hat{\alpha}^{\top}\hat{B}_n^{\top}X_i \leq u_0)$ as an estimator of $\psi(u)$.
Therefore, we obtain
$$
ACM_n^2  \longrightarrow  \int_{0}^{1} B^2(u)du  \quad {\rm in \ distribution}.
$$

In the homoscedastic models, $\sigma_n^2(u)$ is free of $u$ and thus we can  estimate it by
$$
\hat{\sigma}_n^2= \frac{1}{n} \sum_{i=1}^{n} [Y_i-g(\hat{\beta}_n^{\top}X_i, \hat{\theta}_n)]^2.
$$
Now we also have $\psi_n(u_0)=\sigma_n^2F_{\kappa \beta_{0}}(u_0)$ and thus it can be estimated by $\hat{\sigma}_n^2 F_n(u_0)$. Consiquently, $ACM_n^2$ becomes
$$
ACM_n^2 = \frac{1}{\hat{\sigma}_n^2 F_n(u_0)^2} \int_{-\infty}^{u_0} \sup_{\hat{\alpha} \in \mathcal{S}_{\hat{q}}^{+}}  |\hat{T}_n V_{n}(\hat{\alpha}, u)|^2  F_n(du).
$$
For $u_0$, as suggested by Stute and Zhu (2002), we take $99\%$ quantile of $F_n$ in the simulation studies.

\subsection{Numerical studies}
In this subsection we conduct some simulation studies to examine the performance of the proposed test in this paper. From the results,  we set $p=[4n^{1/4}]-5$ with $n=100, 200, 400 \ {\rm and}\ 800$, as used in Fan and Peng (2004). As there are no relevant tests dealing with the case with  divergent dimension, we give comparisons with some existing tests that were developed with fixed dimension as for practical use, they would be workable. 

1. Stute and Zhu's (2002) test  is given by
  $$ T_n^{SZ}=\frac{1}{\hat{\psi}_n(x_0)} \int_{-\infty}^{x_0} |\hat{T}_nR_n^1|^2 \hat{\sigma}_n^2 dF_n, $$
where
\begin{eqnarray*}
R_n^1(u)&=& \frac{1}{\sqrt{n}} \sum_{i=1}^{n} [Y_i-g(\hat{\beta}_n^{\top} X_i, \hat{\theta}_n)]I(\hat{\beta}_n^{\top} X_i \leq u);\\
\hat{T}_n R_n^1(u) &=& R_n^1(u)- \int_{-\infty}^{u} \hat{a}_n(z)^{\top} \hat{A}_n^{-1}(z) \left(\int_{z}^{\infty}\hat{a}_n(v) R_n^1(dv)\right) \hat{\sigma}_n^2(z) F_n(dz).
\end{eqnarray*}
For $\hat{\psi}_n(x_0), \hat{\sigma}_n^2, \hat{a}_n(z), \hat{A}_n^{-1}(z)$, one can refer to their paper for detail.

2. Bierens (1982) proposed an integrated conditional moment (ICM) test which is based on the following statistic:
\begin{equation*}
  ICM_n=\frac{1}{n}\sum_{i=1}^{n} \sum_{j=1}^{n} \hat{e}_i \hat{e}_j  \exp(-\frac{1}{2}|X_i-X_j|)
\end{equation*}
where $\hat{e}_i=Y_i-g(\hat{\beta}_n^{\top} X_i, \hat{\theta}_n)$.

3. Escanciano's (2006) test statistic is defined as
\begin{equation*}
PCvM_n=\frac{1}{n^2}\sum_{i,j,r=1}^{n} \hat{e}_i \hat{e}_j \int_{S^p}I(\beta^{\top}X_i \leq \beta^{\top}X_r)I(\beta^{\top}X_j \leq \beta^{\top}X_r) d\beta
\end{equation*}
with  the critical value determination by the wild bootstrap. More details can be found in Escanciano (2006).

4. Zheng (1996) proposed a locally smoothing test whose  statistic is given by
\begin{equation*}
  T_n^{ZH}=\frac{\sum_{i \neq j} K((X_i-X_j)/h)\hat{e}_i\hat{e}_j}{\{\sum_{i \neq j} 2K^2((X_i-X_j)/h)\hat{e}_i^2\hat{e}_j^2\}^{1/2}}.
\end{equation*}

5.  An adaptive-to-model test defined in Guo et. al. (2016) with the test statistic:
\begin{equation*}
  T_n^{GWZ}=\frac{h^{1/2}\sum_{i \neq j} \hat{e}_i \hat{e}_j \frac{1}{h^{\hat{q}}} K(\hat{B}_n^{\top}(X_i-X_j)/h) }{\{2\sum_{i \neq j} \hat{e}_i^2 \hat{e}_j^2 \frac{1}{h^{\hat{q}}} K^2(\hat{B}_n^{\top}(X_i-X_j)/h)\}^{1/2}}.
\end{equation*}
Here we use the kernel function $K(u)=(15/16)(1-u^2)^2I(|u| \leq 1)$ and the bandwidth $h=1.5n^{1/(4+\hat{q})}$ as in Guo et. al. (2016) and $\hat{B}_n$ is a sufficient dimension estimate of $B$ with an estimated structural dimension $\hat{q}$ of $q$.

The significance levels are set to be $\alpha=0.1$, $0.05$, and $0.01$. The simulation results are based on the averages of $2000$ replications. In the following simulation studies, $a=0$ corresponds to the null while $a\neq 0$ to the alternatives.

$Study$ 1. The data are generated from the following models:
\begin{eqnarray*}
H_{11}:  Y &=& \beta_0^{\top}X+ a \exp(-(\beta_0^{\top}X)^2) +\varepsilon;  \\
H_{12}:  Y &=& \beta_0^{\top}X+ a \cos(0.6\pi \beta_0^{\top}X) +\varepsilon;\\
H_{13}:  Y &=& \beta_1^{\top}X+ a (\beta_2^{\top}X)^2 +\varepsilon; \\
H_{14}:  Y &=& \beta_1^{\top}X+ a \exp(\beta_2^{\top}X)+\varepsilon;
\end{eqnarray*}
where $\beta_0=(1, \cdots, 1)^{\top}/\sqrt{p}$, $\beta_1=(\underbrace{1,\dots,1}_{p_1},0,\dots,0)^{\top}/\sqrt{p_1}$ and $\beta_2=(0,\dots,0,\underbrace{1,\dots,1}_{p_1})/\sqrt{p_1}$ with $p_1=[p/2]$. The predictors $\{X_i, 1\leq i \leq n \}$ are i.i.d. from $N(0, I_p)$ and $\varepsilon$ is Guassian white noise with variance $1$. $H_{12}$ is a high-frequency/oscilating model and the other three are low-frequency models. In $H_{11}$ and $H_{12}$, the structural dimension equals $1$ under both the null and the alternative, while, in $H_{13}$ and $H_{14}$, the structural dimension is $2$ under the alternatives.

The simulation results are reported in Tables~1 to 4. We can see that both $ACM_n^2$ and $T_n^{SZ}$ maintain the significance levels very well. The empirical sizes of $PCvM_n$ are also very close to the significance levels, but slightly more unstable in some cases. $T^{GWZ}_n$ can only maintain the significance level when it is  $\alpha=0.05$. $T_n^{ZH}$ can maintain the significance levels occasionally, but generally, it is  conservative with smaller sizes.  $ICM_n$ is the worst among these tests in both the significance level maintenance and power performance.  According to our experience, when $p$ is smaller than $5$, $ICM_n$ could work well. The powers of $ACM_n^2$, $T_n^{SZ}$, $PCvM_n$ and $T^{GWZ}_n$ are all very high for models $H_{11}$, $H_{13}$ and $H_{14}$. But $T^{GWZ}_n$'s power grows slightly slower than the other three, while, for model $H_{12}$, $T^{GWZ}_n$ beats the other competitors.  These may validate again the empirical experience in this area that locally smoothing tests perform better for high frequency/oscillating  models, while globally smoothing tests work better for low frequency models. Nevertheless, $T_n^{ZH}$, a representative of locally smoothing tests, has very low power for model $H_{12}$. This is because $T_n^{ZH}$ severely suffers from the dimensionality problem, while $T^{GWZ}_n$ uses a dimension reduction technique to greatly alleviate the curse of dimensionality.
$$ \rm{Tables \ 1-4 \ about \ here}$$

The null models are all linear in $Study$~1. We then consider nonlinear hypothetical models in the next simulation study.

$Study$~2. The data are generated from the following models
\begin{eqnarray*}
H_{21}:  Y &=& (\beta_1^{\top}X)^3  + a(\beta_2^{\top}X)^2 +\varepsilon; \\
H_{22}:  Y &=& \exp(\beta_1^{\top}X)+ a(\beta_2^{\top}X)+\varepsilon,
\end{eqnarray*}
where $\beta_1=(\underbrace{1,\dots,1}_{p_1},0,\dots,0)^{\top}/\sqrt{p_1}$ and $\beta_2=(0,\dots,0, \underbrace{1, \dots,1}_{p_1})^{\top}/\sqrt{p_1}$ with $p_1=[p/2]$, $\varepsilon$ is $N(0,1)$, and $X$ is $N(0, I_p)$ independent of $\varepsilon$.

We report the empirical sizes and powers in Tables~5 and 6. For model $H_{21}$, The conclusions  are very similar to those in $Study$~1. For model $H_{22}$, we can see that the empirical sizes of $ACM_n^2$, $T_n^{SZ}$ and $PCvM_n$ are very close to the significance levels, while $T_n^{ZH}$ and $T^{GWZ}_n$ can only control the  level of $\alpha=0.1$. $ICM_n$ is still the worst one. The empirical powers of $T^{GWZ}_n$ and $ACM_n^2$ are higher than the other competitors, while $T_n^{SZ}$'s empirical powers grow very slow in this case. This would  confirm the theoretical result that $T_n^{SZ}$ is not an omnibus test.
$$ \rm{Tables \ 5-6 \ about \ here}$$

Therefore, overall, the proposed test in this paper performs well and can detect different alternatives. Further, the dimension of predictors has less negative impact on its performance.
\subsection{A real data example}
In this subsection we analyze the baseball salary data set that can be obtain through the website \url{http://www4.stat.ncsu.edu/~boos/var.select/baseball.html}. This data set contains 337 Major League Baseball players on the salary $Y$ from the year 1992 and 16 performance measures from the year 1991. The performance measures are $X_1$: Batting average, $X_2$: On-base percentage, $X_3$: runs, $X_4$: hits, $X_5$: doubles, $X_6$: triples, $X_7$: home runs, $X_8$: runs batted in, $X_9$: walks, $X_{10}$: strike-outs, $X_{11}$: stolen bases, and $X_{12}$: errors; and $X_{13}$: Indicator of free agency eligibility, $X_{14}$: Indicators of free agent in 1991/2, $X_{15}$: Indicators of arbitration eligibility, and $X_{16}$: Indicators of arbitration in 1991/2. The dummy variables $X_{13}-X_{16}$ measure the freedom of movement of a player to another team. For easy interpretation, we standardize all variables separately. To obtain the regression relationship between $Y$ and the performance measures $X=(X_1, \cdots, X_{16})^{\top}$, we first test for a linear regression model by the proposed test because the dimension $16\approx (337)^{0.476}$ and in the simplest case with linear model, the proposed test can theoretically handle $p=O(n^{1/2})$. The value of the test statistic is $ACM_n^2=1.3651$ with the $p$-value equal to $0.077$. Since the $p$-value is small although it is larger than, say, $0.05$, an often used significance level, we may consider a more plausible model to  better  fit this dataset. Hence we apply the dimension reduction techniques. Recalling in Section~2.3, we claimed that to estimate the central subspace, the CSE method is used. The estimated structural dimension of this datset is $\hat{q}=1$. This means that $Y$ may be conditionally independent of $X$ given the projected covariate $\hat{\beta}_1^{\top}X$ where
\begin{eqnarray*}
\hat{\beta}_1 &=& (0.0463, -0.1078, 0.0383, 0.2447, -0.0322, -0.0436, 0.0545, 0.2229, 0.1173, -0.1718, \\
                     &&  0.0491, -0.0494, 0.7479, -0.0965, 0.5022, -0.0165)^{\top},
\end{eqnarray*}
is the first direction obtained by CSE. The scatter plot of $Y$ against $\hat{\beta}_1^{\top}X$ is presented in Figure~1(a). It indicates that  a linear regression model for $(Y, X)$ is not reasonable.
$$ \rm Figure \ 1 \ about \ here $$
To further exhaust  possible projected covariates, we consider the second projected covariate $\hat{\beta}_2^{\top}X$ obtained by CSE. The scatter plot of $Y$ against $(\hat{\beta}_1^{\top}X, \hat{\beta}_2^{\top}X)$ is presented in Figure~2.
$$ \rm Figure \ 2 \ about \ here $$
This figure shows that the second projected covariate $\hat{\beta}_2^{\top}X$ has no information in predicting the response $Y$, as the plot along $\hat{\beta}_2^{\top}X$ is almost invariable. This means that the projection of the data onto the subspace $\hat{\beta}_1^{\top}X$ would already contain most of regression information of $(Y, X)$. Figure~1(a) seems to suggest a quadratic polynomial of $\hat{\beta}_1^{\top}X$ to fit the data.
Hence we use the following regression mode:
$$ Y=\theta_1+\theta_2 (\beta^{\top}X)+\theta_3(\beta^{\top}X)^2 +\varepsilon.$$
Figure~1(b) adds the fitted curve on the scatter plot.
The value of the test statistic $ACM_n^2=0.1038$ and the $p$-value is about $0.83$. Therefore the above regression model is plausible.

\section{Discussions}
In this paper, we investigate model checking for regressions when the dimension of predictors diverges to infinity as the sample size tends to infinity. Three remarkable features are worthwhile to discuss. First, although the empirical process is similar to that in Stute and Zhu (2002), it involves much more difficult estimation issues in the construction procedure of test statistics. Second, as the Khmaladze martingale transformation has become an important methodology for model checking as its asymptotically distribution-free property, we suggest another way to construct the transformation, rather than directly targeting the limit of shift terms in the fixed dimension cases. The transformed process still has the same limiting Gaussian process as that with  fixed dimension. This provides us an easy way to handle the cases with divergent dimension. Third, the model adaptation property shows its advantage in maintaining the significance level and enhancing power performance. The research also leaves some unsolved topics. An important topic is about how to relax the condition on the diverging rate of the dimension. In this paper, we cannot have faster rate than $p=o(n^{1/4})$ in general although for some special regression models such as linear models, it can achieve $p=o(n^{1/2}).$ This is mainly because of technical difficulties in estimation. Thus, to attack this problem, we need to improve the asymptotic properties of involved estimators. This is beyond the scope of this paper and deserves further studies.

\section{Appendix}
\subsection{Regularity Conditions}
In this subsection we present some regularity conditions for the theoretical results. Although these conditions may not be the weakest possible, they make technical arguments easy to understand. In the following, $C$ always stands for a constant which may be different in different cases.

First, we give some regularity conditions for the norm consistency of $(\hat{\beta}_n, \hat{\theta}_n)$  to $(\tilde{\beta}_{0}, \tilde{\theta}_0)$ and the decomposition of $\left( \begin{array}{c}
\hat{\beta}_n-\tilde{\beta}_{0}\\
\hat{\theta}_n-\tilde{\theta}_0\\
\end{array}
\right)$.\\
(A1) The matrix $\Sigma_n$ is positive definite and satisfies the following condition
$$ 0< \underline{\lambda} \leq \lambda_{min}(\Sigma_n) \leq \lambda_{max}(\Sigma_n) \leq \overline{\lambda} < \infty  \quad {\rm for \ all} \ {n}, $$
where $\lambda_{min}(\Sigma_n)$ and $\lambda_{max}(\Sigma_n)$ are the smallest and largest eigenvalues of $\Sigma$, respectively.

The first to third derivatives of the regression function $g(\cdot)$ satisfy the conditions: \\
(A2) $E|Y|^4 \leq C$, $E|e|^8 \leq C$; $E|g'_{j}(\tilde{\beta}_{0}, \tilde{\theta}_0, X)|^8 \leq C$;\\
(A3) $|g(\beta^{\top}x, \theta)| \leq F(x)$ with $E F(X)^4 \leq C$ for all $(\beta, \theta)$; \\
(A4) $|g'_{j}(\beta, \theta, x)| \leq F_j(x)$ with $E F_j(X)^4 \leq C$ for all $j$ and $(\beta, \theta)$; \\
(A5) $|g''_{jk}(\beta, \theta, x)| \leq F_{jk}(x)$ with $E F_{jk}(X)^4 \leq C$ for all $j, k$, and $(\beta, \theta)$; \\
(A6) $|g'''_{jkl}(\beta, \theta, x)| \leq F_{jkl}(x)$ with $E F_{jkl}(X)^4 \leq C$ for all $j, k,l$, and $(\beta, \theta)$; \\
where $g'_j(\beta, \theta, x)$ is the $j$-th component of $g'(\beta, \theta, x)$, $g''_{jk}(\beta, \theta, x)$ is the $(j,k)$-element of  $g''(\beta, \theta, x)$, and $g'''_{jkl}(\beta, \theta, x)$ is the $(j,k,l)$-element of  $g'''(\beta, \theta, x)$.

Condition (A1) is similar to the regularity condition on the Fisher information matrix $I_n$ proposed by Fan and Peng (2004), where the Fisher information matrix $I_n$ plays the same role in deriving the asymptotic theory as the matrix $\Sigma_n$ does here. Conditions (A2)-(A6) are standard for nonlinear least squares estimation, see, e.g., Jennrich (1969) and White (1981).

Next, we present some regularity condition for the convergence of the adaptive-to-model residual marked empirical process.\\
(B1) There exists a constant $C$ such that if $\|\beta- \kappa \beta_{0}\| \leq C \sqrt{p\log{n}/n}$, then
$$\mathbb{P}(\{ \beta^{\top}X \leq u \} \vartriangle \{ \kappa \beta_{0}^{\top}X \leq u \} ) \leq \sqrt{2p\log{n}/n},$$
where $\vartriangle $ denotes the symmetric difference of two sets. This condition is given by Zhu (1993) who showed  the existence of distributions satisfying this condition. \\
(B2) If $M_n(u)=E[g'(\beta_{0}, \theta_0, X)I(\kappa \beta_{0}^{\top}X \leq u)]$,
 $\|M_n(u)\|= O(1)$ uniformly in $u$. 
\\
(B3) For any unit non-random vector $\gamma \in \mathbb{R}^p$, there exist $F_{Y}$-integrable functions $h_i(t)$ such that
\begin{eqnarray*}
&&E[\gamma^{\top}XI(Y \leq t)] \longrightarrow h_1(t), \\
&&E[\gamma^{\top}Xf_{Y|X}(t)G(X)] \longrightarrow h_2(t),
\end{eqnarray*}
where $G(X)$ is given by (\ref{3.8}) and $f_{Y|X}$ is the conditional density of $Y$ given $X$.

\subsection{Lemmas}
In this subsection we present some Lemmas that will be needed in proving the propositions and theorems. Since we consider the empirical process with diverging dimension, there are no relevant results available in the literature. Thus, in the following Lemmas, we give the results about the convergence rate of the involved  empirical process, which are different from the classical ones with fixed dimension in the literature.

\begin{lemma}\label{lemma 1}
Suppose $A_n(u)$ is nonsingular for all $u$, then we have
\begin{equation*}
 Cov[T_n V_n^0(s), T_n V_n^0(t)] = Cov[V_n^0(s), V_n^0(t)]=\psi_n(s \wedge t),
\end{equation*}
that is, (\ref{3.5}) and thus (\ref{3.6}) hold.
\end{lemma}

\textbf{Proof.} Assume $t \leq s$. By the definition of $ T_n V_n^0$ and the Fubini Theorem, the left-hand side of~(\ref{3.5}) equals
\begin{eqnarray*}
&&   Cov[V_n^0(s), V_n^0(t)]\\
&&   -\int_{-\infty}^{t} a_n(z)^{\top}A_n(z)^{-1} \int_{z}^{t} a_n(v) \sigma_n^2(v) F_{\kappa \beta_{0}}(dv) \psi_n(dz) \\
&&   -\int_{-\infty}^{t} a_n(z)^{\top}A_n(z)^{-1} \int_{z}^{s} a_n(v) \sigma_n^2(v) F_{\kappa \beta_{0}}(dv) \psi_n(dz) \\
&&   +\int_{-\infty}^{t} \int_{-\infty}^{s} a_n(z_1)^{\top}A_n(z_1)^{-1} A_n(z_1 \vee z_2) A_n(z_2)^{-1} a_n(z_2) \psi_n(dz_1) \psi_n(dz_2).
\end{eqnarray*}
It is easy to see that the sum of the last three terms is equal to zero. Thus we complete the proof. \hfill$\Box$

Next we consider the convergence rate of the involved empirical processes in the diverging dimension. Let $F(x)$ be a fixed function and $\mathcal{F}_n$ be a VC-class of functions with a VC-index $V(\mathcal{F}_n)$ which may depend on $n$. Let $N_i(\epsilon, \mathcal{F}_n, L_i(Q))$ be the covering number of $\mathcal{F}_n$ with respective to the seminorm $L_i(Q)$. See e.g. Pollard (1984) for details.
Suppose $\sup_{\mathcal{F}_n}|f(x)| \leq 1$ for any $n$ and $x \in \mathbb{R}^p$ and
$$
\sup_{Q} N_2(\epsilon, \mathcal{F}_n, L_2(Q)) \leq A_n \epsilon^{-W_n} \quad {\rm for\ all} \quad 0<\epsilon<1.
$$
Set $\tilde{\mathcal{F}}_n=\{F(x)f(x): f(x) \in \mathcal{F}_n\}$. By some elementary calculations, we have
$$
N_1(\sqrt{QF^2} \epsilon, \tilde{\mathcal{F}}_n, L_1(Q)) \leq  N_2(\epsilon, \mathcal{F}_n, L_2(Q)),
$$
whence
$$
\sup_{Q} N_1(\sqrt{QF^2} \epsilon, \tilde{\mathcal{F}}_n, L_1(Q)) \leq A_n \epsilon^{-W_n} \quad {\rm for\ all} \quad 0<\epsilon<1.
$$

\begin{lemma}\label{lemma 2}
Let $\tau_n$ and $\epsilon_n$ be positive sequences. If $E|F|^4 < \infty$ and $Var(P_n Ff)/(4\epsilon_n)^2 \leq 1/2$ for $n$ large enough, then
\begin{eqnarray*}
\mathbb{P} \{\sup_{\mathcal{F}_n} |P_n Ff-PFf|> 8 \epsilon_n \} \leq 8 A_n \tau_n^{W_n} \epsilon_n^{-W_n} \exp(-\frac{1}{2}n\epsilon_n^2/\tau_n^2)+4 \mathbb{P}\{ P_n F^2 \geq \tau_n^2\}
\end{eqnarray*}
where $A_n$ and $W_n$ are constants which may depend on $n$.
\end{lemma}

\textbf{Proof.} The proof is similar to Theorem 37 in Chapter~2 of Pollard (1984) and Theorem~3.1 in Zhu (1993).
Since $Var(P_n Ff)/(4\epsilon_n)^2 \leq 1/2$
for $n$ large enough, by the formula (30) in Chapter~2 of Pollard (1984), we have
$$
\mathbb{P} \{\sup_{\mathcal{F}_n} |P_n Ff-PFf|> 8 \epsilon_n \} \leq 4 \mathbb{P} \{\sup_{\mathcal{F}_n} |P_n^o Ff|> 2 \epsilon_n \},
$$
Conditionally on $\{X_1, \cdots, X_n\}$. Using the same argument as that for proving the inequality (31) in Chapter~2 of Pollard (1984), it follows that
\begin{eqnarray*}
\mathbb{P} \left\{\sup_{\mathcal{F}_n} |P_n^o Ff|> 2 \epsilon_n | X_1, \cdots, X_n \right\}
&\leq& 2 N_1(\epsilon_n, \tilde{\mathcal{F}}_n, L_1(P_n)) \exp \left( -\frac{1}{2} n \epsilon_n^2 /(\max_j P_n g_j^2) \right)\wedge 1  \\
&\leq& 2 A_n (P_nF^2)^{W_n/2} \epsilon_n^{-W_n} \exp \left( -\frac{1}{2} n \epsilon_n^2 /P_n F^2 \right) \wedge 1.
\end{eqnarray*}
Taking  expectation, we obtain that
\begin{eqnarray*}
\mathbb{P} \{\sup_{\mathcal{F}_n} |P_n^o Ff|> 2 \epsilon_n \} \leq 2 A_n \tau_n^{W_n} \epsilon_n^{-W_n} \exp(-\frac{1}{2}n\epsilon_n^2/\tau_n^2)+ \mathbb{P}\{ P_n F^2 \geq \tau_n^2\}.
\end{eqnarray*}
Consequently,
\begin{eqnarray*}
\mathbb{P} \{\sup_{\mathcal{F}_n} |P_n Ff-PFf|> 8 \epsilon_n \} \leq 8 A_n \tau_n^{W_n} \epsilon_n^{-W_n} \exp(-\frac{1}{2}n\epsilon_n^2/\tau_n^2) + 4 \mathbb{P}\{ P_n F^2 \geq \tau_n^2\}.
\end{eqnarray*}
Therefore, we complete the proof.  \hfill$\Box$

\begin{lemma}\label{lemma 3}
If $|g'_j(\beta_{0}, \theta_0, x)| \leq F(x)$ and $E[F(X)]^4 < \infty$, then we have
$$
\sup_{u}\| \frac{1}{n}\sum_{i=1}^{n} g'(\beta_{0}, \theta_0, X_i)I(\kappa \beta_{0}^{\top}X_i \leq u)-M_n(u) \|=o_p(\sqrt{\frac{p}{n}}p^{1/4} \log{n}),
$$
where $M_n(u)=E[g'(\beta_{0}, \theta_0, X)I(\kappa \beta_{0}^{\top}X \leq u)]$.
\end{lemma}

\textbf{Proof.} Fix $\epsilon >0$ and let $\epsilon_n^2= \epsilon^2 (\log{n})^2 \sqrt{p}/n$. We have
\begin{eqnarray*}
&&  \mathbb{P}\left( \sup_{u} \|\frac{1}{n}\sum_{i=1}^{n} g'(\beta_{0}, \theta_0, X_i) I(\kappa \beta_{0}^{\top}X_i \leq u)-M_n(u)\| > 8
\sqrt{p} \epsilon_n  \right) \\
&=& \mathbb{P}\left( \sup_{u} \|\frac{1}{n}\sum_{i=1}^{n} g'(\beta_{0}, \theta_0, X_i)I(\kappa \beta_{0}^{\top}X_i \leq u)-M_n(u)\|^2 > 64p \epsilon_n^2  \right) \\
&=& \mathbb{P}\left( \sup_{u} \sum_{j=1}^{p} |\frac{1}{n}\sum_{i=1}^{n} g'_j(\beta_{0}, \theta_0, X_i)I(\kappa \beta_{0}^{\top}X_i \leq u)-M_{nj}(u) |^2 > 64p \epsilon_n^2  \right)\\
&\leq& \sum_{j=1}^{p} \mathbb{P}\left( \sup_{u} |\frac{1}{n}\sum_{i=1}^{n} g'_j(\beta_{0}, \theta_0, X_i)I(\kappa \beta_{0}^{\top}X_i \leq u)-M_{nj}(u)|^2 > 64 \epsilon_n^2  \right)\\
&=& \sum_{j=1}^{p} \mathbb{P}\left( \sup_{u}  |\frac{1}{n}\sum_{i=1}^{n} g'_j(\beta_{0}, \theta_0, X_i)I(\kappa \beta_{0}^{\top}X_i \leq u)-M_{nj}(u)| > 8 \epsilon_n  \right).
\end{eqnarray*}
For every term in the last sum, we use Lemma~2. Let
\begin{eqnarray*}
\mathcal{F}_{1n}         &=&\{f_{u}(x)=I(\kappa \beta_{0}^{\top}x \leq u): u \in \bar{\mathbb{R}} \}\\
\tilde{\mathcal{F}}_{1n} &=&\{ g'_j(\beta_{0}, \theta_0, x)f_{u}(x): f_u(x) \in \mathcal{F}_{1n} \}.
\end{eqnarray*}
It is easy to see that $\mathcal{F}_{1n} $ is a VC-class with VC-index $V(\mathcal{F}_{1n} )= 2$. By Theorem 2.6.7 in Van Der Vaart and Wellner (1996), we obtain that
\begin{eqnarray*}
&\sup_{Q} N_2(\epsilon, \mathcal{F}_{1n}, L_2(Q)) \leq 2K(16e)^2 \epsilon^{-2},\\
&\sup_{Q} N_1(\epsilon (Q{g'}_j^{2})^{1/2}, \tilde{\mathcal{F}}_{1n}, L_1(Q)) \leq 2K(16e)^2 \epsilon^{-2},
\end{eqnarray*}
where $K$ is a universal constant. Set $A=2K(16e)^2$ and $\tau_n^2=\sqrt{p \log{n}}$. Lemma~2 leads to
\begin{eqnarray*}
\mathbb{P}\left( \sup_{\mathcal{F}_n} |P_n g'_j f_{\beta, u}- Pg'_j f_{\beta, u}| > 8 \epsilon_n  \right)
\leq 8 A \tau_n^2 \epsilon_n^{-2} \exp( -\frac{n\epsilon_n^2}{2\tau_n^2}) + 4 \mathbb{P}\{ P_n F^2 \geq \tau_n^2\},
\end{eqnarray*}
whence
\begin{eqnarray*}
&&  \mathbb{P}\left( \sup_{u} \|\frac{1}{n}\sum_{i=1}^{n} g'(\beta_{0}, \theta_0, X_i) I(\kappa \beta_{0}^{\top}X_i \leq u)-M_n(u)\| > 8
\sqrt{p} \epsilon_n  \right) \\
&\leq& 8 A p \tau_n^2 \epsilon_n^{-2} \exp( -\frac{n\epsilon_n^2}{2\tau_n^2}) + 4 p \mathbb{P}\{ P_n F^2 \geq \tau_n^2\}\\
&\leq& 8 A  \frac{pn}{\epsilon^2 (\log{n})^{3/2}}   \exp(-\frac{1}{2}\epsilon \sqrt{\log{n}} \log{n}) + 4PF^4/\log{n}\\
&=& o(1).
\end{eqnarray*}
Therefore, we obtain the result.    \hfill$\Box$

\begin{lemma}\label{lemma 4}
Let $\mathcal{F}$ be a permissible class of functions with $|f| \leq 1$ and $P|f| \leq \delta$ for all $f \in \mathcal{F}$. Then
$$ \mathbb{P}(\sup_{\mathcal{F}} P_n|f| > 8\delta) \leq 4 P[N_1(\delta, \mathcal{F}, L_1(P_n)) \exp(-n\delta^2) \wedge 1]. $$
\end{lemma}

For the definition of ``a permissible class of functions '', one can refer to Chapter 2 of Pollard (1984) for details.

\textbf{Proof.} This Lemma is a slightly modified version of Lemma~33 in Chapter 2 of Pollard (1984) as we need the result with diverging $p$. But the proof can be very similar and thus   is omitted here.  \hfill$\Box$

\begin{lemma}\label{lemma 5}
Let $\delta_n$ and $\alpha_n$ be positive real valued sequences. Suppose $P|f| \leq \delta_n$ for all $f(x) \in \mathcal{F}_n$ and $Var(P_n F f)/(4\epsilon_n)^2 \leq 1/2$ for $n$ large enough. If $E|F|^8 < \infty$, then
\begin{eqnarray*}
\mathbb{P} \{\sup_{\mathcal{F}_n} |P_n Ff-PFf|> 8 \epsilon_n \}
&\leq& 8A_n \alpha_n^{W_n} \epsilon_n^{-W_n} \exp\{-\frac{1}{2} n \epsilon_n^2/[\alpha_n^2(8\delta_n)^{\frac{3}{4}}]\}+ \\
&& 4\mathbb{P}(P_nF^8 > \alpha_n^8) + 16 A_n \delta_n^{-W_n} \exp(-n\delta_n^2).
\end{eqnarray*}
where $A_n$ and $W_n$ are constants which may depend on $n$.
\end{lemma}

\textbf{Proof.} The proof is similar to Theorem 37 in Chapter~2 of Pollard (1984) and Theorem 3.1 of Zhu (1993). Since $Var(P_n Ff)/(4\epsilon_n)^2 \leq 1/2$, similar to the proof for  Lemma 2, we have
$$
\mathbb{P} \{\sup_{\mathcal{F}_n} |P_n Ff-PFf|> 8 \epsilon_n \} \leq 4 \mathbb{P} \{\sup_{\mathcal{F}_n} |P_n^o Ff|> 2 \epsilon_n \}.
$$
Conditionally on $\{X_1, \cdots, X_n\}$, we obtain
\begin{eqnarray*}
\mathbb{P} \left\{\sup_{\mathcal{F}_n} |P_n^o Ff|> 2 \epsilon_n | X_1, \cdots, X_n \right\}
&\leq& 2 N_1(\epsilon_n, \tilde{\mathcal{F}}_n, L_1(P_n)) \exp \left( -\frac{1}{2} n \epsilon_n^2 /(\max_j P_n F^2 f_j^2) \right)\wedge 1  \\
&\leq& 2 A_n (P_nF^2)^{W_n/2} \epsilon_n^{-W_n} \exp \left( -\frac{1}{2} n \epsilon_n^2 /(\max_j [P_n F^8]^{\frac{1}{4}} [P_n|f_j|]^{\frac{3}{4}}) \right) \wedge 1
\end{eqnarray*}
Take expectation to obtain
\begin{eqnarray*}
\mathbb{P} \{\sup_{\mathcal{F}_n} |P_n^o Ff|> 2 \epsilon_n \}
&\leq& 2A_n \alpha_n^{W_n} \epsilon_n^{-W_n} \exp\{-\frac{1}{2} n \epsilon_n^2/[\alpha_n^2(8\delta_n)^{\frac{3}{4}}]\}+ \mathbb{P}(P_nF^8 > \alpha_n^8) + \\
&&\mathbb{P}\{ \sup_{\mathcal{F}_n} P_n |f| \geq 8\delta_n \}\\
&\leq& 2A_n \alpha_n^{W_n} \epsilon_n^{-W_n} \exp\{-\frac{1}{2} n \epsilon_n^2/[\alpha_n^2(8\delta_n)^{\frac{3}{4}}]\}+ \mathbb{P}(P_nF^8 > \alpha_n^8) + \\
&& 4 A_n \delta_n \exp(-n\delta_n^2).
\end{eqnarray*}
The last inequality is due to Lemma 4. Altogether we complete the proof.   \hfill$\Box$

\begin{lemma}\label{lemma 6}
Suppose $H_0$ and condition (B1) hold. If $E\varepsilon^8 < \infty$ and $p^4/n \to 0$, then we have
$$
\sup_{u} |\frac{1}{\sqrt{n}} \sum_{i=1}^n \varepsilon_i [I(\hat{B}_n^{\top}X_i \leq u)-I(\kappa \beta_{0}^{\top} X_i \leq u)]|=o_p(1)
$$
\end{lemma}

\textbf{Proof.} Fix $\epsilon > 0$ and set $H_C=\{ \beta: \beta \in \mathbb{R}^p, \|\beta\| \leq 1, \|\beta-\kappa \beta_{0}\| \leq C \sqrt{p\log{n}/n}\}$. Since $\|\hat{B}_n-\kappa \beta_{0}\|=O_p(\sqrt{p/n})$, by condition (B1), it suffices to prove
\begin{eqnarray*}
\mathbb{P}\left\{ \sup_{\beta \in H_C} \sup_{u} |\frac{1}{n} \sum_{i=1}^n \varepsilon_i [I(\beta^{\top}X_i \leq u)-I(\kappa \beta_{0}^{\top} X_i \leq u)] | > \frac{8}{\sqrt{n}} \epsilon \right\}  \rightarrow 0,
\end{eqnarray*}
Let
\begin{eqnarray*}
\mathcal{F}_{2n}&=&\{f_{\beta, u}(x)=I(\beta^{\top}x \leq u)-I(\kappa \beta_{0}^{\top}x \leq u): \beta \in H_C, u \in \mathbb{R}\},\\
\mathcal{F}_{2n}^1&=&\{f_{\beta, u}(x)=I(\beta^{\top}x \leq u): \beta \in H_C, u \in \mathbb{R}\},\\
\mathcal{F}_{2n}^2&=&\{f_{u}(x)=I(\kappa \beta_{0}^{\top}x \leq u): u \in \mathbb{R}\}.
\end{eqnarray*}
Then it is easy to see that
$$
N_2(2\epsilon, \mathcal{F}_{2n}, L_2(Q)) \leq N_2(\epsilon, \mathcal{F}_{2n}^1, L_2(Q)) \cdot N_2(\epsilon, \mathcal{F}_{2n}^2, L_2(Q)).
$$
Since $\mathcal{F}_{2n}^1$ and $\mathcal{F}_{2n}^2$ are both VC-classes with the VC-index $p+2$ and $2$ respectively, by Theorem~2.6.7 in Van Der Vaart and Wellner (1996), we obtain that
\begin{eqnarray*}
  \sup_{Q} N_2(\epsilon, \mathcal{F}_{2n}^1, L_2(Q)) &\leq& K(p+2)(16e)^{p+2} \epsilon^{-2(p+1)}, \\
  \sup_{Q} N_2(\epsilon, \mathcal{F}_{2n}^2, L_2(Q)) &\leq& 2K(16e)^2 \epsilon^{-2},
\end{eqnarray*}
whence
$$
\sup_{Q} N_2(\epsilon, \mathcal{F}_{2n}, L_2(Q)) \leq K(p+2)(64e)^{p+2} \epsilon^{-2(p+2)}.
$$
Let $\tilde{\mathcal{F}}_{2n}=\{\varepsilon f_{\beta, u}(x): f_{\beta, u} \in \mathcal{F}_{2n}  \}$. It follows that
$$
\sup_{Q} N_1(\epsilon \sqrt{Q\varepsilon^2}, \tilde{\mathcal{F}}_{2n}, L_1(Q)) \leq K(p+2)(64e)^{p+2} \epsilon^{-2(p+2)}.
$$
Let $\delta_n=\sqrt{2p\log{n}/n}$, $\alpha_n^8=\log{n}$ and $\epsilon_n= \epsilon/\sqrt{n}$. Since
\begin{eqnarray*}
&&P|f_{\beta, u}|=\mathbb{P}(\{ \beta^{\top}X \leq u \} \vartriangle \{ \kappa \beta_{0}^{\top}X \leq u \} ) \leq \delta_n,\\
&&\frac{Var(P_n \varepsilon f_{\beta, u})}{(4\epsilon_n)^2} \leq \frac{(P \varepsilon^4)^{1/2} Pf_{\beta, u}^2}{16 \epsilon^2} \leq \frac{1}{2} \quad {\rm for} \ n {\rm \ large \ enough},
\end{eqnarray*}
by Lemma 5, we have
\begin{eqnarray*}
\mathbb{P}\left\{ \sup_{\mathcal{F}_{2n}}|P_n \varepsilon f_{\beta, u}| > \frac{8}{\sqrt{n}} \epsilon \right\} &\leq& 8K(p+2)(64e)^{p+2} \alpha_n^{2(p+2)} \epsilon_n^{-2(p+2)} \exp(-\frac{n \epsilon_n^2}{2 \alpha_n^2 (8\delta_n)^{\frac{3}{4}}})+ \\
&& 4\mathbb{P}(P_n \varepsilon^8 > \alpha_n^8)+16K(p+2)(64e)^{p+2} \delta_n^{-2(p+2)}\exp(-n\delta_n^2).
\end{eqnarray*}
Since $p^4/n \to 0$, it follows that $\mathbb{P}\left\{ \sup_{\mathcal{F}_{2n}}|P_n f_{\beta, u}| > \frac{8}{\sqrt{n}} \epsilon \right\} \to 0$ which completes our proof.  \hfill$\Box$

Next, we consider the convergence rate of the following process
$$
\frac{1}{n} \sum_{i=1}^n g'(\beta_{0}, \theta_0, X_i) [I(\hat{B}_n^{\top}X_i \leq u)-I(\kappa \beta_{0}^{\top} X_i \leq u)].
$$


\begin{lemma}\label{lemma 7}
Let $\tilde{M}_n(\beta, u)=E\{ g'(\beta_{0}, \theta_0, X)[I(\beta^{\top}X_i \leq u)- I(\kappa \beta_{0}X_i \leq u)]\}$.
Suppose  conditions (A2) and (B1) hold. If $p^4/n \to 0$, then
$$
\sup_{u} \|\frac{1}{n} \sum_{i=1}^n g'(\beta_{0}, \theta_0, X_i) [I(\hat{B}_n^{\top}X_i \leq u)-I(\kappa \beta_{0}^{\top} X_i \leq u)]- \tilde{M}_n(\hat{B}_n, u)\| =o_p(\sqrt{\frac{p^{3/2} \log{n}}{n}}).
$$
\end{lemma}

\textbf{Proof.} Fix $\epsilon >0$ and set $\epsilon_n=\epsilon \sqrt{p^{1/2}\log{n}/n}$, $\delta_n=\sqrt{2p\log{n}/n}$, and $\alpha_n^8=p \log{n}$. Similar to the proof for Lemma 6, it suffices to prove
$$
\mathbb{P}\left\{ \sup_{\beta \in H_C} \sup_{u}|\frac{1}{n} \sum_{i=1}^n g'(\beta_{0}, \theta_0, X) [I(\beta^{\top}X_i \leq u)-I(\kappa \beta_{0}^{\top} X_i \leq u)]- \tilde{M}_n(\beta, u)| > 8\sqrt{p}\epsilon_n \right\}  \rightarrow 0.
$$
By the same argument  for proving Lemma~3, we obtain
\begin{eqnarray*}
&&     \mathbb{P}\left\{ \sup_{\beta \in H_C} \sup_{u}|\frac{1}{n} \sum_{i=1}^n g'(\beta_{0}, \theta_0, X) [I(\beta^{\top}X_i \leq u)-I(\kappa \beta_{0}^{\top} X_i \leq u)]- \tilde{M}_n(\beta, u)| > 8\sqrt{p}\epsilon_n \right\} \\
&\leq& \sum_{j=1}^{p+d} \mathbb{P}\left\{ \sup_{\beta \in H_C} \sup_{u}|\frac{1}{n} \sum_{i=1}^n g'_{j}(\beta_{0}, \theta_0, X) [I(\beta^{\top}X_i \leq u)-I(\kappa \beta_{0}^{\top} X_i \leq u)]-\tilde{M}_{nj}(\beta, u)| > 8\epsilon_n \right\}.
\end{eqnarray*}
For every term in the last sum, we use Lemma 5 to derive the result. Let
\begin{eqnarray*}
\tilde{\mathcal{F}}_{3n} &=& \{ g'_j(\beta_{0}, \theta_0, x)f_{\beta, u}(x): f_{\beta, u}(x) \in \mathcal{F}_{3n} \}, \\
\mathcal{F}_{3n} &=&\{f_{\beta, u}(x)=I(\beta^{\top}x \leq u)-I(\kappa \beta_{0}^{\top}x \leq u): \beta \in H_C, u \in \mathbb{R}\}.
\end{eqnarray*}
Then we have
\begin{eqnarray*}
\sup_{Q} N_2(\epsilon, \mathcal{F}_{3n}, L_2(Q)) &\leq& K(p+2)(64e)^{p+2} \epsilon^{-2(p+2)}, \\
\sup_{Q} N_1(\epsilon \sqrt{Q {g'}_j^2}, \tilde{\mathcal{F}}_{3n}, L_1(Q)) &\leq& K(p+2)(64e)^{p+2} \epsilon^{-2(p+2)},
\end{eqnarray*}
where $K$ is a universal constant free of $n$.

Recall $\epsilon_n=\epsilon \sqrt{p^{1/2}\log{n}/n}$ and $\delta_n =\sqrt{2p\log{n}/n}$. By conditions (A2) and (B1), we have
$$ P|f_{\beta, u}| = \mathbb{P}(\{ \beta^{\top}X \leq u \} \vartriangle \{ \kappa \beta_{0}^{\top}X \leq u \}) \leq  \delta_n; $$
$$ \frac{Var(P_n g'_j f_{\beta, u})}{(4\epsilon_n)^2}  \leq  \frac{P{g'_j}^2}{16 \epsilon^2 p \log{n}}  < \frac{1}{2} \quad {\rm for} \ n \ { \rm large \ enough}. $$
By Lemma 5, we obtain that
\begin{eqnarray*}
&&\mathbb{P}\left\{ \sup_{\beta \in H_C} \sup_{u}|\frac{1}{n} \sum_{i=1}^n g'_{j}(\beta_{0}, \theta_0, X) [I(\beta^{\top}X_i \leq u)-I(\kappa \beta_{0}^{\top} X_i \leq u)]- \tilde{M}_{nj}(\beta, u)| > 8\epsilon_n \right\} \\
&&\leq 8K(p+2)(64e)^{p+2} \alpha_n^{2(p+2)} \epsilon_n^{-2(p+2)} \exp(-\frac{n\epsilon_n^2}{2 \alpha_n^2 (8\delta_n)^{\frac{3}{4}}}) + 4\mathbb{P}(P_n {g'_j}^8 >\alpha_n^8) \\
&& +16K(p+2)(64e)^{p+2} \delta_n^{-2(p+2)} \exp(-n\delta_n^2),
\end{eqnarray*}
whence
\begin{eqnarray*}
&&  \mathbb{P}\left\{ \sup_{\beta \in H_C} \sup_{u}|\frac{1}{n} \sum_{i=1}^n g'(\beta_{0}, \theta_0, X) [I(\beta^{\top}X_i \leq u)-I(\kappa \beta_{0}^{\top} X_i \leq u)]- \tilde{M}_n(\beta, u)| > 8\sqrt{p}\epsilon_n \right\}  \\
&&\leq 8K(p+d)(p+2)(64e)^{p+2} \alpha_n^{2(p+2)} \epsilon_n^{-2(p+2)} \exp(-\frac{n\epsilon_n^2}{2 \alpha_n^2 (8\delta_n)^{\frac{3}{4}}}) + \sum_{j=1}^{p+d} 4\mathbb{P}(P_n {g'_j}^8 >\alpha_n^8) \\
&& +16K(p+d)(p+2)(64e)^{p+2} \delta_n^{-2(p+2)} \exp(-n\delta_n^2),
\end{eqnarray*}
Since $p^4/n \to 0$, it follows that
the right-hand side of the above inequality tends to zero. Hence we complete the proof.   \hfill$\Box$

In the next lemma, we give the convergence rate of the kernel regression function estimator $\hat{r}_n(y)$. Let $(X_1, Y_1), \cdots, (X_n, Y_n)$ be a sample from $(X, Y)$, $f(y)$ be the density function of $Y$ with a support $\mathcal{C}$ and $m_n(y)=r_n(y)f(y)=\{E(X_{11}|Y=y)f(y), \cdots, E(X_{1p}|Y=y)f(y)\}^{\top}$. Suppose that
\begin{eqnarray*}
& 0 < \inf_{y \in \mathcal{C}} f(y) \leq \sup_{y \in \mathcal{C}} f(y) < \infty, \\
& E(X_{1i}^2|Y=y) \leq C  \ {\rm for \ all}\ y, 1 \leq i \leq p.
\end{eqnarray*}
It follows that $m_n(y)=O_p(\sqrt{p})$ uniformly in $y$. Set
$$
\hat{m}_n(y)=\frac{1}{nh}\sum_{j=1}^{n} X_j K(\frac{y-Y_j}{h}) \quad {\rm and} \quad \hat{f}(y)=\frac{1}{nh}\sum_{j=1}^{n} K(\frac{y-Y_j}{h}).
$$
Then $\hat{r}_n(y)=\hat{m}_n(y)/\hat{f}(y)$. Here $K(\cdot)$ is the kernel function and $h$ is a bandwidth.

\begin{lemma}\label{lemma 8}
Suppose the above conditions hold. If $ph^3 \log{n} \to 0$, then we have
\begin{eqnarray*}
\sup_{y} \|\hat{m}_n(y)-m_n(y)\|&=&O_p(\sqrt{\frac{p \log{n}}{nh^2}}) +O_p(\sqrt{p}h^4),\\
\sup_{y} \|\hat{r}_n(y)-r_n(y)\|&=&O_p(\sqrt{\frac{p \log{n}}{nh^2}}) +O_p(\sqrt{p}h^4).
\end{eqnarray*}
\end{lemma}

\textbf{Proof.} Let $r_{ni}(y), \hat{r}_{ni}(y), m_{ni}(y)$, and $\hat{m}_{ni}(y)$ be the $i$-th component of $r_{n}(y), \hat{r}_{n}(y), m_{n}(y)$, and $\hat{m}_{n}(y)$ respectively. For fixed $\epsilon$, set $\epsilon_n^2=\frac{\log{n}}{n} \epsilon^2$, $\delta_n=h$, and $\alpha_n^8=p \log{n}$. Then
\begin{eqnarray*}
\mathbb{P}(\sup_{y} \|\hat{m}_{n}(y)-E\hat{m}_{n}(y)\| > 8\sqrt{\frac{p \log{n}}{nh^2}} \epsilon ) \leq \sum_{i=1}^{p} \mathbb{P}(\sup_{y}| h \hat{m}_{ni}(y)-hE\hat{m}_{ni}(y)|>8\epsilon_n)
\end{eqnarray*}
Define
\begin{eqnarray*}
  \mathcal{F}_{4n} = \left\{ f_{y,h}(u): f_{y,h}(u)=K(\frac{y-u}{h}), y \in \mathcal{C} \right\}.
\end{eqnarray*}
Without loss of generality, assume $|K(x)| \leq 1$ and $f(y) \leq 1$. By the arguments in Example 38 of Chapter~2 of Pollard (1984), we obtain that
$$
\sup_{Q} N_2(\epsilon, \mathcal{F}_n, L_2(Q)) \leq A \epsilon^{-W} \quad {\rm for\ all} \quad 0<\epsilon<1,
$$
where $A$ and $W$ are free of $n$. Let $ \tilde{\mathcal{F}}_{4n}=\{zf_{y,h}(u): f_{y,h}(u) \in \mathcal{F}_{4n}\}$. Then
$$
\sup_{Q} N_1(\sqrt{Qz^2}\epsilon, \tilde{\mathcal{F}}_{4n}, L_1(Q)) \leq A \epsilon^{-W}.
$$
Since
$$
P|f_{y,h}|=\int |K(\frac{y-u}{h})|f(u)du= h \int |K(u)|f(y-uh)du \leq h=\delta_n
$$
and
$$
\frac{Var(P_n zK(\frac{y-u}{h}))}{16 \epsilon_n^2} \leq \frac{EX_{1i}^2K^2(\frac{y-Y}{h})}{16n \epsilon_n^2} \leq \frac{hC \int K^2(u)du}{16n\epsilon_n^2} =\frac{Ch \int K^2(u)du}{\epsilon \log{n}} < \frac{1}{2}
$$
for $n$ large enough,  Lemma~5 yields that
\begin{eqnarray*}
&&\mathbb{P}(\sup_{y}h|\hat{m}_{ni}(y)-E\hat{m}_{ni}(y)|>8\epsilon_n) \\
&&\leq 8 A (p \log{n})^{W/8} \epsilon^{-W}(n/\log{n})^{W/2} \exp\{- \frac{\epsilon^2 \log{n}}{2 (p\log{n})^{1/4} (8h)^{3/4}}\}+4\mathbb{P}(P_n X_{1i}^8 >p \log{n})\\
&& +16A h^{-W} \exp(-nh^2),
\end{eqnarray*}
whence
\begin{eqnarray*}
&&\mathbb{P}(\sup_{y} \|\hat{m}_{n}(y)-E\hat{m}_{n}(y)\| > 8\sqrt{\frac{p \log{n}}{nh^2}} \epsilon )  \\
&& \leq  8 Ap (p \log{n})^{W/8} \epsilon^{-W}(n/\log{n})^{W/2} \exp\{- \frac{\epsilon^2 \log{n}}{2 (p\log{n})^{1/4} (8h)^{3/4}}\}+\sum_{i=1}^{p} 4\mathbb{P}(P_n X_{1i}^8 >p \log{n})\\
&& +16A ph^{-W} \exp(-nh^2).
\end{eqnarray*}
Since $ph^3 \log{n} \to 0$, it is easy to see that the right-hand side of the inequality tends to zero. Thus $\sup_{y} \|\hat{m}_{n}(y)-E\hat{m}_{n}(y)\|=o_p(\sqrt{p \log{n}/(nh^2)})$. By the arguments for proving Lemma~3.3 of Zhu and Fang (1996), we obtain that
$$
\sup_{y}|E\hat{m}_{ni}(y)-m_{ni}(y)| \leq C h^4 \int |K(u)|u^4 du.
$$
Consequently,
$$
\sup_{y}\|E\hat{m}_{n}(y)-m_{n}(y)\| \leq C \sqrt{p} h^4 \int |K(u)|u^4 du.
$$
Thus we obtain the first result. For the second, note that
$$
\|\hat{r}_n(y)-r_n(y)\| \leq \frac{\|\hat{m}_n(y)- E\hat{m}_n(y)\|}{|\hat{f}(y)|} + \| E\hat{m}_n(y) \||\frac{1}{\hat{f}(y)}-\frac{1}{f(y)}| + \|\frac{E\hat{m}_n(y)- m_n(y)}{f(y)}\|
$$
and
$$
\sup_{y} |\hat{f}(y)-f(y)|=O_p(h^4)+O_p(\sqrt{\log{n}/(nh^2)}).
$$
Combining these with the uniformly boundedness of $f(y)$, the proof is concluded.   \hfill$\Box$

\subsection{Proofs of The Propositions and Theorems}
For simplicity of notations, we consider a parametric family of functions $\mathcal{G}=\{g(\beta, \cdot): \beta \in \Theta \subset \mathbb{R}^p\}$. Let $\hat{\beta}_n = \mathop{\rm argmin}\limits_{\beta} \sum_{i=1}^{n}[Y_i-g(\beta, X_i)]^2 $ and $ \tilde{\beta}_{0} = \mathop{\rm argmin}\limits_{\beta} E[Y-g(\beta, X)]^2.$


\textbf{Proof of Proposition 1.} Let $\beta=\tilde{\beta}_{0}+\alpha$ and $F(\alpha)=\sum_{i=1}^{n}[Y_i-g(\tilde{\beta}_{0}+\alpha, X_i)]g'(\tilde{\beta}_{0}+\alpha,
X_i)$. Then it suffices to show that there is a root $\alpha_n$ of $F(\alpha)$ such that $\|\alpha_n\|^2=O_p(p/n)$. Applying the results in (6.3.4) of Ortega and Rheinboldt (1970), it in turn needs to show that $\alpha^{\top}F(\alpha)<0$ for $\|\alpha\|^2=Cp/n$ where $C$ is some large enough constant.

Let $\alpha=\sqrt{p/n}U$ with $\|U\|=C$, and $e_i=Y_i-g(\tilde{\beta}_{0}, X_i)$. Using Taylor's expansion we obtain
\begin{eqnarray*}
\alpha^{\top}F(\alpha)   &=& \sum_{i=1}^{n} \alpha^{\top} g'(\tilde{\beta}_{0}+\alpha, X_i) [Y_i-g(\tilde{\beta}_{0}+\alpha, X_i)]\\
                      &=& \sum_{i=1}^{n} \alpha^{\top} g'(\tilde{\beta}_{0}+\alpha, X_i) e_i- \sum_{i=1}^{n} \alpha^{\top} g'( \tilde{\beta}_{0}+\alpha, X_i) [g(\tilde{\beta}_{0}+\alpha, X_i)- g(\tilde{\beta}_{0}, X_i)]\\
                      &=& \sum_{i=1}^{n} \alpha^{\top}  g'(\tilde{\beta}_{0}, X_i) e_i + \sum_{i=1}^{n} \alpha^{\top} e_i g''(\tilde{\beta}_{0}, X_i) \alpha +\frac{1}{2} \alpha^{\top} \sum_{i=1}^{n} \alpha^{\top} e_i g'''(\beta_{1n}, X_i)\alpha-\\
                      &&  \sum_{i=1}^{n} \alpha^{\top}g'(\tilde{\beta}_{0}, X_i)g'(\tilde{\beta}_{0}, X_i)^{\top}\alpha - \sum_{i=1}^{n} \alpha^{\top}g''(\beta_{2n}, X_i)\alpha g'(\tilde{\beta}_{0}, X_i)^{\top}\alpha -\\
                      &&  \sum_{i=1}^{n} \alpha^{\top} g'(\tilde{\beta}_{0}, X_i) \alpha^{\top}g''(\beta_{3n}, X_i) \alpha-\sum_{i=1}^{n}\alpha^{\top}g''(\beta_{2n}, X_i)\alpha \alpha^{\top}g''(\beta_{3n}, X_i)\alpha\\
                      &=:& A_{1}-A_{2}+A_{3},
\end{eqnarray*}
where $\beta_{1n}, \beta_{2n}, \beta_{3n}$ lie between $\tilde{\beta}_{0}$ and $\tilde{\beta}_{0}+\alpha$ and
\begin{eqnarray*}
  A_{1} &=& \sum_{i=1}^{n} \alpha^{\top}  g'(\tilde{\beta}_{0}, X_i) e_i, \\
  A_{2} &=& \alpha^{\top} \sum_{i=1}^{n} [g'(\tilde{\beta}_{0}, X_i)g'(\tilde{\beta}_0, X_i)^{\top}- e_i g''(\tilde{\beta}_{0}, X_i)]\alpha, \\
  A_{3} &=& \frac{1}{2} \alpha^{\top} \sum_{i=1}^{n} \alpha^{\top} \varepsilon_i g'''(\beta_{1n}, X_i)\alpha - \sum_{i=1}^{n} \alpha^{\top}g''(
            \beta_{2n}, X_i)\alpha g'(\tilde{\beta}_{0}, X_i)^{\top}\alpha- \\
        &&  \sum_{i=1}^{n} \alpha^{\top} g'(\tilde{\beta}_0, X_i) \alpha^{\top}g''(\beta_{3n}, X_i)\alpha - \sum_{i=1}^{n}\alpha^{\top}g''(
            \beta_{2n}, X_i)\alpha \alpha^{\top}g''(\beta_{3n}, X_i)\alpha.
\end{eqnarray*}
Then we have $|A_{1}| \leq \sqrt{p/n} \|U \| \|\sum_{i=1}^{n} g'(\tilde{\beta}_{0}, X_i) e_i\|$. Since $E[g'(\tilde{\beta}_{0}, X_i) e_i]=0 $, it follows that
$$
E\| \sum_{i=1}^{n} g'(\tilde{\beta}_{0}, X_i) e_i \|^2 = n E e_1^2 \| g'(\tilde{\beta}_{0}, X_1) \|^2=n \sum_{j=1}^{p} E[e_1 g'_j( \tilde{\beta}_{0}, X_1)]^2 \leq npC.
$$
Thus $A_{1}=p\|U\|O_p(1)$. Recall that $\Sigma_{n1}=E[g'(\tilde{\beta}_{0}, X)g'(\tilde{\beta}_{0}, X)^{\top}], \Sigma_{n2}= E [e g''(\tilde{\beta}_{0}, X)] $, and $\Sigma_n=\Sigma_{n1}-\Sigma_{n2}$.
Then we decompose the term $A_{2}$ as follows
$$
A_{2} = p U^{\top} \Sigma_n U + p U^{\top} \{ \frac{1}{n} \sum_{i=1}^{n} [g'(\tilde{\beta}_{0}, X_i)g'(\tilde{\beta}_{0}, X_i)^{\top}- e_i g''( \tilde{\beta}_{0}, X_i)]-\Sigma_n \} U.
$$
By condition (A2), we obtain that
\begin{eqnarray*}
   &&  E\| \frac{1}{n} \sum_{i=1}^{n} (g'(\tilde{\beta}_{0}, X_i)g'(\tilde{\beta}_{0}, X_i)^{\top}-\Sigma_{n1}) \|^2\\
   &=& \frac{1}{n^2}\sum_{j,k=1}^{p} E\{\sum_{i=1}^{n}[g'_j(\tilde{\beta}_{0}, X_i)g'_k(\tilde{\beta}_{0}, X_i)- \Sigma_{n1jk})] \}^2 \\
   &=& \frac{1}{n^2}\sum_{j,k=1}^{p} \sum_{i=1}^{n} E[g'_j(\tilde{\beta}_{0}, X_i)g'_k(\tilde{\beta}_{0}, X_i)- \Sigma_{n1jk})]^2 \\
   &\leq& \frac{1}{n}\sum_{j,k=1}^{p} E[g'_j(\tilde{\beta}_{0}, X_1)g'_k(\tilde{\beta}_{0}, X_1)]^2\\
   &\leq& \frac{p^2}{n}C.
\end{eqnarray*}
It follows that $\frac{1}{n} \sum_{i=1}^{n} [g'(\tilde{\beta}_{0}, X_i)g'(\tilde{\beta}_{0}, X_i)^{\top}-\Sigma_{n1}]= \frac{p}{\sqrt{n}}O_p(1)$. By the same argument, we have
$$
\frac{1}{n} \sum_{i=1}^{n} [e_i g''(\tilde{\beta}_{0}, X_i)-\Sigma_{n2}]=\frac{p}{\sqrt{n}}O_p(1).
$$
Therefore $A_{2} = p U^{\top} \Sigma_n U + \frac{p^2}{\sqrt{n}} \|U\|^2 O_p(1) =p U^{\top} \Sigma_n U + p \|U\|^2 o_p(1)$.
For the first term of $A_3$, by the triangle inequality and condition (A6), we have
\begin{eqnarray*}
E|\alpha^{\top} \sum_{i=1}^{n} \alpha^{\top} e_i g'''(\beta_{1n}, X_i)\alpha|
&\leq& \|\alpha\|^3 E\|\sum_{i=1}^{n} e_i g'''(\beta_{1n}, X_i) \| = \|\alpha\|^3 n E\|e_1 g'''(\beta_{1n}, X_1) \| \\
&\leq& n \|\alpha\|^3 (E\|e_1 g'''(\beta_{1n}, X_1) \|^2)^{\frac{1}{2}} \leq n \|\alpha\|^3 (\sum_{j,k,l=1}^{p} E[e_1 g_{jkl}'''( \beta_{1n}, X_1)
]^2)^{\frac{1}{2}}\\
&\leq& \frac{p^3}{\sqrt{n}}\|U\|^3C.
\end{eqnarray*}
For the second term of $A_3$, we have
\begin{eqnarray*}
  E|\sum_{i=1}^{n} \alpha^{\top}g''(\beta_{2n}, X_i)\alpha g'(\tilde{\beta}_{0}, X_i)^{\top}\alpha |
  &\leq&  \sum_{i=1}^{n} E |\alpha^{\top}g''(\beta_{2n}, X_i)\alpha g'(\tilde{\beta}_{0}, X_i)^{\top}\alpha | \\
  &\leq & n \{E[\alpha^{\top}g''(\beta_{2n}, X_1)\alpha]^2\}^{1/2} \{E[g'(\tilde{\beta}_{0}, X_1)^{\top}\alpha]^2\}^{1/2}\\
  &\leq & n \|\alpha\|^3 [E\|g''(\beta_{2n}, X_1) \|^2]^{1/2} [E\|g'(\tilde{\beta}_{0}, X_1) \|^2]^{1/2}\\
  &\leq & \frac{p^3}{\sqrt{n}}\|U\|^3C.
\end{eqnarray*}
By the same argument for the third and forth term of $A_3$, we obtain that $A_3=\frac{p^3}{\sqrt{n}}\|U\|^3O_p(1)+\frac{p^4}{n}\|U\|^4O_p(1)$. Therefore
\begin{eqnarray*}
\alpha^{\top}F(\alpha)    &=&    p\|U\|O_p(1)-p U^{\top} \Sigma_n U + p \|U\|^2 o_p(1)\\
                       &\leq& p\|U\|O_p(1)- p \lambda_{min} (\Sigma_n) \|U\|^2+p \|U\|^2 o_p(1)\\
                       &=&    p\|U\|\{O_p(1)-\lambda_{min} (\Sigma_n) \|U\|+\|U\| o_p(1)\}.
\end{eqnarray*}
If $\|U\|=C$ be large enough, for any $\epsilon > 0$, we have
$$\mathbb{P}(\alpha^{\top}F(\alpha)<0) \geq \mathbb{P}\{O_p(1)-\lambda_{min}(\Sigma_n)\|U\|+\|U\| o_p(1)<0\} \geq 1-\epsilon .$$
Thus our result follows from (6.3.4) of Ortega and Rheinboldt (1970).
\hfill$\Box$

If $g(\beta, X)=\beta^{\top}X$ follows a linear regression model, then $g''(\beta, x)=0$ and $g'''(\beta, x)=0$. According to the proof of Proposition 1, we can obtain the norm consistency of $\hat{\beta}_n$ under the weaker condition $p^2/n \to 0$.

\textbf{Proof of Proposition 2.} We use the same notations as those in the proof of Proposition~1. Let $\Psi_n(\beta)=\sum_{i=1}^{n}[Y_i-g(\beta, X_i)]g'(\beta, X_i)$. Then $\Psi_n(\hat{\beta}_n)=0$. Applying Taylor's expansion around $\tilde{\beta}_{0}$, we obtain
$$ 0 = \Psi_n(\hat{\beta}_n) =\Psi_n(\tilde{\beta}_{0})+\Psi^{'}_n(\tilde{\beta}_{0})(\hat{\beta}_n-\tilde{\beta}_{0})+
\frac{1}{2}(\hat{\beta}_n-\tilde{\beta}_{0})^{\top}\Psi^{''}_n(\beta_{4n})(\hat{\beta}_n-\tilde{\beta}_{0}) $$
where $\beta_{4n}$ lies between $\hat{\beta}_n$ and $\tilde{\beta}_{0}$.
Therefore
$$ \Sigma_n(\hat{\beta}_n-\tilde{\beta}_{0})=\frac{1}{n}\Psi_n(\tilde{\beta}_{0})+ [\Sigma_n+\frac{1}{n}\Psi^{'}_n(\tilde{\beta}_{0})](\hat{\beta}_n-\tilde{\beta}_{0})+
\frac{1}{2n}(\hat{\beta}_n-\tilde{\beta}_{0})^{\top}\Psi^{''}_n(\beta_{4n})(\hat{\beta}_n-\tilde{\beta}_{0}).
$$
Note that
$$
\Sigma_n+\frac{1}{n}\Psi^{'}_n(\tilde{\beta}_{0})=\frac{1}{n} \sum_{i=1}^{n} [\Sigma_{n1}-g'(\tilde{\beta}_{0}, X_i)g'(\tilde{\beta}_{0}, X_i)^{\top}] -\frac{1}{n} \sum_{i=1}^{n} [\Sigma_{n2}-[Y_i-g(\tilde{\beta}_{0}, X_i)] g''(\tilde{\beta}_{0}, X_i)].
$$
Following the same arguments in Proposition~1, we obtain that $\Sigma_n+\frac{1}{n}\Psi^{'}_n(\tilde{\beta}_{0})=\frac{p}{\sqrt{n}}O_p(1)$ and $\Psi^{''}_n(\beta_{4n})=n\sqrt{p^3}O_p(1)$. Since $\| \hat{\beta}_n-\tilde{\beta}_{0} \|=O_p(\sqrt{p/n})$, it follows that
$$ \Sigma_n(\hat{\beta}_n-\tilde{\beta}_{0})=\frac{1}{n}\Psi_n(\tilde{\beta}_{0})+ \frac{\sqrt{p^3}}{n}O_p(1)+\frac{\sqrt{p^5}}{n}O_p(1).
$$
Because $\Sigma_n^{-1}O_p(1)=O_p(1)$, the result follows. Indeed, $\| \Sigma_n^{-1}O_p(1) \|^2= O_p(1)^{\top}\Sigma_n^{-2}O_p(1) \leq \lambda_{\rm max}(\Sigma_n^{-2})\|O_p(1)\|^2=O_p(1).$  \hfill$\Box$

If $g(X, \beta)=\beta^{\top}X$, it is easy to see that $\Psi^{''}_n(\beta_{4n})=0$. Consequently,
$$
 \Sigma_n(\hat{\beta}_n-\tilde{\beta}_{0})=\frac{1}{n}\Psi_n(\tilde{\beta}_{0})+ \frac{\sqrt{p^3}}{n}O_p(1).
$$
Therefore only the convergence rate $p^3/n \to 0$ is needed to obtain the result in Proposition~2.



\textbf{Proof of Proposition 3.} (1) Suppose that $M_n\beta_i=\lambda_i \beta_i$ and $\hat{M}_n\hat{\beta}_i=\hat{\lambda}_i \hat{\beta}_i$ for $1 \leq i \leq p$. Similar to the arguments of Theorem 2.2 in Zhu and Fang (1996), we have
$$ \sqrt{n}(\hat{\lambda}_i- \lambda_i)=\sqrt{n}\beta_i^{\top} (\hat{M}_n-M_n)\beta_i +o_p(1).$$
By Theorem 3 in Zhu et al. (2010b), we obtain that $\sqrt{n}\beta_i^{\top} (\hat{M}_n-M_n)\beta_i$ is asymptotically normal. Thus $\hat{\lambda}_i- \lambda_i=O_p(1/\sqrt{n})$. Following the arguments of Lemma~1 in Tan et al. (2017), we obtain $\mathbb{P}(\hat{q}=1) \to 1$.
Again, by Theorem 2.2 in Zhu and Fang (1996), we obtain
$$ \sqrt{n}(\hat{\beta}_1-\beta_1)=\sqrt{n} \sum_{i=2}^{p} \frac{\beta_i \beta_1^{\top}(\hat{M}_n-M_n) \beta_1}{\lambda_1-\lambda_i} +o_p(1). $$
Note that $\hat{B}_n(1)=\hat{\beta}_1$ and $\beta_1=\kappa \beta_{0}$ under $H_0$. Then we have
$$ \sqrt{n}(\hat{B}_n(1)-\kappa \beta_{0})=\sqrt{n}\beta_1^{\top}(\hat{M}_n-M_n) \beta_1 \sum_{i=2}^{p}\frac{\beta_i}{\lambda_1-\lambda_i}+o_p(1).  $$
Since $\sqrt{n}\beta_1^{\top}(\hat{M}_n-M_n) \beta_1=O_p(1)$ and $\|\sum_{i=2}^{p}\frac{\beta_i}{\lambda_1-\lambda_i}\|^2=O(p)$, it follows that
$\|\sqrt{n}(\hat{B}_n(q)-\kappa \beta_{0})\|=O_p(\sqrt{p})$. \\
(2) Note that $q$ is free of $n$ under $H_1$. The proof is concluded from the argument for proving (1).    \hfill$\Box$


\textbf{Proof of Theorem 3.1.} Under the null hypothesis, we have $\mathbb{P}(\hat{q}= 1)\rightarrow 1$. Thus we need only work on the event $\{\hat{q}=1\}$. It follows that $\hat{\alpha}=1$ and we can rewrite $V_n(\hat{\alpha}, u)$ as
\begin{eqnarray*}
V_n(\hat{\alpha}, u) &=& \frac{1}{\sqrt{n}}\sum_{i=1}^n[Y_i- g(\hat{\beta}^{\top}_{n}X_i, \theta_n)]I(\hat{B}_n^{\top}X_i \leq u)  \\
                     &=& \frac{1}{\sqrt{n}}\sum_{i=1}^n[Y_i- g(\hat{\beta}^{\top}_{n}X_i, \theta_n)]I(\kappa \beta_{0}^{\top} X_i \leq u) +\\
                     &&  \frac{1}{\sqrt{n}}\sum_{i=1}^n[Y_i- g(\hat{\beta}^{\top}_{n}X_i, \theta_n)][I(\hat{B}_n^{\top}X_i \leq u)-I(\kappa
                         \beta_{0}^{\top} X_i\leq u)]\\
                     &=:& V_{n1}+V_{n2}.
\end{eqnarray*}
Let $\gamma= (\beta^{\top}, \theta^{\top})^{\top}$. Then we obtain that
\begin{eqnarray*}
V_{n1} &=& \frac{1}{\sqrt{n}}\sum_{i=1}^n \varepsilon_i I(\kappa \beta_{0}^{\top} X_i \leq u)-
           \frac{1}{\sqrt{n}}\sum_{i=1}^n[g(\hat{\beta}^{\top}_{n}X_i, \theta_n)- g(\beta_{0}^{\top}X_i, \theta_0)]I(\kappa \beta_{0}^{\top}X_i \leq u)\\
       &=& \frac{1}{\sqrt{n}}\sum_{i=1}^n \varepsilon_i I(\kappa \beta_{0}^{\top} X_i \leq u)-
           \frac{1}{\sqrt{n}}\sum_{i=1}^n (\hat{\gamma}_n-\gamma_{0})^{\top} g'(\beta_{n0}, \theta_0, X_i)I(\kappa \beta_{n0}^{\top} X_i \leq u)- \\
       &&  \frac{1}{\sqrt{n}}\sum_{i=1}^n (\hat{\gamma}_n-\gamma_{0})^{\top} g''(\beta_{1n},\theta_{1n}, X_i) (\hat{\gamma}_n-\gamma_{0})
           I(\kappa \beta_{0}^{\top} X_i \leq u) \\
       &=& V_{n11}-V_{n12}-V_{n13}
\end{eqnarray*}
where $(\beta_{1n}, \theta_{1n})$ lies between $(\hat{\beta}_n, \hat{\theta}_n)$ and $(\beta_{0}, \theta_0)$. For the third term $V_{n13}$ in $V_{n1}$, note that
\begin{eqnarray*}
E \sup_{u} \|\sum_{i=1}^{n} g^{''}(\beta_{1n}, \theta_{1n}, X_i)I(\kappa \beta_{0}^{\top} X_i \leq u) \|
&\leq& \sum_{i=1}^{n} E  \sup_{u} \| g^{''}(\beta_{1n}, \theta_{1n}, X_i)I(\kappa \beta_{0}^{\top} X_i \leq u)\| \\
&\leq& \sum_{i=1}^{n} [E \sup_{u} \| g^{''}(\beta_{1n}, \theta_{1n}, X_i)I(\kappa \beta_{0}^{\top} X_i \leq u)\|^2]^{1/2} \\
&\leq& \sum_{i=1}^{n} \left(\sum_{j,k=1}^{p+d} E g_{jk}^{''}(\beta_{1n}, \theta_{1n}, X_i)^2\right)^{1/2}\\
&\leq& C n (p+d).
\end{eqnarray*}
Therefore $V_{n13}=\frac{1}{\sqrt{n}} \frac{p}{n} n(p+d) O_p(1)=o_p(1)$ uniformly in $u$. For $V_{n12} $, recall that $M_n(u)=E[g'(\beta_{0}, \theta_0, X)I(\kappa \beta_{0}^{\top}X \leq u)]$. Then we decompose $V_{n12}$ as follows
$$
V_{n12}=\sqrt{n}(\hat{\gamma}_n-\gamma_{0})^{\top}M_n(u)+\sqrt{n}(\hat{\gamma}_n-\gamma_{0})^{\top} \left(\frac{1}{n} \sum_{i=1}^n g'(\beta_{0}, \theta_0, X_i)I(\kappa \beta_{0}^{\top} X_i \leq u) -M_n(u)\right).
$$
For the second term in $V_{n12}$,
by Lemma 3, we have
$$
\sup_{u} \|\frac{1}{n} \sum_{i=1}^n g'(\beta_{0}, \theta_0, X_i)I(\kappa \beta_{0}^{\top} X_i \leq u) -M_n(u)\| =o_p(\sqrt{\frac{p^{3/2}\log{n}}{n}}).
$$
Conclude that
$$ \sqrt{n}(\hat{\gamma}_n-\gamma_{0})^{\top}\left( \frac{1}{n} \sum_{i=1}^n g'(\beta_{0}, \theta_0, X_i)I(\kappa \beta_{0}^{\top} X_i \leq u) -M_n(u)\right)= \sqrt{p} \sqrt{\frac{p^{3/2}\log{n}}{n}} o_p(1)=o_p(1).
$$
Since $\|M_n(u)\|=O(1)$ uniformly in $u$, by Proposition 2, we have
$$
V_{n12}=M_n(u)^{\top} \Sigma_n^{-1} \frac{1}{\sqrt{n}} \sum_{i=1}^{n}[Y_i-g(\beta_{0}^{\top}X_i, \theta_0)]g'(\beta_{0}, \theta_0, X_i)+o_p(1).
$$
Therefore, we obtain that
\begin{eqnarray}\label{6.1}
  V_{n1} = \frac{1}{\sqrt{n}}\sum_{i=1}^n \varepsilon_i I(\kappa \beta_{0}^{\top} X_i \leq u)- \frac{1}{\sqrt{n}} M_n(u)^{\top} \Sigma_n^{-1} \sum_{i=1}^{n} \varepsilon_i g'(\beta_{0}, \theta_0, X_i)+o_p(1).
\end{eqnarray}

Now we consider the term $V_{n2}$. It can be decomposed as follow
\begin{eqnarray*}
  V_{n2} &=& \frac{1}{\sqrt{n}} \sum_{i=1}^n \varepsilon_i [I(\hat{B}_n^{\top}X_i \leq u)-I(\kappa \beta_{0}^{\top} X_i \leq u)]-\\
         &&  \frac{1}{\sqrt{n}} \sum_{i=1}^n [g(\hat{\beta}_{n}^{\top}X_i, \hat{\theta}_n)- g(\beta_{0}^{\top}X_i, \theta_0)] [I(\hat{B}_n^{\top}X_i
             \leq u)-I(\kappa \beta_{0}^{\top} X_i \leq u)] \\
         &=& \frac{1}{\sqrt{n}} \sum_{i=1}^n \varepsilon_i [I(\hat{B}_n^{\top}X_i \leq u)-I(\kappa \beta_{0}^{\top} X_i \leq u)]-\\
         &&  \frac{1}{\sqrt{n}} \sum_{i=1}^n (\hat{\gamma}_n-\gamma_{0})^{\top} g'(\beta_{0}, \theta_0, X_i) [I(\hat{B}_n^{\top}X_i \leq
             u)-I(\kappa \beta_{0}^{\top} X_i \leq u)]- \\
         &&  \frac{1}{\sqrt{n}} \sum_{i=1}^n (\hat{\gamma}_n-\gamma_{0})^{\top} g''(\beta_{1n},\theta_{1n}, X_i) (\hat{\gamma}_n-\gamma_{0})
             [I(\hat{B}_n^{\top}X_i \leq u)-I(\kappa \beta_{0}^{\top} X_i \leq u)]\\
         &=& V_{n21}-V_{n22}-V_{n23}.
\end{eqnarray*}
By Lemma 6, we obtain that $V_{n21}=o_p(1)$ uniformly in $u$. For the second term $V_{n22}$, let
$$\tilde{M}_n(\beta, u)=E\{g'(\beta_{0}, \theta_0, X)[I(\beta^{\top}X_i \leq u)- I(\kappa \beta_{0}X_i \leq u)] \}.$$
By Lemma 7, we have
$$
\sup_{u}\|\frac{1}{n} \sum_{i=1}^n g'(\beta_{0}, \theta_0, X_i) [I(\hat{B}_n^{\top}X_i \leq u)-I(\kappa \beta_{0}^{\top} X_i \leq u)]- \tilde{M}_n(\hat{B}_n, u)\| =o_p(\sqrt{\frac{p^{3/2} \log{n}}{n}})
$$
Therefore, we derive that
$$
V_{n22}=\sqrt{n}(\hat{\gamma}_n-\gamma_{0})^{\top} \tilde{M}_n(\hat{B}_n, u)+ \sqrt{\frac{p^{5/2} \log{n}}{n}}o_p(1).
$$
Let $M_{n}(\beta, u)=E[g'(\beta_{0}, \theta_0, X)I(\beta^{\top}X_i \leq u)]$. By condition (B1), it is easy to see that
$$\tilde{M}_n(\hat{B}_n, u)=M_{n}(\hat{B}_n, u)-M_{n}(\kappa \beta_{0}, u)=\frac{p^{7/8}\log{n}}{n^{3/8}}  o_p(1).$$
Consequently,
$$ V_{n22} = \frac{p^{11/8}\log{n}}{n^{3/8}}  o_p(1) +\sqrt{\frac{p^{5/2} \log{n}}{n}}o_p(1). $$
It follows that $V_{n22}=o_p(1)$ uniformly in $u$.

Similar to the term $V_{n13}$, we obtain that $V_{n23}=o_p(1)$ uniformly in $u$. Combining these with~(\ref{6.1}), we obtain that
\begin{eqnarray}\label{6.2}
V_n(\hat{\alpha}, u) = \frac{1}{\sqrt{n}}\sum_{i=1}^n \varepsilon_i I(\kappa \beta_{0}^{\top} X_i \leq u)- \frac{1}{\sqrt{n}} M_n(u)^{\top} \Sigma_n^{-1} \sum_{i=1}^{n} \varepsilon_i g'(\beta_{0}, \theta_0, X_i)+o_p(1).
\end{eqnarray}
It is easy to see that the first and second terms of the right-hand side of (\ref{6.2}) are asymptotically tight.

Now we consider the convergence of finite-dimensional distributions. Let $Y_{ni}=(Y_{ni}(u_1), \cdots,\\ Y_{ni}(u_m))^{\top}$ where
$$Y_{ni}(u)=\frac{1}{\sqrt{n}} \varepsilon_i [I(\kappa \beta_{0}^{\top} X_i \leq u)- M_n(u)^{\top} \Sigma_n^{-1} g'(\beta_{0}, \theta_0, X_i)].$$
For any $\delta >0$, we have
$$  \sum_{i=1}^{n} E\|Y_{ni}\|^2 I(\|Y_{ni}\|> \delta)=n E\{\|Y_{n1}\|^2 I(\|Y_{n1}\|> \delta)\} \leq n \{E\|Y_{n1}\|^4\}^{1/2} \{\mathbb{P}(\|Y_{n1}\| > \delta)\}^{1/2}.$$
Since
$$ \mathbb{P}(\|Y_{n1}\| > \delta)=\mathbb{P}(Y_{n1}(u_1)^2+ \cdots + Y_{n1}(u_m)^2 > \delta^2)\leq \sum_{j=1}^{m} \mathbb{P}(Y_{n1}(u_j)^2> \frac{\delta^2}{m}), $$
and
\begin{eqnarray*}
\mathbb{P}(Y_{n1}(u)^2> \frac{\delta^2}{m}) &=&\mathbb{P}(\varepsilon_1^2[I(\kappa \beta_{0}^{\top} X_1 \leq u)-M_n(u)^{\top} \Sigma_n^{-1} g'(\beta_{0}, \theta_0, X_1)]^2 > \frac{n\delta^2}{m} ) \\
&\leq& \frac{2m E\varepsilon_1^2 +2m E\{ \varepsilon_1^2 [M_n(u)^{\top} \Sigma_n^{-1} g'(\beta_{0}, \theta_0, X_1)]^2\}}{n\delta^2}\\
&\leq& \frac{2m E\varepsilon_1^2 +2m \lambda_{max}^2(\Sigma^{-1}) \|M_n(u)\|^2 E\{ \varepsilon_1^2\|g'(\beta_{0}, \theta_0, X_1)\|^2\}}{n\delta^2},
\end{eqnarray*}
it follows that $\mathbb{P}(\|Y_{n1}\| > \delta)=O(p/n)$. For $E\|Y_{n1}\|^4$, it is easy to see that
$$ E\|Y_{n1}\|^4 \leq m [EY_{n1}(u_1)^4+ \cdots + EY_{n1}(u_m)^4]. $$
Since
\begin{eqnarray*}
EY_{n1}(u)^4&=&\frac{1}{n^2} E\{\varepsilon_1^4[I(\kappa \beta_{0}^{\top} X_1 \leq u)-M_n(u)^{\top} \Sigma_n^{-1} g'(\beta_{0}, \theta_0, X_1)]^4\}\\
&\leq& \frac{8}{n^2} \{E [\varepsilon_1^4I(\kappa \beta_{0}^{\top} X_1 \leq u)] +E[\varepsilon_1 M_n(u)^{\top} \Sigma_n^{-1} g'(\beta_{0}, \theta_0, X_1)]^4 \} \\
&\leq& \frac{8}{n^2} \{E [\varepsilon_1^4I(\kappa \beta_{0}^{\top} X_1 \leq u)] + \lambda_{max}^4(\Sigma^{-1}) \|M_n(u)\|^4 E[\varepsilon_1^4 \|g'(\beta_{0}, \theta_0, X_1)\|^4] \} \\
&\leq& \frac{8}{n^2} \{E\varepsilon_1^4 + \lambda_{max}^4(\Sigma^{-1}) \|M_n(u)\|^4
\sum_{j,k=1}^{p+d} E[\varepsilon_1^4 g'_j(\beta_{0}, \theta_0, X_1)^2 g'_k( \beta_{0}, \theta_0, X_1)^2] \},
\end{eqnarray*}
it follows that $EY_{n1}(u)^4=O(p^2/n^2)$. Hence $ \sum_{i=1}^{n} E\|Y_{ni}\|^2 I(\|Y_{ni}\|> \delta) = O(\sqrt{p^3/n})=o(1).$

For the covariance matrix $\sum_{i=1}^{n} Cov(Y_{ni}) $, we only need to consider $\sum_{i=1}^{n} Cov\{Y_{ni}(s), Y_{ni}(t)\}$.
It is easy to see that
\begin{eqnarray*}
&&\sum_{i=1}^{n} Cov\{Y_{ni}(s), Y_{ni}(t)\} \\
&&= E[\varepsilon_1^2 I(\kappa \beta_{0}^{\top} X_1 \leq s \wedge t)]- M_n(s)^{\top} \Sigma_n^{-1} E[\varepsilon_1^2 g'(\beta_{0}, \theta_0, X_1)I(\kappa \beta_{0}^{\top} X_1 \leq t)] \\
&& -M_n(t)^{\top} \Sigma_n^{-1} E[\varepsilon_1^2 g'(\beta_{0}, \theta_0, X_1) I(\kappa \beta_{0}^{\top} X_1 \leq s)] \\
&&+ M_n(s)^{\top} \Sigma_n^{-1}  E[\varepsilon_1^2 g'(\beta_{0}, \theta_0, X_1) g'(\beta_{0}, \theta_0, X_1)^{\top}] \Sigma_n^{-1} M_n(t).
\end{eqnarray*}
Thus $ \sum_{i=1}^{n} Cov\{Y_{ni}(s), Y_{ni}(t)\} = K_n(s, t)$. Since $K_n(s, t) \to K(s, t)$, it follows that $Y_{ni}$ satisfies the conditions of Lindeberg-Feller Central limit theorem. Hence convergence of the finite-dimensional distributions holds. All together we have
\begin{equation*}
 V_n(u) \longrightarrow |V_{\infty}^1(u)|
\end{equation*}
where $V_{\infty}^1(u)$ is a zero mean Gaussian process with covariance function $ K(s,t)$. Hence we complete the proof.   \hfill$\Box$


\textbf{Proof of Theorem 3.2.} Similar to the proof for Theorem~\ref{Theorem 3.1}, we only need to  work on the event $\{\hat{q}=1\}$. Let
\begin{equation*}
V_n^1(\hat{\alpha}, u)=\frac{1}{\sqrt{n}}\sum_{i=1}^n [Y_i- g(\beta_{0}^{\top} X_i,\theta_0)]I(\hat{\alpha}^{\top}\hat{B}_n^{\top} X_i \leq u).
\end{equation*}
On the event $\{ \hat{q}=1 \}$, we have $\mathcal{S}_{\hat{q}}^{+} = \{ 1 \}$ and then $\hat{\alpha}=1$. Consequently, $V_n^1(\hat{\alpha}, u)$ can be rewritten as
\begin{equation*}
V_n^1(\hat{\alpha}, u) = \frac{1}{\sqrt{n}}\sum_{i=1}^n [Y_i- g(\beta_{0}^{\top} X_i,\theta_0)]I(\hat{B}_n^{\top} X_i \leq u).
\end{equation*}
Next we divide  the whole proof of Theorem~\ref{Theorem 3.2} into three parts.

(I) First, to prove that $ \hat{T}_n V_n(\hat{\alpha}, u)- \hat{T}_n V_n^1(\hat{\alpha}, u) =o_p(1)$ uniformly in $u$. Recall that
\begin{eqnarray*}
  \hat{T}_n V_n(\hat{\alpha}, u) = V_n(\hat{\alpha}, u)- \int_{-\infty}^{u} \hat{a}_n(z)^{\top} \hat{A}_n^{-1}(z) \left(\int_{z}^{\infty}
                                   \hat{a}_n(v) V_n(\hat{\alpha}, dv)\right) \hat{\sigma}_n^2(z) F_{\hat{\alpha}}(dz) \\
  \hat{T}_n V_n^1(\hat{\alpha}, u) = V_n^1(\hat{\alpha}, u)- \int_{-\infty}^{u} \hat{a}_n(z)^{\top} \hat{A}_n^{-1}(z) \left(\int_{z}^{\infty}
                                   \hat{a}_n(v) V_n^1(\hat{\alpha}, dv)\right) \hat{\sigma}_n^2(z) F_{\hat{\alpha}}(dz).
\end{eqnarray*}
Since
\begin{equation*}
  V_n(\hat{\alpha}, u)-V_n^1(\hat{\alpha}, u)=-\frac{1}{\sqrt{n}}\sum_{i=1}^n [g(\hat{\beta}^{\top}_n X_i,\hat{\theta}_n)-g(\beta_{0}^{\top} X_i,\theta_0)] I(\hat{B}_n^{\top} X_i \leq u),
\end{equation*}
by the same arguments in the proof of Theorem~\ref{Theorem 3.1}, we obtain that
\begin{equation*}
  V_n(\hat{\alpha}, u)-V_n^1(\hat{\alpha}, u)=-\sqrt{n}(\hat{\gamma}_n-\gamma_{0})^{\top} M_n(u)+ o_p(1)
\end{equation*}
uniformly in $u$. The two integrals in $\hat{T}_n V_n(\hat{\alpha}, u)$ and $\hat{T}_n V_n^1(\hat{\alpha}, u)$ differ by
\begin{eqnarray*}
&&  \int_{-\infty}^{u} \hat{a}_n(z)^{\top} \hat{A}_n^{-1}(z) \left( \int_{z}^{\infty} \hat{a}_n(v) (V_n^1(\hat{\alpha}, dv)-V_n(\hat{\alpha}, dv))
    \right) \hat{\sigma}_n^2(z) F_{\hat{\alpha}}(dz).
\end{eqnarray*}
It equals
\begin{eqnarray*}
&&  \frac{1}{\sqrt{n}}\sum_{i=1}^n \int_{-\infty}^{u} \hat{a}_n(z)^{\top} \hat{A}_n^{-1}(z)\hat{\sigma}_n^2(z) I(\hat{B}_n^{\top} X_i \geq z)
    \hat{a}_n(\hat{B}_n^{\top} X_i)[g(\hat{\beta}^{\top}_n X_i,\hat{\theta}_n)-g(\beta_{0}^{\top} X_i,\theta_0)] F_{\hat{\alpha}}(dz)\\
&=& \frac{1}{\sqrt{n}}\sum_{i=1}^n \int_{-\infty}^{u} \hat{a}_n(z)^{\top} \hat{A}_n^{-1}(z)\hat{\sigma}_n^2(z) I(\hat{B}_n^{\top} X_i \geq z)
    \hat{a}_n(\hat{B}_n^{\top} X_i) [ g'(\hat{\beta}_n, \hat{\theta}_n, X_i)^{\top} (\hat{\gamma}_n-\gamma_{0})+  \\
&&  (\hat{\gamma}_n-\gamma_{0})^{\top} g''(\hat{\beta}_{1n}, \hat{\theta}_1, X_i)(\hat{\gamma}_n-\gamma_{0})- (\hat{\gamma}_n-\gamma_{0})^{\top}
    g''(\hat{\beta}_{2n}, \hat{\theta}_2, X_i)(\hat{\gamma}_n-\gamma_{0}) ] F_{\hat{\alpha}}(dz) \\
&=& \frac{1}{\sqrt{n}}\sum_{i=1}^n \int_{-\infty}^{u} \hat{a}_n(z)^{\top} \hat{A}_n^{-1}(z)\hat{\sigma}_n^2(z) I(\hat{B}_n^{\top} X_i \geq z)
    \hat{a}_n(\hat{B}_n^{\top} X_i) g'(\hat{\beta}_n, \hat{\theta}_n, X_i)^{\top} (\hat{\gamma}_n-\gamma_{0}) F_{\hat{\alpha}}(dz)+o_p(1),
\end{eqnarray*}
where $(\hat{\beta}_{1n}, \hat{\theta}_1)$ and $(\hat{\beta}_{2n}, \hat{\theta}_2)$ both lie between $(\hat{\beta}_n,\hat{\theta}_n)$ and $(\beta_{0},\theta_0)$.
Recall that
$$ \hat{A}_n(z)= \frac{1}{n} \sum_{i=1}^{n} \hat{a}_n(\hat{B}_n^{\top}X_i) g'(\hat{\beta}_n, \hat{\theta}_n, X_i)^{\top} I(\hat{B}_n^{\top}X_i \geq z).$$
Then the two integrals differ by
\begin{eqnarray*}
&&  \sqrt{n} (\hat{\gamma}_n-\gamma_{0})^{\top} \int_{-\infty}^{u} \hat{a}_n(z) \hat{\sigma}_n^2(z) F_{\hat{\alpha}}(dz) +o_p(1)\\
&=& \sqrt{n} (\hat{\gamma}_n-\gamma_{0})^{\top} \int_{-\infty}^{u} a_n(z) \sigma_n^2(z) F_{\kappa \beta_{0}}(dz) +o_p(1).
\end{eqnarray*}
Since $ \int_{-\infty}^{u} a_n(z) \sigma_n^2(z) F_{\kappa \beta_{0}}(dz)=M_n(u)$, it follows that $ \hat{T}_n V_n(\hat{\alpha}, u)- \hat{T}_n V_n^1(\hat{\alpha}, u) =o_p(1)$ uniformly in $u$.

(II) Second, to prove $T_n V_n^1(\hat{\alpha}, u)- \hat{T}_n V_n^1(\hat{\alpha}, u)=o_p(1)$ uniformly in $u$. Indeed,
\begin{eqnarray*}
&&T_n V_n^1(\hat{\alpha}, u)- \hat{T}_n V_n^1(\hat{\alpha}, u) \\
&=& \int_{-\infty}^{u} \hat{a}_n(z)^{\top} \hat{A}_n^{-1}(z) \left(\int_{z}^{\infty}
    \hat{a}_n(v) V_n^1(\hat{\alpha}, dv)\right)\hat{\sigma}_n^2(z)F_{\hat{\alpha}}(dz)-\\
&&  \int_{-\infty}^{u} a_n(z)^{\top} A_n^{-1}(z) \left(\int_{z}^{\infty} a_n(v) V_n^1(\hat{\alpha}, dv)\right) \sigma_n^2(z)
    F_{\kappa \beta_{0}}(dz)\\
&=& \int_{-\infty}^{u} \hat{a}_n(z)^{\top} \hat{A}_n^{-1}(z) \left(\int_{z}^{\infty} \hat{a}_n(v) V_n^1(\hat{\alpha}, dv)\right)\hat{\sigma}_n^2(z)
    F_{\hat{\alpha}}(dz)-\\
&&  \int_{-\infty}^{u}  a_n(z)^{\top} A_n^{-1}(z) \left(\int_{z}^{\infty} a_n(v) V_n^1(\hat{\alpha}, dv)\right) \sigma_n^2(z) F_{\hat{\alpha}}(dz)+\\
&&  \int_{-\infty}^{u} a_n(z)^{\top} A_n^{-1}(z) \left(\int_{z}^{\infty} a_n(v) V_n^1(\hat{\alpha}, dv)\right) \sigma_n^2(z)
    \{F_{\hat{\alpha}}(dz)-F_{\kappa \beta_{0}}(dz)\} \\
&=:& T_{n1}-T_{n2}+T_{n3}.
\end{eqnarray*}
Putting
$$
h_n(z)=a_n(z)^{\top} A_n^{-1}(z) \left(\int_{z}^{\infty} a_n(v) V_n^1(\hat{\alpha}, dv)\right) \sigma_n^2(z),
$$
it follows that
$$
h_n(z)=\frac{1}{\sqrt{n}} \sum_{i=1}^{n} \varepsilon_i a_n(z)^{\top} A_n^{-1}(z) a_n(\kappa\beta_{0}^{\top}X_i) I(\kappa\beta_{0}^{\top}X_i \geq z) \sigma_n^2(z) +o_p(1).
$$
By the uniformly boundedness of $\sigma_n^2(z)$, we have the sequence $\{ h_n(z) \}$ is asymptotically tight. According to Lemma 3.4 in Stute, Thies, and Zhu (1998) and the arguments thereafter, we obtain that $T_{n3}=o_p(1)$ uniformly in $u \in [-\infty, u_0]$. For $T_{n1}-T_{n2}$, since both $a_n(z)$ and $A_n(z)$ depend on $(\beta_{0}, \theta_0)$, we rewrite $a_n(z)$ and $A_n(z)$ as $a_n(\beta_{0}, \theta_0, z)$ and $A_n(\beta_{0}, \theta_0, z)$ respectively and define
$$
l_n(\beta, \theta, u)=\int_{-\infty}^{u} a_n(\beta, \theta, z)^{\top} A_n^{-1}(\beta, \theta, z) \left(\int_{z}^{\infty} a_n(\beta, \theta, v) V_n^1(\hat{\alpha}, dv)\right) \sigma_n^2(z) F_{\hat{\alpha}}(dz).
$$
By the boundedness of $\sigma_n(u)$ and Condition (B1),
we obtain that $l_n(\hat{\beta}_{n}, \hat{\theta}_n, u)-l_n(\beta_{0}, \theta_0, u)=o_p(1)$. By Lemma 8, we show that
$$
\sup_{v}\|\hat{r}_n(v)-r_n(v)\|=O_p(\sqrt{p \log{n}/n^{4/5}}).
$$
Combining this with the uniformly boundedness of $\hat{\sigma}_n^2$, we obtain $T_{n1}-T_{n2}$ tends to zero in probability.

(III) Finally, to prove $T_n V_n^1(\hat{\alpha}, u)-T_n V_n^0(u) =o_p(1)$ uniformly in $u$.
\begin{eqnarray*}
&&  T_n V_n^1(\hat{\alpha}, u)-T_n V_n^0(u)\\
&=& V_n^1(\hat{\alpha}, u)-\int_{-\infty}^{u} a_n(z)^{\top} A_n^{-1}(z) \left(\int_{z}^{\infty} a_n(v) V_n^1(\hat{\alpha}, dv)\right) \psi_n(dz)-\\
&&  \left\{ V_n^0(u)- \int_{-\infty}^{u} a_n(z)^{\top} A_n^{-1}(z) \left(\int_{z}^{\infty} a_n(v) V_n^0(dv)\right) \psi_n(dz)  \right\}.
\end{eqnarray*}
Since
$$
V_n^1(\hat{\alpha}, u)-V_n^0(u)=\frac{1}{\sqrt{n}}\sum_{i=1}^n \varepsilon_i [I(\hat{B}_n^{\top} X_i \leq u)-I(\kappa \beta_{0}^{\top} X_i \leq u)],
$$
by the same argument in Theorem~\ref{Theorem 3.1}, we obtain that $V_n^1(\hat{\alpha}, u)-V_n^0(u)=o_p(1)$ uniformly in $u$.
For the integrals in $T_n V_n^1(\hat{\alpha}, u)-T_n V_n^0(u)$, note that the two integrals differ by
\begin{eqnarray*}
&&  \int_{-\infty}^{u} a_n(z)^{\top} A_n^{-1}(z) \left\{\int_{z}^{\infty} a_n(v) [V_n^1(\hat{\alpha}, dv)-V_n^0(dv)]\right\} \psi_n(dz) \\
&=& \frac{1}{\sqrt{n}}\sum_{i=1}^n \varepsilon_i \int_{-\infty}^{u} a_n(z)^{\top} A_n^{-1}(z) a_n(\hat{B}_n^{\top} X_i)I(\hat{B}_n^{\top} X_i \geq z)
    \psi_n(dz) -\\
&&  \frac{1}{\sqrt{n}}\sum_{i=1}^n \varepsilon_i \int_{-\infty}^{u} a_n(z)^{\top} A_n^{-1}(z) a_n(\kappa \beta_{0}^{\top} X_i)I(\kappa
    \beta_{0}^{\top} X_i \geq z) \psi_n(dz).
\end{eqnarray*}
Since $\| \hat{B}_n -\kappa \beta_{0}\|=O_p(\sqrt{p/n})$, similar to the arguments in Lemma 6, the difference between the two integrals in $T_n V_n^1(\hat{\alpha}, u)-T_n V_n^0(u)$ tends to zero. Hence $T_n V_n^1(\hat{\alpha}, u)-T_n V_n^0(u)= o_p(1)$ uniformly in $u$. All together we conclude that
$$\sup_{ \hat{\alpha} \in \mathcal{S}_{\hat{q}}^{+}} |\hat{T}_n V_{n}(\hat{\alpha}, u)| \rightarrow |V_{\infty}(u)|$$
in distribution.  \hfill$\Box$



\textbf{Proof of Proposition 4.} Let $Y=g(\beta_{0}^{\top}X, \theta_0)+\varepsilon$, $\alpha_t=E[XI(Y \leq t)]$, $\tilde{\alpha}_t=E[XI(Y_n \leq t)]$, $M_n =\int \alpha_t \alpha_t^{\top} F_{Y}(dt)$, and $\tilde{M}_n =\int \tilde{\alpha}_t \tilde{\alpha}_t^{\top} F_{Y_n}(dt)$. Then the space ${\rm span}(M_n) \in \mathcal{S}_{Y|X}$ and the space  ${\rm span}(\tilde{M}_n) \in \mathcal{S}_{Y_n|X}$. If we show that $\sqrt{n} \gamma^{\top} (\hat{M}_n-M_n) \gamma$ is asymptotically normal for any unit vector $\gamma$, the result of this proposition follows from the exact arguments for proving Proposition 3.

We now prove the above asymptotic normality. Under $H_{1n}$, we have
$$\tilde{\alpha}_t=E\{X F_{Y|X}(t-\frac{1}{\sqrt{n}}G(X))\},$$
where $F_{Y|X}$ is the conditional distribution of $Y$ given $X$. By Taylor's expansion, we derive
\begin{eqnarray*}
\tilde{\alpha}_t =\alpha_t  -\frac{1}{\sqrt{n}} E\{ X f_{Y|X}(t)G(X) \} +\frac{1}{2n} E\{ Xf'_{Y|X}(\xi_t(X))G(X)^2\}.
\end{eqnarray*}
Here $\xi_t(X)$ lies between $t-\frac{1}{\sqrt{n}}G(X)$ and $t$ and $f_{Y|X}$ is the conditional density function of $Y$ given $X$. Therefore,
\begin{eqnarray*}
\tilde{M}_n &=& \int \alpha_t\alpha_t^{\top} F_{Y_n}(dt) - \frac{1}{\sqrt{n}}\int \alpha_t E[X^{\top}f_{Y|X}(t)G(X)] F_{Y_n}(dt) - \\
&&\frac{1}{\sqrt{n}}\int  \{E[Xf_{Y|X}(t)G(X)]\} \alpha_t^{\top} F_{Y_n}(dt) +O_p(\frac{p}{n}).
\end{eqnarray*}
Note that $F_{Y_n}(t)=F_{Y}(t)-\frac{1}{\sqrt{n}} E[G(X)f_{Y|X}(t)]+\frac{1}{2n}E[f'_{Y|X}(\xi_t(X))G(X)^2]$. Consequently,
\begin{eqnarray*}
\tilde{M}_n = M_n - \frac{1}{\sqrt{n}}\int \{ \alpha_t E[X^{\top}f_{Y|X}(t)G(X)] + E[Xf_{Y|X}(t)G(X)] \alpha_t^{\top} \} F_{Y}(dt) +O_p(\frac{p}{n}).
\end{eqnarray*}
By Theorem 3 in Zhu et al. (2010b), we have $\sqrt{n} \gamma^{\top} (\hat{M}_n-\tilde{M}_n) \gamma$ is asymptotically normal. By condition (B3) in Appendix, $\sqrt{n} \gamma^{\top} (\hat{M}_n-M_n) \gamma$ is also asymptotically normal.      \hfill$\Box$

\textbf{Proof of Proposition 5.} The proof is similar to that  for proving Propositions~1 and 2 with $e_i =\varepsilon_i$ and $\Sigma_n=E [g'(\beta_{0}, \theta_0, X)g'(\beta_{0}, \theta_0, X)^{\top}]$.  \hfill$\Box$


\textbf{Proof of Theorem 3.3.} (1) Under $H_1$, Proposition 1 asserts that $P(\hat{q}=q) \to 1$. Thus we only need work on the event $\{\hat{q}=q\}$. It follows that $\sup_{ \hat{\alpha} \in \mathcal{S}_{\hat{q}}^{+}} |\hat{T}_n V_{n}(\hat{\alpha}, u)|= \sup_{\alpha \in \mathcal{S}_{q}^{+}} |\hat{T}_n V_{n}(\alpha, u)|$.

Putting
\begin{eqnarray*}
\tilde{V}^1_n(\alpha, u)=\frac{1}{\sqrt{n}} \sum_{i=1}^{n}[Y_i-g(\tilde{\beta}_{0}^{\top}X_i, \tilde{\theta}_0)]I(\alpha^{\top}\hat{B}_n^{\top}X_i \leq u),\\
\tilde{V}^0_n(\alpha, u)=\frac{1}{\sqrt{n}} \sum_{i=1}^{n}[Y_i-g(\tilde{\beta}_{0}^{\top}X_i, \tilde{\theta}_0)]I(\alpha^{\top} B^{\top} X_i \leq u)
\end{eqnarray*}
and
\begin{eqnarray*}
r_n(\alpha, z) &=& E(X|\alpha^{\top}B^{\top}X=z); \\
\sigma_n^2(\alpha, z) &=& E((Y-g(\tilde{\beta}_{0}X, \tilde{\theta}_0))^2|\alpha^{\top}B^{\top}X=z); \\
a_n(\alpha, z) &=& \{g'_1(z/\|\tilde{\beta}_{0}\|, \tilde{\theta}_0)r_n(\alpha, z)^{\top}/ \sigma_n^2(\alpha, z), g'_2(z/\|\tilde{\beta}_{0}\|,
                   \tilde{\theta}_0)^{\top} / \sigma_n^2(\alpha, z) \}^{\top}; \\
A_n(\alpha, z) &=& E\{a_n(\alpha, \alpha^{\top}B^{\top}X)g'(\tilde{\beta}_{0}, \theta_0, X)^{\top}I(\alpha^{\top}B^{\top} X \geq z)\}.
\end{eqnarray*}
Following the arguments in Theorem~\ref{Theorem 3.2}, we obtain that
$$
\hat{T}_n V_n(\alpha, u)-  T_n \tilde{V}_n^1(\alpha, u)=o_p(1),
$$
where
\begin{eqnarray*}
T_n\tilde{V}_n^1(\alpha, u)=\tilde{V}_n^1(\alpha, u)- \int_{-\infty}^{u} a_n(\alpha, z)^{\top} A_n^{-1}(\alpha, z) \left(\int_{z}^{\infty}
                               a_n(\alpha, v)\tilde{V}_n^1 (\alpha, dv)\right) \sigma_n^2(\alpha, z) F_{\alpha}(dz)
\end{eqnarray*}
and $F_{\alpha}$ is the cumulative distribution function of $\alpha^{\top}B^{\top}X$.
Consider
\begin{eqnarray*}
&&  T_n \tilde{V}_n^1(\alpha, u)-T_n \tilde{V}_n^0(\alpha, u) \\
&=& \tilde{V}_n^1(\alpha, u)- \int_{-\infty}^{u} a_n(\alpha, z)^{\top} A_n^{-1}(\alpha, z) \left(\int_{z}^{\infty}
    a_n(\alpha,v)\tilde{V}_n^1(\alpha,dv) \right) \sigma_n^2(\alpha, z) F_{\alpha}(dz) \\
&&  \tilde{V}_n^0(\alpha, u)-\int_{-\infty}^{u} a_n(\alpha, z)^{\top} A_n^{-1}(\alpha, z) \left(\int_{z}^{\infty}
    a_n(\alpha,v)\tilde{V}_n^0(\alpha,dv) \right) \sigma_n^2(\alpha, z) F_{\alpha}(dz).
\end{eqnarray*}
Since
\begin{eqnarray*}
\frac{1}{\sqrt{n}} (\tilde{V}_n^1(\alpha, u)- \tilde{V}_n^0(\alpha, u)) &=& \frac{1}{n} \sum_{i=1}^{n}[Y_i-g(\tilde{\beta}_{0}^{\top}X_i, \tilde{\theta}_0)] \{I(\alpha^{\top}\hat{B}_n^{\top} X_i \leq u)-I(\alpha^{\top} B^{\top} X_i \leq u)\} \\
&=& \frac{1}{n} \sum_{i=1}^{n} \varepsilon_i \{I(\alpha^{\top}\hat{B}_n^{\top} X_i \leq u)-I(\alpha^{\top} B^{\top} X_i \leq u)\}+\\
&&  \frac{1}{n} \sum_{i=1}^{n}[G(X_i)-g(\tilde{\beta}_{0}^{\top}X_i, \tilde{\theta}_0)] \{I(\alpha^{\top}\hat{B}_n^{\top} X_i \leq u)-I(\alpha^{\top} B^{\top} X_i \leq u)\},
\end{eqnarray*}
it follows that $\frac{1}{\sqrt{n}} (\tilde{V}_n^1(\alpha, u)- \tilde{V}_n^0(\alpha, u)) =o_p(1)$. For the two integrals in $T_n \tilde{V}_n^1(\alpha, u)-T_n \tilde{V}_n^0(\alpha, u)$, we have
\begin{eqnarray*}
&&  \frac{1}{\sqrt{n}} \int_{-\infty}^{u} a_n(\alpha, z)^{\top} A_n^{-1}(\alpha, z) \left(\int_{z}^{\infty} a_n(\alpha,v)
    (\tilde{V}_n^1(\alpha,dv)-\tilde{V}_n^1(\alpha,dv)) \right) \sigma_n^2(\alpha, z)  F_{\alpha}(dz)\\
&=& \frac{1}{n} \sum_{i=1}^{n} \varepsilon_i \int_{-\infty}^{u} a_n(\alpha, z)^{\top} A_n^{-1}(\alpha, z) a_n(\alpha, \alpha^{\top}\hat{B}_n^{\top}X_i)
    I(\alpha^{\top}\hat{B}_n^{\top} X_i \geq z) \sigma_n^2(\alpha, z) F_{\alpha}(dz)-\\
&&  \frac{1}{n} \sum_{i=1}^{n} \varepsilon_i \int_{-\infty}^{u} a_n(\alpha, z)^{\top} A_n^{-1}(\alpha, z) a_n(\alpha, \alpha^{\top} B^{\top} X_i)
    I(\alpha^{\top} B^{\top} X_i \geq z) \sigma_n^2(\alpha, z) F_{\alpha}(dz)+\\
&&  \frac{1}{n} \sum_{i=1}^{n} [G(X_i)-g(\tilde{\beta}_{0}^{\top}X_i, \tilde{\theta}_0)] \int_{-\infty}^{u} a_n(\alpha, z)^{\top} A_n^{-1}(\alpha, z)
    a_n(\alpha, \alpha^{\top}\hat{B}_n^{\top}X_i) I(\alpha^{\top}\hat{B}_n^{\top} X_i \geq z) \sigma_n^2(\alpha, z) F_{\alpha}(dz)-\\
&&  \frac{1}{n} \sum_{i=1}^{n} [G(X_i)-g(\tilde{\beta}_{0}^{\top}X_i, \tilde{\theta}_0)] \int_{-\infty}^{u} a_n(\alpha, z)^{\top} A_n^{-1}(\alpha, z)
    a_n(\alpha, \alpha^{\top} B^{\top} X_i) I(\alpha^{\top} B^{\top} X_i \geq z) \sigma_n^2(\alpha, z) F_{\alpha}(dz)\\
&=& o_p(1).
\end{eqnarray*}
Therefore, we obtain that
$$
\frac{1}{\sqrt{n}} (\hat{T}_n V_n(\alpha, u)-T_n \tilde{V}_n^0(\alpha, u))=o_p(1).
$$
Note that
\begin{eqnarray*}
\frac{1}{\sqrt{n}} T_n \tilde{V}_n^0(\alpha, u)
&=&\frac{1}{n} \sum_{i=1}^{n}[Y_i-g(\tilde{\beta}_{0}^{\top}X_i, \tilde{\theta}_0)] I(\alpha^{\top} B^{\top} X_i \leq u)-\\
&& \frac{1}{n} \sum_{i=1}^{n} \varepsilon_i \int_{-\infty}^{u} a_n(\alpha, z)^{\top} A_n^{-1}(\alpha, z) a_n(\alpha, \alpha^{\top} B^{\top} X_i) I(\alpha^{\top} B^{\top} X_i \geq z) \sigma_n^2(\alpha, z) F_{\alpha}(dz)-\\
&& \frac{1}{n} \sum_{i=1}^{n} [G(X_i)-g(\tilde{\beta}_{0}^{\top}X_i, \tilde{\theta}_0)] \int_{-\infty}^{u} \{a_n(\alpha, z)^{\top} A_n^{-1}(\alpha, z) a_n(\alpha, \alpha^{\top} B^{\top} X_i) \times \\
&& I(\alpha^{\top} B^{\top} X_i \geq z) \sigma_n^2(\alpha, z)\} F_{\alpha}(dz).
\end{eqnarray*}
It follows that
\begin{eqnarray*}
\frac{1}{\sqrt{n}} T_n \tilde{V}_n^0(\alpha, u) = \tilde{G}_{1n}(\alpha, u) - \tilde{G}_{2n}(\alpha, u) +o_p(1) \longrightarrow \tilde{G}_1(\alpha, u) - \tilde{G}_2(\alpha, u)
\end{eqnarray*}
where
\begin{eqnarray*}
\tilde{G}_{1n}(\alpha, u) &=& E \{[G(X)-g(\tilde{\beta}_{0}^{\top}X, \tilde{\theta}_0)]I(\alpha^{\top} B^{\top} X \leq u)\} \\
\tilde{G}_{2n}(\alpha, u) &=& E \{[G(X)-g(\tilde{\beta}_{0}^{\top}X, \tilde{\theta}_0)] \int_{-\infty}^{u} a_n(\alpha, z)^{\top}A_n^{-1}(\alpha, z) a_n(\alpha, \alpha^{\top} B^{\top} X) \times \\
&&  I(\alpha^{\top} B^{\top} X \geq z) \sigma_n^2(\alpha, z) F_{\alpha}(dz)\}
\end{eqnarray*}
Therefore, we obtain that
$$
\frac{1}{\sqrt{n}} \sup_{ \hat{\alpha} \in \mathcal{S}_{\hat{q}}^{+}} |\hat{T}_n V_{n}(\hat{\alpha}, u)| \longrightarrow |L(u)|
$$
where $L(u)$ is an nonzero function.

(2) We use the same notations as in the arguments of Theorem~\ref{Theorem 3.2}. Under the local alternatives~(\ref{3.8}), by Proposition~3, we have $\mathbb{P}\{\hat{q}=1 \} \to 1$. Thus we just work on this event $\{ \hat{q}=1 \}$. Hence $\mathcal{S}_{\hat{q}}^{+}=\{ 1 \}$ and $\sup_{\hat{\alpha} \in \mathcal{S}_{\hat{q}}^{+}} |\hat{T}_n V_{n}(\hat{\alpha}, u)|=|\hat{T}_n V_{n}(\hat{\alpha}, u)| $.

Following the same arguments for Theorem~\ref{Theorem 3.2}, we obtain that
$$
\hat{T}_n V_n(\hat{\alpha}, u)- T_n V_n^1(\hat{\alpha}, u) =o_p(1)
$$
Next, we consider $T_n V_n^1(\hat{\alpha}, u)-T_n V_n^0(u)$. Recall that
\begin{eqnarray*}
V_n^0(u) = \frac{1}{\sqrt{n}}\sum_{i=1}^n [Y_i- g(\beta_{0}^{\top} X_i,\theta_0)]I(\kappa \beta_{0}^{\top} X_i \leq u), \\
V_n^1(\hat{\alpha}, u) = \frac{1}{\sqrt{n}}\sum_{i=1}^n [Y_i- g(\beta_{0}^{\top} X_i,\theta_0)]I(\hat{B}_n^{\top} X_i \leq u).
\end{eqnarray*}
Under $H_{1n}$, we have
\begin{eqnarray*}
V_n^1(\hat{\alpha}, u)-V_n^0(u) &=& \frac{1}{\sqrt{n}}\sum_{i=1}^n \varepsilon_i [I(\hat{B}_n^{\top} X_i \leq u) - I(\kappa \beta_{0}^{\top} X_i \leq
                                    u)]+\\
                                &&  \frac{1}{n}\sum_{i=1}^n G(X_i) [I(\hat{B}_n^{\top} X_i \leq u)-I(\kappa \beta_{0}^{\top} X_i \leq u)].
\end{eqnarray*}
Then $V_n^1(\hat{\alpha}, u)-V_n^0(u)=o_p(1)$. For the integrals in $T_n V_n^1(\hat{\alpha}, u)-T_n V_n^0(u)$, since
\begin{eqnarray*}
&&  \int_{-\infty}^{u} a_n(z)^{\top} A_n^{-1}(z) \left\{\int_{z}^{\infty} a_n(v) [V_n^1(\hat{\alpha}, dv)-V_n^0(dv)]\right\} \psi_n(dz) \\
&=& \frac{1}{\sqrt{n}}\sum_{i=1}^n \varepsilon_i \int_{-\infty}^{u} a_n(z)^{\top} A_n^{-1}(z) a_n(\hat{B}_n^{\top} X_i)I(\hat{B}_n^{\top} X_i \geq z)
    \psi_n(dz) -\\
&&  \frac{1}{\sqrt{n}}\sum_{i=1}^n \varepsilon_i \int_{-\infty}^{u} a_n(z)^{\top} A_n^{-1}(z) a_n(\kappa \beta_{0}^{\top} X_i)I(\kappa
    \beta_{0}^{\top} X_i \geq z) \psi_n(dz)+\\
&&  \frac{1}{n}\sum_{i=1}^n G(X_i) \int_{-\infty}^{u} a_n(z)^{\top} A_n^{-1}(z) a_n(\hat{B}_n^{\top} X_i)I(\hat{B}_n^{\top} X_i \geq z)
    \psi_n(dz)-\\
&&  \frac{1}{n}\sum_{i=1}^n G(X_i) \int_{-\infty}^{u} a_n(z)^{\top} A_n^{-1}(z) a_n(\kappa \beta_{0}^{\top} X_i)I(\kappa
    \beta_{0}^{\top} X_i \geq z) \psi_n(dz),
\end{eqnarray*}
by the same arguments for Theorem~\ref{Theorem 3.2}, we have
$$
\int_{-\infty}^{u} a_n(z)^{\top} A_n^{-1}(z) \left\{\int_{z}^{\infty} a_n(v) [V_n^1(\hat{\alpha}, dv)-V_n^0(dv)]\right\} \psi_n(dz)=o_p(1).
$$
Hence we obtain that $T_n V_n^1(\hat{\alpha}, u)-T_n V_n^0(u)=o_p(1)$.

To complete the proof, it remains to derive the asymptotic distribution of $T_n V_n^0(u)$. Under the alternatives, note that
$$
V_n^0(u)=\frac{1}{\sqrt{n}}\sum_{i=1}^n \varepsilon_i I(\kappa \beta_{0}^{\top} X_i \leq u) +\frac{1}{n}\sum_{i=1}^n G(X_i) I(\kappa \beta_{0}^{\top} X_i \leq u).
$$
It follows that
\begin{eqnarray*}
T_n V_n^0(u) &=& \frac{1}{\sqrt{n}}\sum_{i=1}^n \varepsilon_i I(\kappa \beta_{0}^{\top} X_i \leq u) +\frac{1}{n}\sum_{i=1}^n G(X_i)
                 I(\kappa \beta_{0}^{\top} X_i \leq u)- \\
             &&  \frac{1}{\sqrt{n}}\sum_{i=1}^n \varepsilon_i \int_{-\infty}^{u} a_n(z)^{\top} A_n^{-1}(z) a_n(\kappa \beta_{0}^{\top} X_i)
                 I(\kappa \beta_{0}^{\top} X_i \geq z) \psi_n(dz)- \\
             &&  \frac{1}{n}\sum_{i=1}^n G(X_i) \int_{-\infty}^{u} a_n(z)^{\top} A_n^{-1}(z) a_n(\kappa \beta_{0}^{\top} X_i)
                 I(\kappa \beta_{0}^{\top} X_i \geq z) \psi_n(dz)
\end{eqnarray*}
By Glivenko-Cantelli Theorem, we have
\begin{eqnarray*}
&&  \frac{1}{n}\sum_{i=1}^n G(X_i) I(\kappa \beta_{0}^{\top} X_i \leq u) = E[G(X) I(\kappa \beta_{0}^{\top} X \leq u)] + o_p(1),\\
&&  \frac{1}{n}\sum_{i=1}^n G(X_i) \int_{-\infty}^{u} a_n(z)^{\top} A_n^{-1}(z) a_n(\kappa \beta_{0}^{\top} X_i)
    I(\kappa \beta_{0}^{\top} X_i \geq z)\psi_n(dz) \\
&=& E \left( G(X) \int_{-\infty}^{u} a_n(z)^{\top} A_n^{-1}(z) a_n(\kappa \beta_{0}^{\top} X)
    I(\kappa \beta_{0}^{\top} X \geq z) \psi_n(dz)\right)+o_p(1).
\end{eqnarray*}
Since $E[G(X) I(\kappa \beta_{0}^{\top} X \leq u)] \to G_1(u)$ and
$$E \left( G(X) \int_{-\infty}^{u} a_n(z)^{\top} A_n^{-1}(z) a_n(\kappa \beta_{0}^{\top} X) I(\kappa \beta_{0}^{\top} X \geq z) \psi_n(dz)\right) \to G_2(u),$$
we conclude that
$$
T_n V_n^0(u) \longrightarrow V_{\infty}(u) + G_1(u)-G_2(u) \quad {\rm in \ distribution},
$$
where $V_{\infty}(u)$ is a zero-mean Gaussian process given by (\ref{3.6}).     \hfill$\Box$

\newpage
\begin{table}[ht!]\caption{Empirical sizes and powers of $ACM_n^2$, $T_n^{SZ}$, $PCvM_n$, $ICM_n$, $T_n^{ZH}$ and $T_n^{GWZ}$ for $H_0$ vs. $H_{11}$ in Study 1.}
\centering
{\small\scriptsize\hspace{12.5cm}
\renewcommand{\arraystretch}{1}\tabcolsep 0.5cm
\begin{tabular}{*{20}{c}}
\hline
&\multicolumn{1}{c}{a}&\multicolumn{1}{c}{n=100}&\multicolumn{1}{c}{n=200}&\multicolumn{1}{c}{n=400}&\multicolumn{1}{c}{n=800}\\
&&\multicolumn{1}{c}{p=7}& \multicolumn{1}{c}{p=10}  & \multicolumn{1}{c}{p=12}&\multicolumn{1}{c}{p=16}\\
\hline
$ACM_n^2, \alpha=0.10$   &0.0   &0.0970 &0.0905 &0.0890 &0.1020\\
                         &0.5   &0.8650 &0.9915 &1.0000 &1.0000\\
$ACM_n^2, \alpha=0.05$   &0.0   &0.0500 &0.0530 &0.0500 &0.0505\\
                         &0.5   &0.7770 &0.9810 &1.0000 &1.0000\\
$ACM_n^2, \alpha=0.01$   &0.0     &0.0085 &0.0105 &0.0115 &0.0130\\
                         &0.5     &0.5620 &0.9095 &0.9975 &1.0000\\
\hline
$T_n^{SZ}, \alpha=0.10$   &0.0   &0.0915 &0.0995 &0.1060 &0.0985\\
                          &0.5   &0.8675 &0.9865 &1.0000 &1.0000\\

$T_n^{SZ}, \alpha=0.05$   &0.0   &0.0510 &0.0470 &0.0420 &0.0495\\
                          &0.5   &0.7825 &0.9795 &1.0000 &1.0000\\

$T_n^{SZ}, \alpha=0.01$   &0.0     &0.0120 &0.0090 &0.0120 &0.0100\\
                          &0.5     &0.5290 &0.9065 &0.9990 &1.0000\\
\hline
$PCvM_n, \alpha=0.10$     &0.0     &0.1140 &0.1220 &0.0980 &0.1190\\
                          &0.5     &0.8850 &0.9880 &1.0000 &1.0000\\
$PCvM_n, \alpha=0.05$     &0.0     &0.0480    &0.0590    &0.0650    &0.0490\\
                          &0.5     &0.8110    &0.9860    &1.0000    &1.0000\\
$PCvM_n, \alpha=0.01$     &0.0      &0.0150    &0.0100    &0.0110    &0.0090\\
                          &0.5       &0.6190    &0.9310    &0.9970    &1.0000\\
\hline
$ICM_n, \alpha=0.10$      &0.0     &0.0390 &0.0010 &0.0000 &0.0000\\
                          &0.5     &0.5490 &0.2910 &0.1760 &0.0000\\
$ICM_n, \alpha=0.05$     &0.0      &0.0070 &0.0000 &0.0000 &0.0000\\
                         &0.5      &0.3900 &0.0910 &0.0180 &0.0000\\
$ICM_n, \alpha=0.01$     &0.0      &0.0000 &0.0000 &0.0000 &0.0000\\
                         &0.5      &0.1220 &0.0060 &0.0020 &0.0000\\
\hline
$T_n^{ZH}, \alpha=0.10$      &0.0     &0.0805 &0.0950 &0.1055 &0.1060\\
                             &0.5     &0.2240 &0.2205 &0.2420 &0.2430\\
$T_n^{ZH}, \alpha=0.05$      &0.0     &0.0305 &0.0300 &0.0330 &0.0310\\
                             &0.5     &0.1460 &0.1285 &0.1445 &0.0980\\
$T_n^{ZH}, \alpha=0.01$      &0.0     &0.0015 &0.0020 &0.0025 &0.0025\\
                             &0.5     &0.0420 &0.0210 &0.0225 &0.0150\\
\hline
$T_n^{GWZ}, \alpha=0.10$      &0.0     &0.0710 &0.0755 &0.0850 &0.0830\\
                              &0.5     &0.8170 &0.9795 &1.0000 &1.0000\\
$T_n^{GWZ}, \alpha=0.05$      &0.0     &0.0525 &0.0430 &0.0585 &0.0475\\
                              &0.5     &0.7690 &0.9690 &1.0000 &1.0000\\
$T_n^{GWZ}, \alpha=0.01$      &0.0     &0.0220 &0.0170 &0.0205 &0.0170\\
                              &0.5     &0.6510 &0.9455 &0.9995 &1.0000\\
\hline
\end{tabular}
}
\end{table}

\newpage
\begin{table}[ht!]\caption{Empirical sizes and powers of $ACM_n^2$, $T_n^{SZ}$, $PCvM_n$, $ICM_n$, $T_n^{ZH}$ and $T_n^{GWZ}$ for $H_0$ vs. $H_{12}$ in Study 1.}
\centering
{\small\scriptsize\hspace{12.5cm}
\renewcommand{\arraystretch}{1}\tabcolsep 0.5cm
\begin{tabular}{*{20}{c}}
\hline
&\multicolumn{1}{c}{a}&\multicolumn{1}{c}{n=100}&\multicolumn{1}{c}{n=200}&\multicolumn{1}{c}{n=400}&\multicolumn{1}{c}{n=800}\\
&&\multicolumn{1}{c}{p=7}& \multicolumn{1}{c}{p=10}  & \multicolumn{1}{c}{p=12}&\multicolumn{1}{c}{p=16}\\
\hline
$ACM_n^2, \alpha=0.10$   &0.0   &0.1010 &0.0925 &0.1055 &0.0900\\
                         &0.5   &0.2550 &0.5135 &0.9190 &1.0000\\
$ACM_n^2, \alpha=0.05$   &0.0   &0.0520 &0.0465 &0.0445 &0.0515\\
                         &0.5   &0.1445 &0.3225 &0.7550 &1.0000\\
$ACM_n^2, \alpha=0.01$   &0.0   &0.0095 &0.0090 &0.0120 &0.0070\\
                         &0.5   &0.0460 &0.1060 &0.3485 &0.9140\\
\hline
$T_n^{SZ}, \alpha=0.10$   &0.0   &0.0980 &0.0990 &0.0865 &0.0930\\
                          &0.5   &0.2630 &0.5265 &0.9240 &1.0000\\
$T_n^{SZ}, \alpha=0.05$   &0.0   &0.0530 &0.0480 &0.0515 &0.0495\\
                          &0.5   &0.1760 &0.3235 &0.7350 &0.9970\\
$T_n^{SZ}, \alpha=0.01$   &0.0     &0.0100 &0.0060 &0.0085 &0.0105\\
                          &0.5     &0.0470 &0.1145 &0.3580 &0.9350\\
\hline
$PCvM_n, \alpha=0.10$     &0.0     &0.1080 &0.1170 &0.1230 &0.1000\\
                          &0.5     &0.2560 &0.3390 &0.5160 &0.7590\\
$PCvM_n, \alpha=0.05$     &0.0     &0.0530 &0.0590 &0.0440 &0.0700\\
                          &0.5     &0.1470 &0.2320 &0.4080 &0.6250\\
$PCvM_n, \alpha=0.01$     &0.0       &0.0130 &0.0130 &0.0080 &0.0130\\
                          &0.5       &0.0450 &0.1020 &0.2010 &0.4080\\
\hline
$ICM_n, \alpha=0.10$      &0.0     &0.0370 &0.0000 &0.0000 &0.0000\\
                          &0.5     &0.1950 &0.0330 &0.0020 &0.0000\\
$ICM_n, \alpha=0.05$      &0.0      &0.0110 &0.0000 &0.0000 &0.0000\\
                          &0.5      &0.0790 &0.0020 &0.0000 &0.0000\\
$ICM_n, \alpha=0.01$      &0.0      &0.0020 &0.0000 &0.0000 &0.0000\\
                          &0.5      &0.0110 &0.0000 &0.0000 &0.0000\\
\hline
$T_n^{ZH}, \alpha=0.10$      &0.0     &0.0805 &0.0830 &0.0800 &0.1095\\
                             &0.5     &0.1630 &0.1515 &0.1825 &0.1665\\
$T_n^{ZH}, \alpha=0.05$      &0.0     &0.0325 &0.0350 &0.0320 &0.0330\\
                             &0.5     &0.0755 &0.0775 &0.0940 &0.0615\\
$T_n^{ZH}, \alpha=0.01$      &0.0     &0.0045 &0.0015 &0.0035 &0.0035\\
                             &0.5     &0.0155 &0.0095 &0.0125 &0.0060\\
\hline
$T_n^{GWZ}, \alpha=0.10$      &0.0     &0.0820 &0.0725 &0.0810 &0.0745\\
                              &0.5     &0.6765 &0.9460 &1.0000 &1.0000\\
$T_n^{GWZ}, \alpha=0.05$      &0.0     &0.0495 &0.0500 &0.0500 &0.0535\\
                              &0.5     &0.6035 &0.9335 &0.9995 &1.0000\\
$T_n^{GWZ}, \alpha=0.01$      &0.0     &0.0190 &0.0165 &0.0180 &0.0210\\
                              &0.5     &0.4660 &0.8705 &0.9980 &1.0000\\
\hline
\end{tabular}
}
\end{table}

\newpage
     \begin{table}[ht!]\caption{Empirical sizes and powers of $ACM_n^2$, $T_n^{SZ}$, $PCvM_n$, $ICM_n$, $T_n^{ZH}$ and $T_n^{GWZ}$ for $H_0$ vs. $H_{13}$ in Study 1.}
\centering
{\small\scriptsize\hspace{12.5cm}
\renewcommand{\arraystretch}{1}\tabcolsep 0.5cm
\begin{tabular}{*{20}{c}}
\hline
&\multicolumn{1}{c}{a}&\multicolumn{1}{c}{n=100}&\multicolumn{1}{c}{n=200}&\multicolumn{1}{c}{n=400}&\multicolumn{1}{c}{n=800}\\
&&\multicolumn{1}{c}{p=7}& \multicolumn{1}{c}{p=10}  & \multicolumn{1}{c}{p=12}&\multicolumn{1}{c}{p=16}\\
\hline
$ACM_n^2, \alpha=0.10$   &0.00   &0.0985 &0.1050 &0.1085 &0.1090\\
                         &0.25   &0.7130 &0.9410 &0.9955 &1.0000\\
$ACM_n^2, \alpha=0.05$   &0.00   &0.0500 &0.0455 &0.0435 &0.0450\\
                         &0.25   &0.5970 &0.8945 &0.9980 &1.0000\\
$ACM_n^2, \alpha=0.01$   &0.00   &0.0095 &0.0090 &0.0095 &0.0090\\
                         &0.25   &0.3470 &0.7225 &0.9840 &1.0000\\
\hline
$T_n^{SZ}, \alpha=0.10$   &0.00   &0.0960 &0.1055 &0.1060 &0.0960\\
                          &0.25   &0.7190 &0.9405 &0.9975 &1.0000\\
$T_n^{SZ}, \alpha=0.05$   &0.00   &0.0505 &0.0420 &0.0470 &0.0495\\
                          &0.25   &0.5940 &0.8980 &0.9945 &1.0000\\
$T_n^{SZ}, \alpha=0.01$   &0.00   &0.0080 &0.0125 &0.0095 &0.0115\\
                          &0.25   &0.3310 &0.7190 &0.9705 &0.9995\\
\hline
$PCvM_n, \alpha=0.10$     &0.00   &0.1030 &0.0980 &0.1140 &0.1210\\
                          &0.25   &0.7180 &0.9500 &0.9970 &1.0000\\
$PCvM_n, \alpha=0.05$     &0.00   &0.0580 &0.0600 &0.0440 &0.0570\\
                          &0.25   &0.6160 &0.8980 &0.9970 &1.0000\\
$PCvM_n, \alpha=0.01$     &0.00   &0.0060 &0.0150 &0.0080 &0.0070\\
                          &0.25   &0.3870 &0.7360 &0.9780 &1.0000\\
\hline
$ICM_n, \alpha=0.10$      &0.00   &0.0290 &0.0010 &0.0000 &0.0000\\
                          &0.25   &0.1590 &0.0190 &0.0030 &0.0000\\
$ICM_n, \alpha=0.05$      &0.00   &0.0110 &0.0000 &0.0000 &0.0000\\
                          &0.25   &0.0590 &0.0010 &0.0000 &0.0000\\
$ICM_n, \alpha=0.01$      &0.00   &0.0010 &0.0000 &0.0000 &0.0000\\
                          &0.25   &0.0140 &0.0000 &0.0000 &0.0000\\
\hline
$T_n^{ZH}, \alpha=0.10$   &0.00   &0.0765 &0.0810 &0.0940 &0.0970\\
                          &0.25   &0.1135 &0.1185 &0.1400 &0.1305\\
$T_n^{ZH}, \alpha=0.05$   &0.00   &0.0275 &0.0310 &0.0315 &0.0340\\
                          &0.25   &0.0730 &0.0485 &0.0745 &0.0625\\
$T_n^{ZH}, \alpha=0.01$   &0.00   &0.0030 &0.0020 &0.0030 &0.0010\\
                          &0.25   &0.0055 &0.0060 &0.0080 &0.0030\\
\hline
$T_n^{GWZ}, \alpha=0.10$  &0.00   &0.0800 &0.0735 &0.0770 &0.0765\\
                          &0.25   &0.4580 &0.7430 &0.9795 &0.9995\\
$T_n^{GWZ}, \alpha=0.05$  &0.00   &0.0510 &0.0505 &0.0540 &0.0490\\
                          &0.25   &0.3840 &0.6660 &0.9465 &1.0000\\
$T_n^{GWZ}, \alpha=0.01$  &0.00   &0.0200 &0.0225 &0.0235 &0.0240\\
                          &0.25   &0.2590 &0.5570 &0.9040 &0.9995\\
\hline
\end{tabular}
}
\end{table}

\newpage
\begin{table}[ht!]\caption{Empirical sizes and powers of $ACM_n^2$, $T_n^{SZ}$, $PCvM_n$, $ICM_n$, $T_n^{ZH}$ and $T_n^{GWZ}$ for $H_0$ vs. $H_{14}$ in Study 1.}
\centering
{\small\scriptsize\hspace{12.5cm}
\renewcommand{\arraystretch}{1}\tabcolsep 0.5cm
\begin{tabular}{*{20}{c}}
\hline
&\multicolumn{1}{c}{a}&\multicolumn{1}{c}{n=100}&\multicolumn{1}{c}{n=200}&\multicolumn{1}{c}{n=400}&\multicolumn{1}{c}{n=800}\\
&&\multicolumn{1}{c}{p=7}& \multicolumn{1}{c}{p=10}  & \multicolumn{1}{c}{p=12}&\multicolumn{1}{c}{p=16}\\
\hline
$ACM_n^2, \alpha=0.10$    &0.00   &0.1130 &0.1000 &0.0970 &0.0955\\
                          &0.25   &0.9825 &1.0000 &1.0000 &1.0000\\
$ACM_n^2, \alpha=0.05$    &0.00   &0.0520 &0.0460 &0.0545 &0.0490\\
                          &0.25   &0.9525 &1.0000 &1.0000 &1.0000\\
$ACM_n^2, \alpha=0.01$    &0.00   &0.0110 &0.0090 &0.0075 &0.0105\\
                          &0.25   &0.8680 &0.9950 &1.0000 &1.0000\\
\hline
$T_n^{SZ}, \alpha=0.10$   &0.00   &0.1090 &0.0970 &0.0910 &0.1090\\
                          &0.25   &0.9805 &0.9990 &1.0000 &1.0000\\
$T_n^{SZ}, \alpha=0.05$   &0.00   &0.0475 &0.0490 &0.0460 &0.0555\\
                          &0.25   &0.9605 &0.9995 &1.0000 &1.0000\\
$T_n^{SZ}, \alpha=0.01$   &0.00   &0.0095 &0.0115 &0.0075 &0.0090\\
                          &0.25   &0.8700 &0.9970 &1.0000 &1.0000\\
\hline
$PCvM_n, \alpha=0.10$     &0.00   &0.0950 &0.1130 &0.1110 &0.1040\\
                          &0.25   &0.9960 &1.0000 &1.0000 &1.0000\\
$PCvM_n, \alpha=0.05$     &0.00   &0.0580 &0.0540 &0.0570 &0.0540\\
                          &0.25   &0.9690 &0.9990 &1.0000 &1.0000\\
$PCvM_n, \alpha=0.01$     &0.00   &0.0140 &0.0170 &0.0080 &0.0150\\
                          &0.25   &0.8730 &0.9980 &1.0000 &1.0000\\
\hline
$ICM_n, \alpha=0.10$      &0.00   &0.0290 &0.0010 &0.0000 &0.0000\\
                          &0.25   &0.5680 &0.2420 &0.1330 &0.0000\\
$ICM_n, \alpha=0.05$      &0.00   &0.0050 &0.0000 &0.0000 &0.0000\\
                          &0.25   &0.3670 &0.0740 &0.0120 &0.0000\\
$ICM_n, \alpha=0.01$      &0.00   &0.0010 &0.0000 &0.0000 &0.0000\\
                          &0.25   &0.1060 &0.0040 &0.0000 &0.0000\\
\hline
$T_n^{ZH}, \alpha=0.10$   &0.00   &0.0700 &0.0910 &0.0875 &0.0985\\
                          &0.25   &0.2420 &0.2125 &0.2680 &0.2210\\
$T_n^{ZH}, \alpha=0.05$   &0.00   &0.0320 &0.0295 &0.0325 &0.0380\\
                          &0.25   &0.1145 &0.1195 &0.1410 &0.1145\\
$T_n^{ZH}, \alpha=0.01$   &0.00   &0.0015 &0.0045 &0.0050 &0.0035\\
                          &0.25   &0.0335 &0.0230 &0.0220 &0.0095\\
\hline
$T_n^{GWZ}, \alpha=0.10$  &0.00   &0.0780 &0.0805 &0.0815 &0.0830\\
                          &0.25   &0.8645 &0.9935 &1.0000 &1.0000\\
$T_n^{GWZ}, \alpha=0.05$  &0.00   &0.0455 &0.0560 &0.0540 &0.0625\\
                          &0.25   &0.8405 &0.9870 &1.0000 &1.0000\\
$T_n^{GWZ}, \alpha=0.01$  &0.00   &0.0210 &0.0195 &0.0225 &0.0195\\
                          &0.25   &0.7285 &0.9735 &1.0000 &1.0000\\
\hline
\end{tabular}
}
\end{table}

\newpage
\begin{table}[ht!]\caption{Empirical sizes and powers of $ACM_n^2$, $T_n^{SZ}$, $PCvM_n$, $ICM_n$, $T_n^{ZH}$ and $T_n^{GWZ}$ for $H_0$ vs. $H_{21}$ in Study 2.}
\centering
{\small\scriptsize\hspace{12.5cm}
\renewcommand{\arraystretch}{1}\tabcolsep 0.5cm
\begin{tabular}{*{20}{c}}
\hline
&\multicolumn{1}{c}{a}&\multicolumn{1}{c}{n=100}&\multicolumn{1}{c}{n=200}&\multicolumn{1}{c}{n=400}&\multicolumn{1}{c}{n=800}\\
&&\multicolumn{1}{c}{p=7}& \multicolumn{1}{c}{p=10}  & \multicolumn{1}{c}{p=12}&\multicolumn{1}{c}{p=16}\\
\hline
$ACM_n^2, \alpha=0.10$    &0.00   &0.1075 &0.0965 &0.0910 &0.1035\\
                          &0.25   &0.6185 &0.8980 &0.9955 &1.0000\\
$ACM_n^2, \alpha=0.05$    &0.00   &0.0520 &0.0490 &0.0495 &0.0570\\
                          &0.25   &0.4895 &0.8185 &0.9925 &1.0000\\
$ACM_n^2, \alpha=0.01$    &0.00   &0.0100 &0.0085 &0.0100 &0.0115\\
                          &0.25   &0.2505 &0.5920 &0.9450 &0.9995\\
\hline
$T_n^{SZ}, \alpha=0.10$   &0.00   &0.0935 &0.0935 &0.1070 &0.1055\\
                          &0.25   &0.7005 &0.9120 &0.9965 &1.0000\\
$T_n^{SZ}, \alpha=0.05$   &0.00   &0.0515 &0.0425 &0.0460 &0.0445\\
                          &0.25   &0.5600 &0.8505 &0.9940 &1.0000\\
$T_n^{SZ}, \alpha=0.01$   &0.00   &0.0080 &0.0100 &0.0060 &0.0100\\
                          &0.25   &0.3180 &0.6680 &0.9665 &1.0000\\
\hline
$PCvM_n, \alpha=0.10$     &0.00   &0.1150 &0.0910 &0.1090 &0.1050\\
                          &0.25   &0.7080 &0.9320 &0.9990 &1.0000\\
$PCvM_n, \alpha=0.05$     &0.00   &0.0560 &0.0480 &0.0570 &0.0430\\
                          &0.25   &0.6230 &0.9080 &0.9960 &1.0000\\
$PCvM_n, \alpha=0.01$     &0.00   &0.0080 &0.0120 &0.0100 &0.0090\\
                          &0.25   &0.3810 &0.7230 &0.9820 &1.0000\\
\hline
$ICM_n, \alpha=0.10$      &0.00   &0.0180 &0.0010 &0.0000 &0.0000\\
                          &0.25   &0.1220 &0.0060 &0.0000 &0.0000\\
$ICM_n, \alpha=0.05$      &0.00   &0.0040 &0.0000 &0.0000 &0.0000\\
                          &0.25   &0.0470 &0.0010 &0.0000 &0.0000\\
$ICM_n, \alpha=0.01$      &0.00   &0.0000 &0.0000 &0.0000 &0.0000\\
                          &0.25   &0.0070 &0.0000 &0.0000 &0.0000\\
\hline
$T_n^{ZH}, \alpha=0.10$   &0.00   &0.1100 &0.1020 &0.0960 &0.1110\\
                          &0.25   &0.1420 &0.1370 &0.1550 &0.1545\\
$T_n^{ZH}, \alpha=0.05$   &0.00   &0.0400 &0.0410 &0.0365 &0.0390\\
                          &0.25   &0.0710 &0.0700 &0.0610 &0.0550\\
$T_n^{ZH}, \alpha=0.01$   &0.00   &0.0045 &0.0035 &0.0040 &0.0035\\
                          &0.25   &0.0140 &0.0075 &0.0065 &0.0035\\
\hline
$T_n^{GWZ}, \alpha=0.10$  &0.00   &0.1135 &0.1045 &0.1115 &0.1240\\
                          &0.25   &0.5275 &0.8140 &0.9860 &0.9995\\
$T_n^{GWZ}, \alpha=0.05$  &0.00   &0.0790 &0.0760 &0.0775 &0.0750\\
                          &0.25   &0.4625 &0.7300 &0.9610 &1.0000\\
$T_n^{GWZ}, \alpha=0.01$  &0.00   &0.0340 &0.0345 &0.0310 &0.0305\\
                          &0.25   &0.3175 &0.6015 &0.9295 &0.9985\\
\hline
\end{tabular}
}
\end{table}

\newpage
\begin{table}[ht!]\caption{Empirical sizes and powers of $ACM_n^2$, $T_n^{SZ}$, $PCvM_n$, $ICM_n$, $T_n^{ZH}$ and $T_n^{GWZ}$ for $H_0$ vs. $H_{22}$ in Study 2.}
\centering
{\small\scriptsize\hspace{12.5cm}
\renewcommand{\arraystretch}{1}\tabcolsep 0.5cm
\begin{tabular}{*{20}{c}}
\hline
&\multicolumn{1}{c}{a}&\multicolumn{1}{c}{n=100}&\multicolumn{1}{c}{n=200}&\multicolumn{1}{c}{n=400}&\multicolumn{1}{c}{n=800}\\
&&\multicolumn{1}{c}{p=7}& \multicolumn{1}{c}{p=10}  & \multicolumn{1}{c}{p=12}&\multicolumn{1}{c}{p=16}\\
\hline
$ACM_n^2, \alpha=0.10$    &0.0   &0.1180 &0.1190 &0.1095 &0.1060\\
                          &0.5   &0.2255 &0.3090 &0.4805 &0.7390\\
$ACM_n^2, \alpha=0.05$    &0.0   &0.0575 &0.0550 &0.0585 &0.0530\\
                          &0.5   &0.1295 &0.1895 &0.3030 &0.5790\\
$ACM_n^2, \alpha=0.01$    &0.0   &0.0110 &0.0135 &0.0115 &0.0120\\
                          &0.5   &0.0325 &0.0605 &0.1155 &0.2830\\
\hline
$T_n^{SZ}, \alpha=0.10$   &0.0   &0.1110 &0.1075 &0.0980 &0.1010\\
                          &0.5   &0.1335 &0.1480 &0.1580 &0.1920\\
$T_n^{SZ}, \alpha=0.05$   &0.0   &0.0650 &0.0535 &0.0550 &0.0550\\
                          &0.5   &0.0755 &0.0970 &0.0835 &0.1195\\
$T_n^{SZ}, \alpha=0.01$   &0.0   &0.0085 &0.0140 &0.0095 &0.0120\\
                          &0.5   &0.0205 &0.0285 &0.0180 &0.0330\\
\hline
$PCvM_n, \alpha=0.10$     &0.0   &0.1110 &0.1160 &0.1010 &0.1180\\
                          &0.5   &0.2370 &0.3480 &0.4730 &0.6630\\
$PCvM_n, \alpha=0.05$     &0.0   &0.0470 &0.0560 &0.0690 &0.0510\\
                          &0.5   &0.1310 &0.2000 &0.2760 &0.4450\\
$PCvM_n, \alpha=0.01$     &0.0   &0.0070 &0.0100 &0.0240 &0.0100\\
                          &0.5   &0.0430 &0.0580 &0.0930 &0.1700\\
\hline
$ICM_n, \alpha=0.10$      &0.0   &0.0200 &0.0000 &0.0000 &0.0000\\
                          &0.5   &0.0980 &0.0140 &0.0030 &0.0020\\
$ICM_n, \alpha=0.05$      &0.0   &0.0050 &0.0000 &0.0000 &0.0000\\
                          &0.5   &0.0210 &0.0020 &0.0000 &0.0000\\
$ICM_n, \alpha=0.01$      &0.0   &0.0000 &0.0000 &0.0000 &0.0000\\
                          &0.5   &0.0000 &0.0000 &0.0000 &0.0000\\
\hline
$T_n^{ZH}, \alpha=0.10$   &0.0   &0.0940 &0.0915 &0.0985 &0.1135\\
                          &0.5   &0.1325 &0.1455 &0.1625 &0.1455\\
$T_n^{ZH}, \alpha=0.05$   &0.0   &0.0445 &0.0365 &0.0410 &0.0380\\
                          &0.5   &0.0690 &0.0765 &0.0770 &0.0545\\
$T_n^{ZH}, \alpha=0.01$   &0.0   &0.0050 &0.0035 &0.0020 &0.0020\\
                          &0.5   &0.0125 &0.0090 &0.0070 &0.0040\\
\hline
$T_n^{GWZ}, \alpha=0.10$  &0.0   &0.1015 &0.1020 &0.0995 &0.1125\\
                          &0.5   &0.2380 &0.3745 &0.5450 &0.8265\\
$T_n^{GWZ}, \alpha=0.05$  &0.0   &0.0615 &0.0675 &0.0670 &0.0580\\
                          &0.5   &0.1700 &0.2750 &0.4560 &0.7725\\
$T_n^{GWZ}, \alpha=0.01$  &0.0   &0.0240 &0.0270 &0.0290 &0.0335\\
                          &0.5   &0.1015 &0.1655 &0.3360 &0.6260\\
\hline
\end{tabular}
}
\end{table}

\begin{figure}
\centering
  \includegraphics[width=14cm,height=18cm]{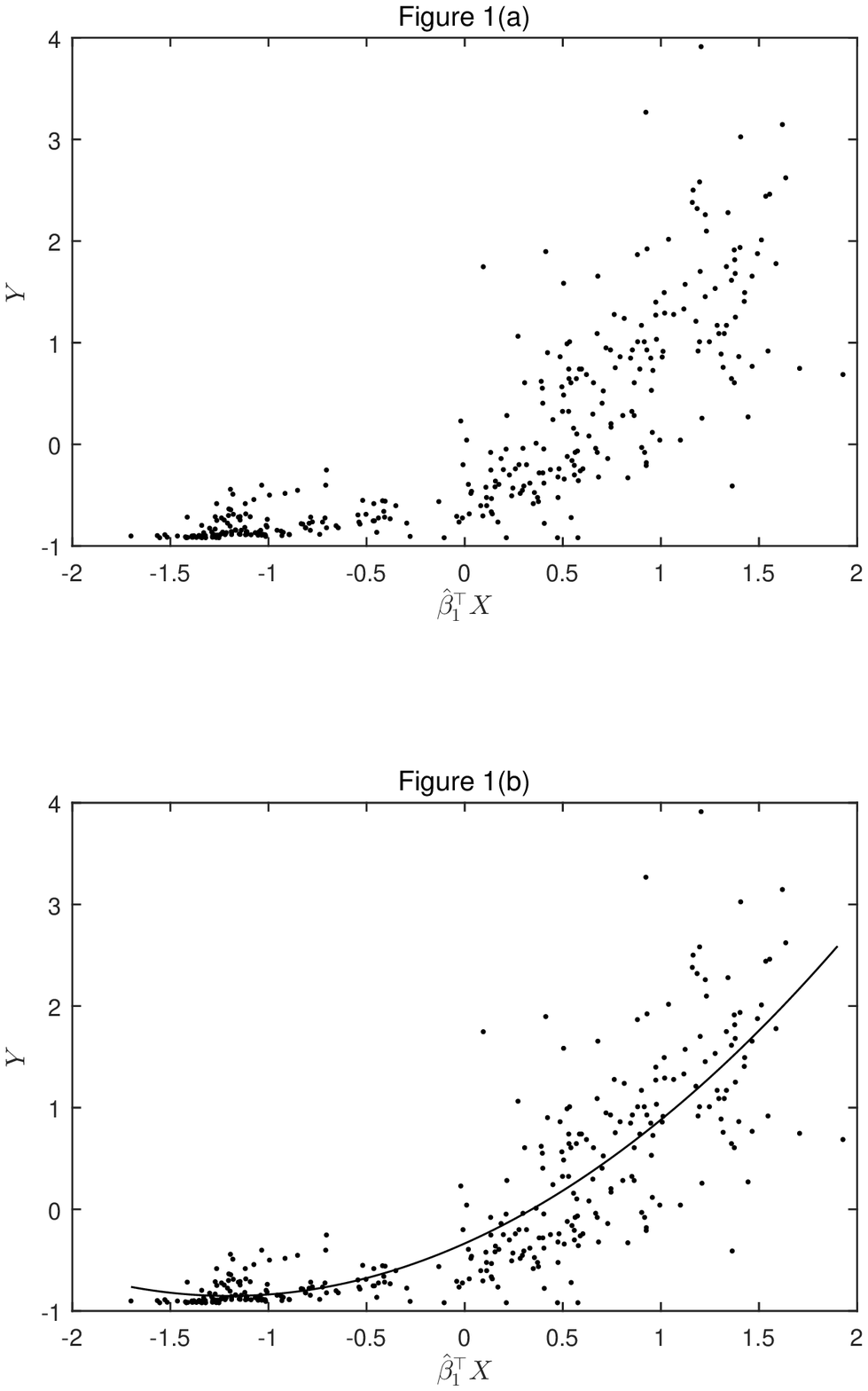}
    \caption{Scatter plots of the response $Y$ against the projected covariate $\hat{\beta}_1^{\top}X$ and the fitted quadratic polynomial curve where the direction $\hat{\beta}_1$ is obtained by CSE.}\label{Figure 1}
\end{figure}

\begin{figure}
\centering
  \includegraphics[width=14cm,height=12cm]{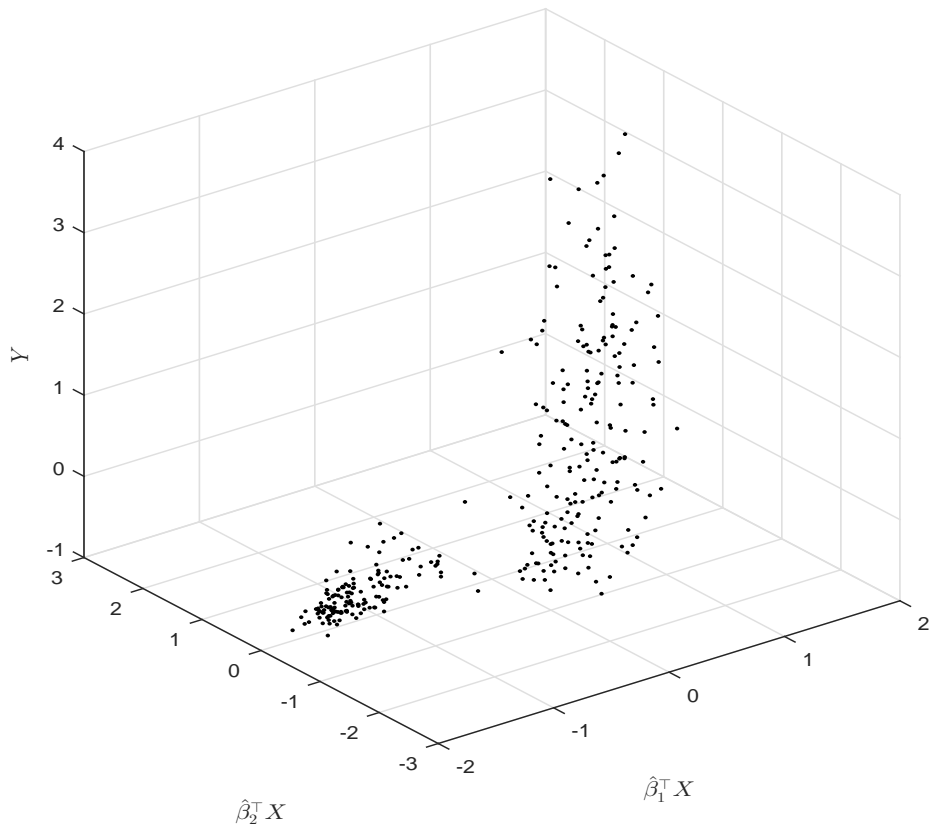}
    \caption{Scatter plot of the response $Y$ against the projected covariates $(\hat{\beta}_1^{\top}X, \hat{\beta}_2^{\top}X)$ where the directions $(\hat{\beta}_1, \hat{\beta}_2) $ are obtained by CSE.}\label{Figure 2}
\end{figure}


\end{document}